# Microscopic Theory for Multiple Light Scattering in Magnetic Fields

Bart A. van Tiggelen[1], Roger Maynard[1] and Theo M. Nieuwenhuizen[2]
[1] *Laboratoire de Physique Numérique des Systèmes Complexes,*
*Université Joseph Fourier/CNRS,*
*B.P. 166, 38042 Grenoble Cedex 09, France.*
[2] *Van der Waals-Zeeman Institute,*
*Valckenierstraat 65 - 67, 1092 AN Amsterdam, the Netherlands.*

## Abstract

We present a microscopic theory for multiple light scattering occurring in inhomogeneous 3D media subject to an external magnetic field. Magneto-optical effects (the Faraday effect and the Cotton-Mouton effect) occur inside the small inhomogeneities. We thereby take into account the spatial anisotropy, time-reversal-symmetry breaking and birefringence caused by the magnetic field, and discuss the consequences for the diffusion tensor and the polarization characteristics of the diffuse light. We will frequently compare our findings to a similar phenomenon in dilute polyatomic gases: the Beenakker-Senftleben effect. Coherent Backscattering and the field-field correlator $\left\langle E_i\left(\mathbf{B}\right)\overline{E}_j\left(\mathbf{0}\right)\right\rangle$ are addressed, which have both been obtained experimentally by the group of Maret *etal.*. All modifications in transport theory due to the magnetic field exhibit a rather sensitive dependence on the scattering phase shift of the individual scatterers.

## I. PHYSICAL CONTEXT

Multiple scattering of light is a fascinating topic having much overlap with other branches of physics such as solid-state physics, atomic physics and astrophysics. Most interesting aspects are due to the persisting role of interference in multiple scattering of light, together with the relative ease to observe and manipulate it in controlled laboratory experiments. For electrons this is more difficult due to phase-destroying mechanisms such as electron-phonon coupling that are not so easy to eliminate. The interference in multiple light scattering manifests itself in - by now rather well understood - phenomena as Coherent Backscattering [1] and universal fluctuations [2]. Other applications, such as localization of light, are still part of speculation.

One perhaps unexpected conclusion is that interference in multiple light scattering is in fact very difficult to suppress. Experimentalists spend hard times in undoing the effect of speckles in light measurements and extracting an average value of some transport variable. Absorption is known to suppress multiple light scattering but not the relative importance of interference. A finite coherence length of incident light does not destroy all interference and some experiments, like Coherent Backscattering can even be done with sunlight.

The only known way to manipulate interference externally is by applying a magnetic field. In the solid state a magnetic field breaks the charge symmetry between holes and electrons, destroying the Weak Localization correction and giving rise to a negative magneto-resistance. In the case of light, magneto-optical effects in the dielectric constant have a similar impact on Coherent Backscattering of light. The physics of the latter, constructive interference of time-reversed waves, is sometimes believed to be similar to the Weak Localization of electrons [3]. The dominant magneto-optical effect is the Faraday effect. Mathematically it is due to an anti-symmetric term in the dielectric tensor linear in the external field vector $\mathbf{B}$ [4]. In a homogeneous medium it is well known that this causes a rotation of the polarization vector of a linearly polarized plane wave proportional to traversed length and magnetic field. Equivalently, two opposite circularly polarized light beams achieve different speeds. As such it is often a mean to determine interstellar magnetic fields.

The study of the Faraday effect in inhomogeneous media is much more recent. The inclusion of Faraday rotation into a transport equation of light has first been carried out by plasma astrophysicists [5] in order to understand the combination of emission and absorption of polarized synchrotron radiation in optically thick ionized plasmas. However, in that case the light scattering is inelastic. In sharp contast to the light processes in plasmas, elastic scattering in disordered dielectric media conserves the phase so that interference can take place. After initiating work by Golubenstev [6], MacKintosh & John [7] found, in their rather profound study, that the Coherent Backscattering of light will be suppressed by the time-reversal symmetry breaking character of the Faraday effect. As has been shown by Martinez and Maynard [8] [9], elastic multiple light scattering from inhomogeneities in a dielectric background with Faraday active material is - under some approximations - extremely suited for numerical simulation. These simulations also work for the cases of higher magnetic fields and the inhomogeneities replaced by genuine Mie spheres, both of them being difficult to attack analytically.

Experimentally the Coherent Backscattering of light in magnetic fields is studied extensively by Erbacher, Lenke and Maret [10] [11]. Some of these results will be discussed from a theoretical point of view in section V. In these experiments it is difficult to separate between "medium" and "scatterers". Nevertheless, roughly speaking, they can be divided into two kinds. In some of them the magneto-optical effects occur in the medium and not in the immersed particles. This is the situation envisaged in the theory of MacKintosh & John and the numerical simulations of Martinez & Maynard. In other work, the magneto-optical effects occur inside the particles and not in the surrounding medium. For this case no theory exists and the present paper aims to fill this gap. We will show that this case allows - in the conventional Boltzmann approximation of transport theory - an analytical solution incorporating all aspects of the magneto-optical effects in multiple scattering of light. Most of them, like the Cotton-Mouton effect and the anisotropic diffusion tensor, have not been discussed before.

In pursuing analogies with other branches of physics we wish to mention the Beenakker-Senftleben effect in dilute gases [12] [13] [14]. This effect embodies the field-dependent and anisotropic nature of transport coefficients for paramagnetic and even diamagnetic polyatomic gases in an external magnetic field. This is caused by the precession of the magnetic moment of the molecules between the collisions. This rotation tends to average out the nonspherical collision cross-section between the molecules, and thereby lowers transport properties like viscosity and heat conductivity. The magnetic influence on the scattering cross-section itself is known not to be responsible for the Beenakker-Senftleben effect, since it can be shown to be negligible [15].

In our case it is the electric polarization vector that rotates between the collisions of the light with the dielectric particles. If one assumes that the cross-section is unaffected by the magnetic field indeed a great resemblance with the Beenakker-Senftleben effect turns up. One property of this phenomenon, the presence of a one-parameter scaling variable $\omega_p \tau$ ($\omega_p$ the precession frequency and $\tau$ the mean free collision time), has also been seen in multiple light scattering experiments [11]. Although some differences will nevertheless show up, we will demonstrate that the basic transport variable for multiple light scattering - the diffusion constant - becomes anisotropic in a magnetic field, quite similarly as in the Beenakker-Senftleben effect. Even an *anti-symmetric* part of the diffusion constant appears, generating a transverse current, perpendicular to both density gradient and magnetic field [16]. An anisotropic diffusion tensor (signifying a transverse current) is best known for conduction electrons and ionized plasmas [5], where the Lorentz force gives rise to the Hall-conductivity (and a less well-known transverse heat conductivity called the Righi-Leduc effect). On the other hand, in previous treatments of magneto-optical effects, the direction of the magnetic field was not noticed to play a role in the diffusive regime.

The outline of the paper is as follows. In the next section we derive the scattering properties of one magneto-optical particle and define most of our notation. In section III we obtain the average electric field amplitude and find the dispersion relation $\omega(\mathbf{k}, \sigma)$ of a coherent beam with wave vector $\mathbf{k}$ and helicity $\sigma$.

In section IV we calculate the incoherent energy density by summing - following conventional techniques - the Ladder diagrams in the Boltzmann approximation. In the diffusion approximation we obtain the diffusion tensor for the light in the magneto-optical medium. The associated eigenfunction - the "Goldstone mode" - turns out to be anisotropic in polarization space. The translation of bulk results to finite media is crucial to compare to experiments. It will turn out that role of the "skin layers" [17] for incoming and outgoing light is very important. Due to the anomalous step length distribution in these skin layers, the Stokes parameters in diffuse transmission will not be equal to zero in a magnetic field. We will show that the magnetic field can induce oscillations in the Stokes parameters as a function of the slab thickness and the field strength. Similar oscillations have been reported on the base of numerical work [8], but have not yet been observed.

In section V we will discuss Coherent Backscattering by considering the most-crossed diagrams.

We find the dephasing length due to the change of reciprocity relations (the Faraday effect). The absence of the usual reciprocity also causes the diffusion constant featuring in the most-crossed diagrams to be different from the one in the Ladder diagrams, as is also known to be true for electrons in the solid state [18] [19]. Attention will be paid to polarization properties of Coherent Backscattering, as well as to the anisotropy of the line shape for various directions of the magnetic field. Again, we shall rely on a strict diffusion approximation, and ignore (eight) non-diffusive eigenvalues of the most-crossed diagrams. So far, this has also been done in the interpretational analyses of the data [11]. Let us recall that the angular behavior in the center of the backscattering peak is due to long-range diffusion only.

In section VI the correlation function $\left\langle E_i\left(\mathbf{B}\right)\overline{E}_j\left(\mathbf{0}\right)\right\rangle$ is considered, that is the correlation function of the electric field with and without external magnetic field (in transmission). In the $C_1$ approximation this correlation function is related to the intensity correlator $\left\langle I\left(0\right)I\left(\mathbf{B}\right)\right\rangle$ which is experimentally obtainable [10]. We find a difference in phase between $E\left(0\right)$ and $E\left(\mathbf{B}\right)$ (showing up as a complex eigenvalue near zero) giving rise to a phase factor in the field correlation function itself. The field correlation function $\left\langle E_i\left(\mathbf{B}\right)\overline{E}_j\left(\mathbf{0}\right)\right\rangle$ has not yet been measured but recent advances in heterodyne detection methods may facilitate this.

Although the model is perhaps too simplistic to provide a quantitative theoretical picture for present experiments, it is the simplest approach starting directly from Maxwell's equations. Along the way, no approximations other than the usual ones will be made (independent scattering approximation, diffusion approximation). Therefore we hope that this paper will guide to interpret existing data qualitatively and perhaps will even initiate future experiments.

We want to draw attention to Appendix A where we define our tensor notation.

## II. SINGLE SCATTERING

In this section we derive the scattering properties of one magneto-optical dielectric scatterer located in vacuum. By taking the scatterer *pointlike* we shall be able to obtain these scattering properties in analytic form so that they can be used for multiple scattering in principle without further approximations [20] [21]. In this model we are even able to add a scattering resonance, being characterized by a relatively large cross-section, together with a delay and out-of-phase response of the scattered wave. The phase function of this model is only influenced by anisotropy in polarization indices caused by the magnetic field, and not by anisotropy in wavenumbers as common for Mie scatterers.

Throughout this paper we set $c_0 = 1/\sqrt{\varepsilon_0} = 1$. The refractive index - a tensor of rank two - of a particle in an external magnetic field is assumed to be given by,

$$\mathbf{n}\left(\mathbf{B}\right) = m\mathbf{I}+\frac{V_0}{\omega}\mathbf{\Phi} + M\mathbf{BB} \tag{1}$$

In this formula is $m$ the normal isotropic index of refraction, $V_0$ is the Verdet constant of the Faraday effect (we added a frequency factor $\omega$ to be consistent with the experimental definition and to give it the proper dimension) which is described by the anti symmetric hermitean tensor $\Phi_{ij}=i\epsilon_{ijk}B_k$. The third term is uniaxial birefringent and associated with the Cotton-Mouton effect [4]. The dielectric tensor $\boldsymbol{\varepsilon} = \mathbf{n}^2$ is given by,

$$\boldsymbol{\varepsilon}\left(\mathbf{B}\right) = \left(m^2 + \frac{V_0^2 B^2}{\omega^2}\right)\mathbf{I} + \frac{2mV_0}{\omega}\mathbf{\Phi} + \left(2mM - \frac{V_0^2}{\omega^2} + M^2B^2\right)\mathbf{BB} \equiv \varepsilon\mathbf{I} + \varepsilon_F\mathbf{\Phi} + \varepsilon_M\mathbf{BB} \tag{2}$$

In typical optical experiments [11] at room temperature the Verdet constant is 90°/mm/T. In a field of 15 T the relative magneto-optical perturbation of the dielectric constant $V_0 B/\omega \simeq 0.002$ for one single scatterer is still small. It seems therefore rather safe to ignore orders higher than $B^3$ in the dielectric tensor. Both the Faraday (order $B$) and the Cotton-Mouton effect (order $B^2$) remain present and no physics is lost. Note, however, that in multiple (and even resonant single ) scattering the magneto-optical effects may augment since they are basically proportional to the length of the traversed path.

Two properties of the dielectric tensor are of fundamental importance for the rest of this paper. First, when all material parameters $\varepsilon$, $V_0$ and $M$ are real-valued, the dielectric tensor is hermitean: $\boldsymbol{\varepsilon}(\mathbf{B}) = \boldsymbol{\varepsilon}(\mathbf{B})^*$. This is crucial for multiple scattering since it will finally guarantee long-range diffusion. The second property is caused by the anti symmetric term only: $\boldsymbol{\varepsilon}(\mathbf{B}) = \boldsymbol{\varepsilon}(-\mathbf{B})^t \neq \boldsymbol{\varepsilon}(\mathbf{B})^t$. This property destroys time-reversal symmetry and modifies reciprocity relations. As such it is quite similar to the Lorentz force in electronic systems where the symmetry between electrons and holes will be broken.

A point scatterer with dielectric constant $\boldsymbol{\varepsilon}(\mathbf{B})$ located at position $\mathbf{r}$ in vacuum is characterized by the hermitean operator,

$$\boldsymbol{\varepsilon}(\mathbf{B}, \mathbf{r}) = \mathbf{I} + v\left[\boldsymbol{\varepsilon}(\mathbf{B}) - 1\right]\; |\mathbf{r}\rangle\langle\mathbf{r}| \;, \tag{3}$$

where $v$ can be interpreted as a typical physical volume associated with the point scatterer. The abstract Dirac notation $|\mathbf{r}\rangle\langle\mathbf{r}|$ is used for a local interaction with the profile of a delta-distribution $\delta(\mathbf{r})$. The $t$-matrix of this particle is most conveniently obtained by noting that the Helmholtz equation for the electric field at frequency $\omega$ resembles a Schrödinger equation with potential $\mathbf{V}(\mathbf{r},\omega) = [\mathbf{I} - \varepsilon(\mathbf{B}, \mathbf{r})]\,\omega^2$ and energy $\omega^2$. As a result the $t$-operator - here also a second rank tensor - is given by the Born series,

$$\mathbf{t}(\mathbf{B},\omega) = \mathbf{V}(\omega) + \mathbf{V}(\omega)\cdot\mathbf{G}_0(\omega, \mathbf{p})\cdot\mathbf{V}(\omega) + \cdots \;.$$

Here $\mathbf{G}_0(\omega, \mathbf{p})$ is the free Helmholtz Green's function $\mathbf{G}_0(\omega, \mathbf{p}) = 1/\left[\omega^2 - p^2\boldsymbol{\Delta}_\mathbf{p}\right]$, with $(\boldsymbol{\Delta}_\mathbf{p})_{ij} = \delta_{ij} - p_i p_j/p^2$. For a point scatterer the Born series can be transformed into an ordinary geometric series [21]. Using the matrix identity ($|\mathbf{v}| = 1$),

$$\frac{1}{P\mathbf{I} + Q\,(i\boldsymbol{\epsilon}\cdot\mathbf{v}) + R\mathbf{v}\mathbf{v}} = \frac{\mathbf{v}\mathbf{v}}{P + R} + \frac{(\mathbf{I} - \mathbf{v}\mathbf{v})\,R - Q\,(i\boldsymbol{\epsilon}\cdot\mathbf{v})}{P^2 - Q^2} \;,$$

it follows that,

$$\mathbf{t}_{\mathbf{p}\mathbf{p}'}(\mathbf{B},\omega) = (t_0 - t_3)\,\mathbf{I} + t_1\boldsymbol{\Phi} + t_2\mathbf{B}\mathbf{B} \;, \tag{4}$$

in which $t_0$ is the normal Rayleigh $t$-matrix with phase-shift $\alpha(\omega)$ ,

$$t_0(\omega) = \frac{-4\pi\Gamma\omega^2}{\omega_0^2 - \omega^2 - \frac{2}{3}i\Gamma\omega^3} \equiv -\frac{6\pi}{\omega}\exp(i\alpha)\sin\alpha \;. \tag{5}$$

At low frequencies the phase shift goes to zero proportional to $\omega^3$. Near the resonance $\omega_0$ the phase shift changes rapidly from $-\pi/2$ towards $\pi/2$. Up to orders $B^2$ the other $t$-matrices are given by,

$$\begin{aligned} t_1(\omega, B) &= -\frac{\omega}{6\pi}\mu t_0^2 \\ t_2(\omega, B) &= \frac{\omega}{6\pi}(\varpi - \zeta)\,t_0^2 - \left(\frac{\omega}{6\pi}\right)^2\mu^2 t_0^3 \\ t_3(\omega, B) &= -\frac{\omega}{6\pi}\zeta t_0^2 - \left(\frac{\omega}{6\pi}\right)^2\mu^2 t_0^3 \end{aligned} \tag{6}$$

For future use we define the dimensionless variables $\eta_i \equiv t_i/(-\mathrm{Im}\, t_0)$ and $\xi_i \equiv t_i/t_0$. In Eq. (6) we introduced three other dimensionless variables which will be used in the rest of this paper,

$$\mu \equiv \frac{6\pi\varepsilon_F B}{(\varepsilon-1)^2\omega^3 v} \;;\zeta \equiv \frac{6\pi\varepsilon_F^2 B^2}{(\varepsilon-1)^3\omega^3 v} \;;\varpi \equiv -\frac{6\pi\varepsilon_M B^2}{(\varepsilon-1)^2\omega^3 v}\ . \tag{7}$$

The absolute value of the magnetic field has been included here and from now one we denote by $\mathbf{B}$ a normalized field with $|\mathbf{B}| = 1$. The dimensionless variables $\mu$, $\zeta$ and $\varpi$ will finally determine the role of magneto-optical effects in multiple scattering. Their magnitude is discussed separately in Appendix C. The variable $\mu$ can be argued to be the dominant one, and essentially equal to the variable $q_F\ell \sim V_{\mathrm{eff}} B\ell$ ($\ell$ is some mean free path) introduced by Lenke and Maret [11]. Since the Faraday effect in our case occurs inside the particles, the effective medium Verdet constant is proportional to the number density of the particles. For that reason $q_F\ell$ is independent on the number density (in the dilute regime).

The $t$-matrix does not depend on incoming and outgoing wavenumbers due to its point like origin. In this case we encounter anisotropy only in polarization indices and not in wavenumbers. In a magnetic field the resonant scattering matrix mimics the classical Lorentz model for the atomic, diamagnetic origin of the Faraday effect [11].

The Optical Theorem is an identity for the $t$-matrix that follows from energy conservation and is a mathematical consequence of $\boldsymbol{\varepsilon} = \boldsymbol{\varepsilon}^*$. For our point object it is readily checked that,

$$\begin{aligned} \mathbf{t}(\mathbf{B},\omega) - \mathbf{t}^*(\mathbf{B},\omega) &= \mathbf{t}(\mathbf{B},\omega)\cdot\sum_{\mathbf{p}} \Delta\mathbf{G}_0(\omega,\mathbf{p})\cdot\mathbf{t}^*(\mathbf{B},\omega) \\ &= \frac{\omega}{3\pi i}\,\mathbf{t}(\mathbf{B},\omega)\cdot\mathbf{t}^*(\mathbf{B},\omega)\ . \end{aligned} \tag{8}$$

In this paper we will frequently denote the anti-hermitean second-rank tensor $\mathbf{A} - \mathbf{A}^*$ by $\Delta\mathbf{A}$. The summation sign $\sum_{\mathbf{p}}$ is a short notation for $\int \mathrm{d}^3\mathbf{p}/(2\pi)^3$.

## III. THE AVERAGE GREEN'S FUNCTION

In this section we consider an infinite medium randomly filled (number density $n$) with the magneto-optical scatterers discussed in the previous section. The average field is determined by the average (Dyson) Green's function which - by translation symmetry - takes the form in momentum space,

$$\left\langle \frac{1}{\varepsilon(\mathbf{B},\mathbf{r})\,\omega^2 - p^2\boldsymbol{\Delta}_{\mathbf{p}}} \right\rangle_{\mathbf{pp}'} = \frac{\delta(\mathbf{p}-\mathbf{p}')}{\omega^2 - p^2\boldsymbol{\Delta}_{\mathbf{p}} - \boldsymbol{\Sigma}(\omega,\mathbf{p},\mathbf{B})} \equiv \mathbf{G}(\mathbf{p},\mathbf{B})\,\delta(\mathbf{p}-\mathbf{p}')\ . \tag{9}$$

In the independent scattering approximation the self-energy $\boldsymbol{\Sigma}(\omega,\mathbf{p},\mathbf{B})$ is given by $n\mathbf{t}_{\mathbf{pp}}(\omega,\mathbf{B})$. We note that the tensors $\boldsymbol{\Sigma}(\omega,\mathbf{p},\mathbf{B})$ and $\boldsymbol{\Delta}_{\mathbf{p}}$ do not commute. Using Eqs. (9) and (4) a rather lengthy tensor analysis reveals that,

$$\mathbf{G}(\mathbf{p},\mathbf{B}) = \frac{1}{a^2 - \Sigma_1^2}\left[a + \Sigma_1\boldsymbol{\Phi} + \frac{\mathbf{R}(\mathbf{B},\mathbf{p}) + \mathbf{S}(\mathbf{B},\mathbf{p})}{\det \mathbf{G}^{-1}(\mathbf{B},\mathbf{p})}\right]\ , \tag{10}$$

with

$$\mathbf{R}(\mathbf{B},\mathbf{p}) = -\left(\Sigma_1^2 - a\Sigma_2\right)\left(a^2 + ap^2 - \Sigma_1^2\right)\mathbf{BB} + a\left(\Sigma_1^2 - a\Sigma_2\right)(\mathbf{B}\cdot\mathbf{p})(\mathbf{pB}+\mathbf{Bp})$$
$$-a^2(a-\Sigma_2)\,\mathbf{pp} - a\Sigma_1(a-\Sigma_2)\left[(\mathbf{\Phi}\cdot\mathbf{p})\,\mathbf{p} - \mathbf{p}(\mathbf{\Phi}\cdot\mathbf{p})\right] + \Sigma_1^2(a-\Sigma_2)(\mathbf{\Phi}\cdot\mathbf{p})(\mathbf{\Phi}\cdot\mathbf{p})\ ,$$
$$\mathbf{S}(\mathbf{B},\mathbf{p}) = -\Sigma_1\left(\Sigma_1^2 - a\Sigma_2\right)(\mathbf{B}\cdot\mathbf{p})\left[\mathbf{B}(\mathbf{\Phi}\cdot\mathbf{p}) - (\mathbf{\Phi}\cdot\mathbf{p})\,\mathbf{B}\right]\ ,$$
$$\det \mathbf{G}^{-1}(\mathbf{B},\mathbf{p}) = (a-\Sigma_2)\left(a^2 + ap^2 - \Sigma_1^2\right) - (\mathbf{B}\cdot\mathbf{p})^2\left(\Sigma_1^2 - a\Sigma_2\right)\ . \tag{11}$$

We abbreviated

$$a = \omega^2 - p^2 - (\Sigma_0 - \Sigma_3)\ ,$$

and defined the various components of the self energy as $\Sigma_i \equiv nt_i$ where the $t_i$ have been defined in Eq. (6)

The complex dispersion law is determined by the poles of the Green's function, that is

$$\det \mathbf{G}^{-1}(\mathbf{B},\mathbf{p}) = 0\ , \tag{12}$$

usually called Fresnel's equation [4]. With the explicit expression for the determinant given above we find two solutions in the upper sheet[1] . If we write $z^2 = \omega^2 - \Sigma_0 + \Sigma_3$ and $\cos\theta = \mathbf{B}\cdot\widehat{\mathbf{p}}$, these are explicitly,

$$z^2 - \mathbf{p}_\pm^2 = \frac{\left(\Sigma_1^2 + z^2\Sigma_2\right)\sin^2\theta \pm \sqrt{\left(\Sigma_1^2 + z^2\Sigma_2\right)^2\sin^4\theta + 4\Sigma_1^2\left(z^2 - \Sigma_2\cos^2\theta\right)\left(z^2\cos^2\theta - \Sigma_2\right)}}{2\left(z^2 - \Sigma_2\cos^2\theta\right)}$$
$$= \frac{1}{2}\Sigma_2\sin^2\theta \pm \sqrt{\frac{1}{4}\Sigma_2^2\sin^4\theta + \Sigma_1^2\cos^2\theta} + \mathcal{O}\left(n^2\right)$$
$$= \pm\Sigma_1(\mathbf{B}\cdot\widehat{\mathbf{p}}) + \mathcal{O}\left(B^2, n^2\right)\ . \tag{13}$$

In the third equality all birefringent terms are ignored and the propagating modes obey a dispersion relation similar as found in Ref. [7]. However, for directions nearly perpendicular to the magnetic field or for larger fields, birefringence cannot be ignored and the modes achieve an "ordinary" and "extra-ordinary" character. We emphasize that $\Sigma_1$ is a complex number and thus modifies both wavenumber $k$ and scattering mean free path $\ell$ according to $p_\pm = k_\pm + i/2\ell_\pm$. In former treatments [6] [7] [8] the difference between $\ell_+$ and $\ell_-$ caused by the Faraday effect (even without Cotton-Mouton birefringence) has not been taken into account. This will cause many different results compared to previous work. We show the dispersion law for the low-frequency case, Fig. 1, and the resonant situation, Fig. 2. In the low-frequency regime the difference in wavenumber between the two modes is very small, and the Faraday rotation is almost entirely contained in the mean free path. As evident from Fig. 3 the accumulated phase shift between two successive collisions can nevertheless be large, because the mean free path is large at low frequencies.

The following exact identity can be derived directly from the Dyson equation (9),

$$\mathbf{G}\left(\mathbf{p}+\frac{\mathbf{q}}{2}\right) - \mathbf{G}^*\left(\mathbf{p}-\frac{\mathbf{q}}{2}\right) = \mathbf{G}\left(\mathbf{p}+\frac{\mathbf{q}}{2}\right)\cdot\left[\Delta\mathbf{\Sigma} + 2\mathbf{p}\cdot\mathbf{q} - \mathbf{pq} - \mathbf{qp}\right]\cdot\mathbf{G}^*\left(\mathbf{p}-\frac{\mathbf{q}}{2}\right)\ .$$

Using the more compact Liouville notation (Appendix A) the becomes,

---

[1]Frequencies for which $z^2 - \Sigma_2\cos^2\theta = 0$ need special attention. They have one propagating solution given by $a \sim \Sigma_1^2/z^2$ and one non-propagating solution $z^2 = \Sigma_2(z)$, with $\widehat{\mathbf{p}}$ along $\mathbf{B}$.

$$\mathbf{G}\left(\omega+\frac{\Omega}{2},\mathbf{p}+\frac{\mathbf{q}}{2}\right)-\mathbf{G}^*\left(\omega-\frac{\Omega}{2},\mathbf{p}-\frac{\mathbf{q}}{2}\right)=\left\{\mathbf{G}\left(\omega+\frac{\Omega}{2},\mathbf{p}+\frac{\mathbf{q}}{2}\right)\mathbf{G}^*\left(\omega-\frac{\Omega}{2},\mathbf{p}-\frac{\mathbf{q}}{2}\right)\right\}\cdot$$
$$\cdot\left[\Delta\mathbf{\Sigma}+2\mathbf{p}\cdot\mathbf{q}-\mathbf{p}\mathbf{q}-\mathbf{q}\mathbf{p}-2\omega\Omega+\mathcal{O}(n\Omega)\right]\;. \tag{14}$$

in which $\{\mathbf{AB}\}_{ijkl}$ is considered as the four-rank tensor $A_{ij}B_{lk}$. Identity (14) has also been generalized for a finite frequency $\Omega$ difference between the two Green's tensors on the lefthand side. This becomes necessary to discuss dynamics in multiple scattering.

## IV. THE AVERAGE INTENSITY

The propagation of the average intensity - for electromagnetic waves more generally the complete set of nine Stokes correlations $\left\langle E_i\overline{E}_j\right\rangle$ - is characterized by the Ladder diagrams (Fig. 4). In the Boltzmann approximation the Ladder diagrams are formally equivalent to the solution of the Bethe-Salpeter transport equation. If the $t$-matrices do not depend on incoming and outgoing momenta (as is true for our case), the Ladder diagrams for an infinite medium generate a simple geometric series in the four-rank tensor,

$$\mathcal{Q}^{\pm}(\omega,\mathbf{q},\mathbf{B},\Omega)=\sum_{\mathbf{p}}\mathbf{G}\left(\omega+\frac{\Omega}{2},\mathbf{p}+\frac{\mathbf{q}}{2},\mathbf{B}\right)\mathbf{G}^*\left(\omega-\frac{\Omega}{2},\mathbf{p}-\frac{\mathbf{q}}{2},\pm\mathbf{B}\right)\cdot n\;\mathbf{t}\left(\omega+\frac{\Omega}{2},\mathbf{B}\right)\mathbf{t}^*\left(\omega-\frac{\Omega}{2},\pm\mathbf{B}\right)\;. \tag{15}$$

Explicitly,

$$\mathcal{Q}^{\pm}_{ijkl}(\omega,\mathbf{q},\mathbf{B},\Omega)=\sum_{\mathbf{p}}G_{is}\left(\omega+\frac{\Omega}{2},\mathbf{p}+\frac{\mathbf{q}}{2},\mathbf{B}\right)(G^*)_{tk}\left(\omega-\frac{\Omega}{2},\mathbf{p}-\frac{\mathbf{q}}{2},\pm\mathbf{B}\right)$$
$$nt_{sj}\left(\omega+\frac{\Omega}{2},\mathbf{B}\right)(t^*)_{lt}\left(\omega-\frac{\Omega}{2},\pm\mathbf{B}\right)\;. \tag{16}$$

For later use, the bottom line of the Ladder diagram is allowed to have reversed direction of the magnetic field. The Ladder diagrams sum up to,

$$\mathcal{L}^{\pm}(\omega,\mathbf{q},\mathbf{B},\Omega)=n\;\mathbf{t}\left(\omega+\frac{\Omega}{2},\mathbf{B}\right)\mathbf{t}^*\left(\omega-\frac{\Omega}{2},\pm\mathbf{B}\right)\cdot\frac{1}{1-\mathcal{Q}^{\pm}(\omega,\mathbf{q},\mathbf{B},\Omega)}\;. \tag{17}$$

Usually, this can be worked out by decomposing $\mathcal{Q}(\Omega,\mathbf{q})=\mathcal{Q}_0+\delta\mathcal{Q}(\Omega,\mathbf{q})$ into the various eigenfunctions and eigenvalues of $\mathcal{Q}_0$ thereby treating $\delta\mathcal{Q}(\Omega,\mathbf{q})$ as a Rayleigh-Schrödinger perturbation [7] [22]. We want to note here that, due to the magnetic field, $\mathcal{Q}_0$ is not a normal operator and does therefore not necessarily allow an orthogonal set of eigenfunctions. For this reason Rayleigh-Schrödinger perturbation theory cannot simply be applied here. In fact, the tensor $\mathcal{Q}_0$ is not necessarily diagonalizable, and at most isomorphic to a $9\times 9$ Gauss-Jordan matrix. If we let $\Delta\mathbf{G}\equiv\Delta\mathbf{G}(\mathbf{r}=0)=\sum_{\mathbf{p}}\Delta\mathbf{G}(\mathbf{p})$ then Eqs. (8) and (14) imply that,

$$\mathcal{Q}_0^+\cdot|\Delta\mathbf{G}\rangle=|\Delta\mathbf{G}\rangle\;,$$
$$\langle\Delta\mathbf{\Sigma}|\cdot\mathcal{Q}_0^+=\langle\Delta\mathbf{\Sigma}|\;. \tag{18}$$

Since in a magnetic field $\Delta\mathbf{\Sigma}\neq\mathbf{\Delta G}$, this means that $\mathcal{Q}_0^+$ has - on the basis of energy conservation - an eigenvalue 1 with geometric multiplicity 1 but with algebraic multiplicity 2. In our case the magneto-optical effects only occur in the particles. As a result the “effective medium” described by $\Delta\mathbf{G}$ is only influenced indirectly and $\Delta\mathbf{G}_{ik}= (\omega/3\pi i)\,\delta_{ik} + \mathcal{O}\,(n)$. It is convenient to normalize both eigenfunctions and write

$$
\begin{aligned}
|\mathbf{r}\rangle &= \frac{1}{\sqrt{3}}\,|\mathbf{I}\rangle \\
|\mathbf{l}\rangle &= \frac{i}{\|\Delta\mathbf{\Sigma}\|}\,|\Delta\mathbf{\Sigma}\rangle \\
&= \frac{1}{\sqrt{A}}\left[(1+\mathrm{Im}\ \eta_3)\,|\mathbf{I}\rangle - \mathrm{Im}\ \eta_1\,|\mathbf{\Phi}\rangle - \mathrm{Im}\ \eta_2\,|\mathbf{BB}\rangle\right]\ , \qquad (19)
\end{aligned}
$$

where $\eta_i$ was defined in the first section, and $A = 3\,(1+2\mathrm{Im}\ \eta_3)+2\,(\mathrm{Im}\ \eta_1)^2-2\mathrm{Im}\ \eta_2$; $\|\mathbf{C}\| \equiv \langle\mathbf{C}|\mathbf{C}\rangle^{1/2}$ denotes the norm of the matrix $\mathbf{C}$ (Appendix A).

The eigenvalue 1 of $\mathcal{Q}_0^+$ implies that $\mathcal{L}^+(\mathbf{q})$ will be singular in the hydrodynamic limit $\mathbf{q}\to\mathbf{0}$. This is the origin of long range diffusion. Due to the modification of reciprocity relations in a magnetic field, $\mathcal{Q}_0^-$ has no longer an eigenvalue 1 and - as it turns out later - neither an algebraic multiplicity 2. The tensor $\mathcal{Q}_0^-$ is intimately related to Coherent Backscattering and is in that sense also an "observable". It will be discussed in section IV.

In what follows we will focus on the diffusive 2D eigenspace of $\mathcal{Q}_0^+$ and ignore 7 other dimensions in polarization space in which the propagation is non-diffusive. Let $\mathbf{N}^+$ be the $2\times 2$ Gram matrix of these vectors, constructed from the scalar product defined in Appendix A. For simplicity we first set $\Omega = 0$, and consider the perturbation in the variable $\mathbf{q}$ only. To this end we let

$$
\mathbf{P}^+(\mathbf{q}) = \begin{pmatrix} \langle\mathbf{l}|\ \mathcal{Q}^+(\mathbf{q})\ |\mathbf{l}\rangle & \langle\mathbf{r}|\ \mathcal{Q}^+(\mathbf{q})\ |\mathbf{l}\rangle \\ \langle\mathbf{l}|\ \mathcal{Q}^+(\mathbf{q})\ |\mathbf{r}\rangle & \langle\mathbf{r}|\ \mathcal{Q}^+(\mathbf{q})\ |\mathbf{r}\rangle \end{pmatrix} \equiv \mathbf{N}^+ - \begin{pmatrix} \mathbf{q}\cdot\mathbf{D}_{11}\cdot\mathbf{q} & \kappa^+(B)+\mathbf{q}\cdot\mathbf{D}_{12}\cdot\mathbf{q} \\ \mathbf{q}\cdot\mathbf{D}_{21}\cdot\mathbf{q} & \mathbf{q}\cdot\mathbf{D}_{22}\cdot\mathbf{q} \end{pmatrix}\ ,
$$

in which

$$
\kappa^+(B) = \langle\mathbf{r}|\ 1-\mathcal{Q}_0^+\ |\mathbf{l}\rangle \sim B^2\ . \qquad (20)
$$

The Ladder tensor (17) can now be written as

$$
\mathcal{L}^+(\mathbf{q}) = n\mathbf{t}\mathbf{t}^*\cdot\left(\ |\mathbf{r}\rangle,\ \ |\mathbf{l}\rangle\ \right)\cdot\frac{1}{\mathbf{N}^+ - \mathbf{P}^+(\mathbf{q})}\cdot\begin{pmatrix}\langle\mathbf{r}| \\ \langle\mathbf{l}|\end{pmatrix}\ .
$$

The matrix $\mathbf{N}^+ - \mathbf{P}^+(\mathbf{q}=\mathbf{0})$ is obviously in Gauss-Jordan form. After inverting the matrix $\mathbf{N}^+ - \mathbf{P}^+(\mathbf{q})$, the diffusive mode can be found by looking for the divergent entry as $\mathbf{q}\to\mathbf{0}$ . To associate a diffusion tensor to the diffusive mode, dynamics has to introduced in the form of a Laplace frequency $\Omega$ as indicated in Fig. 4. (For $\Omega = 0$ the diffusion tensor can never be obtained). The dynamical expression can be obtained without explicit calculation using the tensor identity (14) for $\Omega \neq 0$ . The result is,

$$
\mathcal{L}^+(\mathbf{q},\Omega) = n\mathbf{t}\mathbf{t}^*\cdot\ |\mathbf{r}\rangle\ \frac{1}{-6i\omega\Omega\left(\|\Delta\mathbf{\Sigma}\|\sqrt{3}\right)^{-1} + \ \mathbf{q}\cdot\mathbf{D}_{21}\cdot\mathbf{q}}\ \langle\mathbf{l}| = \frac{3\pi}{\omega^2}\frac{\|\Delta\mathbf{\Sigma}\|^2}{6}\frac{|\mathbf{l}\rangle\langle\mathbf{l}|}{-i\Omega + \mathbf{q}\cdot\mathbf{D}_B^+(\mathbf{B})\cdot\mathbf{q}}\ , \qquad (21)
$$

where the Boltzmann diffusion tensor is defined as,

$$
\mathbf{q}\cdot\mathbf{D}_B^+(\mathbf{B})\cdot\mathbf{q} \equiv \frac{\|\Delta\mathbf{\Sigma}\|\sqrt{3}}{6\omega}\ \mathbf{q}\cdot\mathbf{D}_{21}\cdot\mathbf{q} = \frac{i}{6\omega}\ \langle\Delta\mathbf{\Sigma}|\ \delta\mathcal{Q}^+(\mathbf{q},\mathbf{B})\ |\mathbf{I}\rangle\ . \qquad (22)
$$

In the dynamic treatment we did not incorporate $\Omega$-expansions of scattering matrices $\mathbf{t}(\omega \pm \Omega/2)$. This becomes necessary if we want to be consistent with the correct conserved quantity [23] which is not $\left\langle E_i E_j^* \right\rangle$ as would follow from Eq. (21) but rather $\langle \varepsilon_{ik} E_k E_i^* \rangle$. This will modify the transport velocity in the diffusion tensor, which is necessarily a scalar. Therefore the diffusion tensor is here basically (one-third times) the "transport mean free path tensor". Keeping in mind this aspect, we will nevertheless continue to use the more common word "diffusion tensor". Expression (22) can be shown to be equivalent to a DC Kubo-type formula for the "electromagnetic" conductivity tensor (i.e. density of states times diffusion tensor), applied to the special case of point scatterers.

The lefthand eigenfunction $|\mathbf{\Delta\Sigma}\rangle$ must be a linear combination of the complete set of orthogonal eigenfunctions $\{|i\rangle\}_{i=1,\ldots,9}$ of the hermitean tensor $\mathcal{Q}_0\left(\mathbf{B}=\mathbf{0}, \mathbf{q}=\mathbf{0}, \Omega=0\right)$ obtained in Refs. [7] and [22]. A contribution of any of the last six of these will cause the diffusion tensor to be anisotropic. In fact $|\mathbf{\Phi}\rangle$ is equivalent to $|7\rangle$ and $|\mathbf{BB}\rangle$ is a linear combination of the first four (see also section IV.C.3.2).

It is convenient to consider the symmetry of the tensor $\mathcal{L}^{\pm}$ implied by reciprocity. Reciprocity of *both* wave and conjugate wave in the intensity requires the Ladder tensor $\mathcal{L}^{+}$ to obey, in general,

$$\mathcal{L}^{\pm}_{ijkl}\left(\mathbf{q}, \mathbf{B}, \Omega\right) = \mathcal{L}^{\pm}_{jilk}\left(-\mathbf{q}, -\mathbf{B}, \Omega\right) \ . \tag{23}$$

By applying this symmetry to the diffusion approximation found in (21) we conclude that the diffusive mode must satisfy $\bar{\mathbf{l}}(\mathbf{B}) = \mathbf{l}(-\mathbf{B})$ and the diffusion tensor $\mathbf{D}_B^{+}(\mathbf{B}) = \mathbf{D}_B^{+}(-\mathbf{B})$. The representation of the diffusive mode of $\mathcal{L}$ (identified as the lefthand eigenfunction) must be a linear combination of the tensors $\mathbf{I}, \mathbf{\Phi}$ and $\mathbf{BB}$, with *real-valued* coefficients. This will seen to be different for the diffusive mode of $\mathcal{L}^{-}$. The general form of the diffusion tensor is given by,

$$\mathbf{D}_B^{+}(\mathbf{B}) = D_{\mathrm{iso}}(B)\,\mathbf{I} + \ D_{\mathrm{ani}}(B)\,\mathbf{BB} \ . \tag{24}$$

The coefficients are only dependent on the absolute value of $\mathbf{B}$. We emphasize that the Ladder-diagram (21) only provides the *symmetric* part of the diffusion tensor. The *anti-symmetric* part is proportional to the tensor $-i\mathbf{\Phi}_{ij}{=}\epsilon_{ijk}B_k$ and responsible for a transverse Hall-type current, can be found only by looking directly at the current [16]. More generally, reciprocity guarantees the "Onsager relation" $\mathbf{D}_B^{+}(\mathbf{B}) = \left[\mathbf{D}_B^{+}(-\mathbf{B})\right]^{\mathrm{transpose}}$ which requires the anti-symmetric part to be odd in the magnetic field.

A reciprocity relation between $\mathcal{L}^{\pm}$ and $\mathcal{C}^{\pm}$ can be obtained by transposing only the conjugate wave. From $\boldsymbol{\varepsilon}(\mathbf{B}) = \boldsymbol{\varepsilon}^t(-\mathbf{B})$ it follows that,

$$\mathcal{C}^{\pm}_{ijkl}\left(\mathbf{p}+\mathbf{p}', \mathbf{B}, \Omega\right) = \mathcal{L}^{\mp}_{ijlk}\left(\mathbf{q}, \mathbf{B}, \Omega\right) \ , \tag{25}$$

where the tensor $\mathcal{C}^{\pm}$ is given by the sum of the most-crossed diagrams (Fig. 4). Thus this reciprocity relation relates two sets of topologically different diagrams.

The regime for which the diffusion approximation discussed here becomes valid can be characterized by the criterion,

$$\left|\det\left[\mathbf{N}^{+}{-}\mathbf{P}^{+}(\mathbf{q}, \mathbf{B})\right]\right| \ll \left|\kappa^{+}(B)\right| \ .$$

This is equivalent to $\mathbf{q}^2 \ll \left[1 + 2\left(\widehat{\mathbf{B}}\cdot\widehat{\mathbf{q}}\right)^2\right] \ell^2/5$ . A similar criterion holds to guarantee the irrelevance of all other modes [7].

In the next subsections we will calculate explicitly the four-rank tensor $\mathcal{Q}^{\pm}(\omega, \mathbf{q}, \mathbf{B})$ for our model and from that in particular the diffusion tensor. Next we translate the Ladder tensor to a slab geometry by adding boundary conditions. We will discuss polarization properties (Stokes parameters) due to the impact of the magnetic field on both the diffusive and the non-diffusive modes in transmission.

### A. Calculation of Ladder Tensor

To find the explicit form of the four-rank tensor $\mathcal{Q}^{\pm}(\omega, \mathbf{q}, \mathbf{B})$ defined in Eq. (15) one must perform momentum integrals over the product of two Green's tensors (9). The exact expression (10) is quite cumbersome for general multiple scattering calculations. In the Boltzmann approximation one considers only the "most divergent terms", that is the terms that become divergent for very low density. Keeping in mind that in all momentum integrals in the next sections the various divergences occur near the dispersion law ($a \approx 0$) we can consider $a \sim n$ . Furthermore $\Sigma_i \sim n$. Another dramatic simplification can be obtained by consequently expanding in $\mathbf{B}$ and $\mathbf{B}^2$. Altogether this replaces the Green's function by the expression,

$$\mathbf{G}(\mathbf{B}, \mathbf{p}) \rightarrow \left(\frac{1}{a_0} + \frac{\Sigma_1^2}{a_0^3} - \frac{\Sigma_3}{a_0^2}\right) \mathbf{\Delta_p} - \frac{\Sigma_1}{a_0^2} \mathbf{L}(\mathbf{B}, \widehat{\mathbf{p}}) - \left(\frac{\Sigma_1^2}{a_0^3} - \frac{\Sigma_2}{a_0^2}\right) \mathbf{H}(\mathbf{B}, \widehat{\mathbf{p}}) - \frac{\Sigma_1^2}{a_0^3} \mathbf{V}(\mathbf{B}, \widehat{\mathbf{p}}) \ , \tag{26}$$

in which $a_0 = \omega^2 - p^2 - \Sigma_0$ and

$$\begin{aligned} \mathbf{L}(\mathbf{B}, \widehat{\mathbf{p}}) &= -\mathbf{\Phi} + (\mathbf{\Phi} \cdot \widehat{\mathbf{p}}) \widehat{\mathbf{p}} - \widehat{\mathbf{p}} (\mathbf{\Phi} \cdot \widehat{\mathbf{p}}) \ , \\ \mathbf{H}(\mathbf{B}, \widehat{\mathbf{p}}) &= \mathbf{B}\mathbf{B} + (\mathbf{B} \cdot \widehat{\mathbf{p}})^2 \widehat{\mathbf{p}}\widehat{\mathbf{p}} - (\mathbf{B} \cdot \widehat{\mathbf{p}}) (\mathbf{B}\widehat{\mathbf{p}} + \widehat{\mathbf{p}}\mathbf{B}) \ , \\ \mathbf{V}(\mathbf{B}, \widehat{\mathbf{p}}) &= -(\mathbf{\Phi} \cdot \widehat{\mathbf{p}}) (\mathbf{\Phi} \cdot \widehat{\mathbf{p}}) \ , \end{aligned}$$

are three transverse tensors of rank two. Only the first is linear in the field and anti-symmetric, signifying the Faraday rotation of a spherical wave; $\mathbf{H}$ and $\mathbf{V}$ are birefringent parts quadratic in the field that preserve reciprocity. In the simplification procedure outlined above, the longitudinal part of the Green's function has been lost. As a result an ultraviolet singularity has disappeared and $\mathbf{G}(\mathbf{p} \rightarrow \infty) \sim 1/p^2$. Leading orders in the number density of the calculations in the next sections will not be affected. In what follows powers in the magnetic field higher than two will frequently be ignored without explicitly saying so.

The radial momentum integrals for $\mathbf{q} = \mathbf{0}$ all take a form frequently encountered in speckle calculations [24],

$$I(n, m) \equiv \sum_{\mathbf{p}} \frac{1}{a_0^n (\overline{a}_0)^m} = \frac{1}{\pi} \frac{(n+m-2)!}{(n-1)!(m-1)!} \frac{i}{2^{n+m} \omega^{n+m-2}} (-1)^{m-1} (-i)^{n+m-1} \ell^{n+m-1} \ . \tag{27}$$

The $\mathbf{q}$-expansion can be done using the radial momentum integral,

$$\begin{aligned} \sum_p \frac{1}{a_0 (\mathbf{p} + \mathbf{q}/2)^n [\overline{a}_0 (\mathbf{p} - \mathbf{q}/2)]^m} &= I(n, m) + \mathcal{O}(\widehat{\mathbf{p}} \cdot \mathbf{q}) \\ &- (\widehat{\mathbf{p}} \cdot \mathbf{q})^2 \frac{1}{\pi} \frac{(n+m)!}{(n-1)!(m-1)!} \frac{1}{2^{n+m+1} \omega^{n+m-2}} (-1)^m (-i)^{n+m} \ell^{n+m+1} \ . \end{aligned} \tag{28}$$

Here $\ell = -\omega / \mathrm{Im}\, \Sigma_0$ denotes the scattering mean free path in the absence of a magnetic field. Angular averages have to be performed for rank-two, rank-four and rank-six tensors. We find,

$$
\begin{aligned}
\langle \widehat{p}_i \widehat{p}_j \rangle &= \tfrac{1}{3}\delta_{ij} \\
\langle \widehat{p}_i \widehat{p}_j \widehat{p}_k \widehat{p}_l \rangle &= \tfrac{1}{15}\left(\delta_{ij}\delta_{kl} + \delta_{ik}\delta_{jl} + \delta_{il}\delta_{jk}\right) \equiv \mathcal{V}_{ijkl} \\
\langle \widehat{p}_i \widehat{p}_j \widehat{p}_k \widehat{p}_l \widehat{p}_m \widehat{p}_n \rangle &= \tfrac{1}{7}\left(\delta_{ij}\mathcal{V}_{klmn} + \delta_{ik}\mathcal{V}_{jlmn} + \delta_{il}\mathcal{V}_{jkmn} + \delta_{im}\mathcal{V}_{jknl} + \delta_{in}\mathcal{V}_{jklm}\right) \equiv \mathcal{W}_{ijklmn}
\end{aligned} \tag{29}
$$

Using this information the integrals can be carried out straightforwardly and we will only quote the final result,

$$
\begin{aligned}
\mathcal{A}^{\pm}_{ijkl}(\omega, \mathbf{B}, \mathbf{q} = \mathbf{0}) &\equiv \frac{4\pi}{\ell}\sum_{\mathbf{p}} G_{ij}(\omega, \mathbf{p}, \mathbf{B})\, G^{*}_{lk}(\omega, \mathbf{p}, \pm\mathbf{B}) = \\
&= \left(1 - \frac{1}{2}\text{Re}\ \eta_1^2 - \text{Im}\ \eta_3\right)\mathcal{T}_{ijkl} + \frac{i}{15}\eta_1 \mathcal{F}_{ijkl} \mp \frac{i}{15}\overline{\eta}_1 \mathcal{F}_{lkji} + \frac{1}{2}\left(\frac{1}{2}\eta_1^2 - i\eta_2\right)\mathcal{G}_{ijkl} \\
&+ \quad \frac{1}{2}\left(\frac{1}{2}\overline{\eta}_1^2 + i\overline{\eta}_2\right)\mathcal{G}_{lkji} + \frac{1}{4}\eta_1^2 \mathcal{J}_{ijkl} + \frac{1}{4}\overline{\eta}_1^2 \mathcal{J}_{lkji} \pm \frac{1}{2}\left|\eta_1\right|^2 \mathcal{S}_{ijkl}.
\end{aligned} \tag{30}
$$

Here,

$$
\begin{aligned}
\mathcal{T}_{ijkl} &= \tfrac{1}{15}\left(6\delta_{ij}\delta_{kl} + \delta_{ik}\delta_{jl} + \delta_{il}\delta_{jk}\right) , \\
\mathcal{F}_{ijkl} &= -\Phi_{ij}\delta_{kl} - \tfrac{1}{2}\left(\Phi_{il}\delta_{jk} - \delta_{il}\Phi_{jk} - \delta_{ik}\Phi_{jl} + \Phi_{ik}\delta_{jl}\right) , \\
\mathcal{G}_{ijkl} &= \tfrac{1}{15}\delta_{ij}\delta_{kl} - \tfrac{1}{7}\left(\mathcal{V}_{ijkl} + B_iB_m\mathcal{V}_{mjkl} + B_jB_m\mathcal{V}_{mikl} + B_kB_m\mathcal{V}_{mijl} + B_lB_m\mathcal{V}_{mijk}\right) , \\
&\quad + \tfrac{1}{15}\left(4B_iB_j\delta_{kl} + B_jB_l\delta_{ik} + B_iB_k\delta_{jl} + B_iB_l\delta_{jk} + B_jB_k\delta_{il}\right) , \\
\mathcal{J}_{ijkl} &= \tfrac{4}{15}\left(\delta_{ij}\delta_{kl} - B_iB_j\delta_{kl}\right) + \tfrac{1}{15}\left(\Phi_{ki}\Phi_{lj} + \Phi_{li}\Phi_{kj}\right) , \\
\mathcal{S}_{ijkl} &= -\tfrac{1}{5}\Phi_{ij}\Phi_{lk} + \tfrac{2}{15}\Phi_{ik}\Phi_{lj} + \tfrac{2}{15}\Phi_{il}\Phi_{jk} + \tfrac{2}{15}\left(\delta_{ik}\delta_{jl} - \delta_{il}\delta_{jk}\right) , \\
&\quad + \tfrac{1}{15}\left(\delta_{jk}B_iB_l + \delta_{il}B_jB_k - \delta_{jl}B_iB_k - \delta_{ik}B_jB_l\right) ,
\end{aligned}
$$

It is convenient to make an isomorphic transformation to the 9-dimensional vector space of $3 \times 3$ matrices. We will restrict ourselves to the three-dimensional subspace spanned by the hermitean matrices $\{\mathbf{I},\ \mathbf{\Phi},\ \mathbf{BB}\}$, in which the long range diffusion occurs. The tensor $\mathcal{Q}^{\pm}(\omega, \mathbf{q} = \mathbf{0}, \mathbf{B})$ is closed in this subspace. This base is not orthonormal and has Gram matrix,

$$
\mathbf{M} = \begin{pmatrix} 3 & 0 & 1 \\ 0 & 2 & 0 \\ 1 & 0 & 1 \end{pmatrix} .
$$

With respect to this base the tensor $\mathcal{A}^{\pm}$ for $\mathbf{q} = \mathbf{0}$ takes the form,

$$
A^{\pm}_{ij}(\omega, \mathbf{B}, \mathbf{q} = \mathbf{0}) \equiv \langle i|\, \mathcal{A}(\omega, \mathbf{B}, \mathbf{q} = \mathbf{0})\, |j\rangle = \frac{4\pi}{\ell}\langle i| \sum_{\mathbf{p}} \mathbf{G}(\omega, \mathbf{p}, \mathbf{B})\, \mathbf{G}^{*}(\omega, \mathbf{p}, \pm\mathbf{B})\, |j\rangle , \tag{31}
$$

with,

$$
\begin{aligned}
&A^{\pm}_{11} = 2 - \tfrac{1}{3}\text{Re}\ \eta_1^2 - 2\text{Im}\ \eta_3 + \tfrac{2}{3}\text{Im}\ \eta_2 \pm \tfrac{1}{3}\left|\eta_1\right|^2 \\
&A^{\pm}_{12} = A^{\pm}_{21} = -\tfrac{1}{3}i\left(\eta_1 \mp \overline{\eta}_1\right) \\
&A^{\pm}_{22} = \tfrac{2}{3} - \tfrac{1}{5}\text{Re}\ \eta_1^2 + \tfrac{2}{15}\text{Im}\ \eta_2 - \tfrac{2}{3}\text{Im}\ \eta_3 \pm \tfrac{1}{5}\left|\eta_1\right|^2 \\
&A^{\pm}_{23} = A^{\pm}_{32} = -\tfrac{1}{15}i\left(\eta_1 \mp \overline{\eta}_1\right) \\
&A^{\pm}_{13} = A^{\pm}_{31} = \tfrac{2}{3} - \tfrac{1}{15}\text{Re}\ \eta_1^2 + \tfrac{8}{15}\text{Im}\ \eta_2 - \tfrac{2}{3}\text{Im}\ \eta_3 \pm \tfrac{1}{15}\left|\eta_1\right|^2 \\
&A^{\pm}_{33} = \tfrac{8}{15} - \tfrac{1}{35}\text{Re}\ \eta_1^2 + \tfrac{16}{35}\text{Im}\ \eta_2 - \tfrac{8}{15}\text{Im}\ \eta_3
\end{aligned}
$$

This matrix is symmetric and for the "+" choice even hermitean. We will also need the four-rank tensor $n\mathbf{t}(\mathbf{B})\,\mathbf{t}(\pm\mathbf{B})^{*}$ with respect to this base. If we denote,

$$U_{ij}^{\pm}(\omega,\mathbf{B}) \equiv \frac{\ell}{6\pi}\langle i|\,\mathbf{t}(\omega,\mathbf{B})\,\mathbf{t}^*(\omega,\pm\mathbf{B})\,|j\rangle \ , \tag{32}$$

we obtain the symmetric (and for the "+" choice again hermitean) matrix,

$$\begin{array}{ll} U_{11}^{\pm} = 3 - 6\mathrm{Re}\ \xi_3 + 2\mathrm{Re}\ \xi_2 \pm 2\left|\xi_1\right|^2 & U_{12}^{\pm} = U_{21}^{\pm} = 2\left(\xi_1 \pm \overline{\xi}_1\right) \\ U_{22}^{\pm} = 2 - 4\mathrm{Re}\ \xi_3 \pm 2\left|\xi_1\right|^2 & U_{23}^{\pm} = U_{32}^{\pm} = 0 \\ U_{33}^{\pm} = \ 1 - 2\mathrm{Re}\ \xi_3 + 2\mathrm{Re}\ \xi_2 & U_{13}^{\pm} = U_{31}^{\pm} = 1 - 2\mathrm{Re}\ \xi_3 + 2\mathrm{Re}\ \xi_2 \end{array}$$

With respect to this base the Ladder sum becomes,

$$\mathcal{L}^{\pm}(\omega,\mathbf{B},\mathbf{q}) = n\mathbf{t}(\mathbf{B})\,\mathbf{t}(\pm\mathbf{B})^* \cdot \Big(\ |\mathbf{I}\rangle,\ \ |\mathbf{\Phi}\rangle,\ |\mathbf{BB}\rangle\ \Big)\cdot\frac{1}{\mathbf{L}^{\pm}(\mathbf{q})}\cdot\begin{pmatrix}\langle\mathbf{I}|\\ \langle\mathbf{\Phi}|\\ \langle\mathbf{BB}|\end{pmatrix} \ , \tag{33}$$

where $\mathbf{L}^{\pm}(\mathbf{q}) = \mathbf{M}-\frac{3}{2}\ \mathbf{A}^{\pm}(\mathbf{q})\cdot\mathbf{M}^{-1}\cdot\mathbf{U}^{\pm}$ . For $\mathbf{q}=\mathbf{0}$ we find,

$$\begin{array}{rcl} L_{11}^{\pm} &=& \frac{1}{2}\mathrm{Re}\ \eta_1^2 + 3\mathrm{Im}\ \eta_3 - \ \mathrm{Im}\ \eta_2 \mp \frac{1}{2}\left|\eta_1\right|^2 \\ && +\ 6\mathrm{Re}\ \xi_3 \mp 2\left|\xi_1\right|^2 - 2\mathrm{Re}\ \xi_2 + \frac{i}{2}\left(\eta_1 \mp \overline{\eta}_1\right)\left(\xi_1 \pm \overline{\xi}_1\right) \stackrel{\pm}{=} 0 \\ L_{12}^{\pm} &=& \frac{1}{2}i\left(\eta_1 \mp \overline{\eta}_1\right) - 2\left(\xi_1 \pm \overline{\xi}_1\right) \\ L_{21}^{\pm} &=& \frac{1}{2}i\left(\eta_1 \mp \overline{\eta}_1\right) - \left(\xi_1 \pm \overline{\xi}_1\right) \stackrel{\pm}{=} 0 \\ L_{22}^{\pm} &=& 1 + \frac{3}{10}\mathrm{Re}\ \eta_1^2 - \frac{1}{5}\mathrm{Im}\ \eta_2 + \ \mathrm{Im}\ \eta_3 \mp \frac{3}{10}\left|\eta_1\right|^2 \\ && +\ 2\mathrm{Re}\ \xi_3 \mp \left|\xi_1\right|^2 + \frac{2}{5}i\left(\eta_1 \mp \overline{\eta}_1\right)\left(\xi_1 \pm \overline{\xi}_1\right) \\ L_{23}^{\pm} &=& \frac{1}{10}i\left(\eta_1 \mp \overline{\eta}_1\right) \\ L_{32}^{\pm} &=& \frac{1}{10}i\left(\eta_1 \mp \overline{\eta}_1\right) - \frac{1}{5}\left(\xi_1 \pm \overline{\xi}_1\right) \\ L_{13}^{\pm} &=& \frac{1}{10}\mathrm{Re}\ \eta_1^2 - \frac{4}{5}\mathrm{Im}\ \eta_2 + \ \mathrm{Im}\ \eta_3 \mp \frac{1}{10}\left|\eta_1\right|^2 - 2\mathrm{Re}\ \xi_2 + 2\mathrm{Re}\ \xi_3 \\ L_{31}^{\pm} &=& \frac{1}{10}\mathrm{Re}\ \eta_1^2 - \frac{4}{5}\ \mathrm{Im}\ \eta_2 + \mathrm{Im}\ \eta_3 \mp \frac{1}{10}\left|\eta_1\right|^2 \\ && +2\mathrm{Re}\ \xi_3 \mp \frac{1}{5}\left|\xi_1\right|^2 - \frac{8}{5}\mathrm{Re}\ \xi_2 + \frac{i}{10}\left(\eta_1 \mp \overline{\eta}_1\right)\left(\xi_1 \pm \overline{\xi}_1\right) \stackrel{\pm}{=} 0 \\ L_{33}^{\pm} &=& \frac{1}{5} + \frac{3}{70}\mathrm{Re}\ \eta_1^2 - \frac{24}{35}\mathrm{Im}\ \eta_2 + \ \frac{4}{5}\mathrm{Im}\ \eta_3 - \frac{8}{5}\mathrm{Re}\ \xi_2 + \frac{8}{5}\mathrm{Re}\ \xi_3 \end{array}$$

This matrix is not symmetric. Some entries as well as the determinant for the "+" choice are zero on the basis of the Optical Theorem (8).

An extensive calculation - of which we shall not mention any intermediate results and simply quote the final result - yields the $\mathbf{q}$-dependent part of $\mathbf{L}^{\pm}(\mathbf{q})$ with respect to the same base

$$\delta\mathbf{L}^{\pm}(\mathbf{q}) = \ell^2\left[\mathbf{S}^{\pm}q^2 + \mathbf{T}^{\pm}\left(\mathbf{B}\cdot\mathbf{q}\right)^2\right] \tag{34}$$

in which the isotropic part reads

$$
\begin{aligned}
S_{11}^{\pm} &= 1 - 3\mathrm{Re}\ \eta_1^2 - 3\mathrm{Im}\ \eta_3 + \tfrac{6}{5}\mathrm{Im}\ \eta_2 \pm \tfrac{3}{5}\left|\eta_1\right|^2 \\
&\quad -2\mathrm{Re}\ \xi_3 \pm \tfrac{3}{5}\left|\xi_1\right|^2 + \tfrac{4}{5}\mathrm{Re}\ \xi_2 - \tfrac{3}{10}i\left(\eta_1 \mp \overline{\eta}_1\right)\left(\xi_1 \pm \overline{\xi}_1\right) \\
S_{12}^{\pm} &= -\tfrac{3}{10}i\left(\eta_1 \mp \overline{\eta}_1\right) + \tfrac{3}{5}\left(\xi_1 \pm \overline{\xi}_1\right) \\
S_{21}^{\pm} &= -\tfrac{3}{10}i\left(\eta_1 \mp \overline{\eta}_1\right) + \tfrac{1}{5}\left(\xi_1 \pm \overline{\xi}_1\right) \\
S_{22}^{\pm} &= \tfrac{1}{5} - \tfrac{3}{5}\mathrm{Re}\ \eta_1^2 + \tfrac{6}{5}\mathrm{Im}\ \eta_2 - \tfrac{3}{5}\ \mathrm{Im}\ \eta_3 \pm \tfrac{177}{35}\left|\eta_1\right|^2 \\
&\quad -\tfrac{2}{5}\mathrm{Re}\ \xi_3 \pm \tfrac{1}{5}\left|\xi_1\right|^2 - \tfrac{3}{14}i\left(\eta_1 \mp \overline{\eta}_1\right)\left(\xi_1 \pm \overline{\xi}_1\right) \\
S_{23}^{\pm} &= -\tfrac{3}{35}i\left(\eta_1 \mp \overline{\eta}_1\right) \\
S_{32}^{\pm} &= -\tfrac{3}{35}i\left(\eta_1 \mp \overline{\eta}_1\right) + \tfrac{2}{35}\left(\xi_1 \pm \overline{\xi}_1\right) \\
S_{13}^{\pm} &= \tfrac{2}{5} - \tfrac{6}{35}\mathrm{Re}\ \eta_1^2 - \tfrac{6}{5}\mathrm{Im}\ \eta_3 + \tfrac{36}{35}\mathrm{Im}\ \eta_2 \pm \tfrac{6}{35}\left|\eta_1\right|^2 - \tfrac{4}{5}\mathrm{Re}\ \xi_3 + \tfrac{4}{5}\mathrm{Re}\ \xi_2 \\
S_{31}^{\pm} &= \tfrac{2}{5} - \tfrac{6}{35}\mathrm{Re}\ \eta_1^2 - \tfrac{6}{5}\mathrm{Im}\ \eta_3 + \tfrac{36}{35}\mathrm{Im}\ \eta_2 \pm \tfrac{6}{35}\left|\eta_1\right|^2 \\
&\quad -\tfrac{4}{5}\mathrm{Re}\ \xi_3 \pm \tfrac{2}{35}\left|\xi_1\right|^2 + \tfrac{24}{35}\mathrm{Re}\ \xi_2 - \tfrac{i}{105}\left(\eta_1 \mp \overline{\eta}_1\right)\left(\xi_1 \pm \overline{\xi}_1\right) \\
S_{33}^{\pm} &= \tfrac{12}{35} - \tfrac{4}{35}\mathrm{Re}\ \eta_1^2 + \tfrac{32}{35}\mathrm{Im}\ \eta_2 - \tfrac{36}{35}\mathrm{Im}\ \eta_3 + \tfrac{24}{35}\mathrm{Re}\ \xi_2 - \tfrac{24}{35}\mathrm{Re}\ \xi_3
\end{aligned}
$$

and the anisotropic part,

$$
\begin{aligned}
T_{11}^{\pm} &= -\tfrac{3}{5}\mathrm{Im}\ \eta_2 \pm \tfrac{6}{5}\left|\eta_1\right|^2 \pm \tfrac{1}{5}\left|\xi_1\right|^2 - \tfrac{2}{5}\mathrm{Re}\ \xi_2 - \tfrac{3}{5}i\left(\eta_1 \mp \overline{\eta}_1\right)\left(\xi_1 \pm \overline{\xi}_1\right) \\
T_{12}^{\pm} &= -\tfrac{3}{5}i\left(\eta_1 \mp \overline{\eta}_1\right) + \tfrac{1}{5}\left(\xi_1 \pm \overline{\xi}_1\right) \\
T_{21}^{\pm} &= -\tfrac{3}{5}i\left(\eta_1 \mp \overline{\eta}_1\right) + \tfrac{2}{5}\left(\xi_1 \pm \overline{\xi}_1\right) \\
T_{22}^{\pm} &= \tfrac{2}{5} - \tfrac{6}{5}\mathrm{Re}\ \eta_1^2 + \tfrac{3}{35}\mathrm{Im}\ \eta_2 - \tfrac{6}{5}\mathrm{Im}\ \eta_3 \pm \tfrac{48}{35}\left|\eta_1\right|^2 \\
&\quad -\tfrac{4}{5}\mathrm{Re}\ \xi_3 \pm \tfrac{2}{5}\left|\xi_1\right|^2 - \tfrac{39}{70}i\left(\eta_1 \mp \overline{\eta}_1\right)\left(\xi_1 \pm \overline{\xi}_1\right) \\
T_{23}^{\pm} &= -\tfrac{3}{70}i\left(\eta_1 \mp \overline{\eta}_1\right) \\
T_{32}^{\pm} &= -\tfrac{3}{70}i\left(\eta_1 \mp \overline{\eta}_1\right) + \tfrac{1}{35}\left(\xi_1 \pm \overline{\xi}_1\right) \\
T_{13}^{\pm} &= -\tfrac{1}{5} - \tfrac{3}{35}\mathrm{Re}\ \eta_1^2 + \tfrac{3}{5}\mathrm{Im}\ \eta_3 - \tfrac{24}{35}\mathrm{Im}\ \eta_2 \pm \tfrac{3}{35}\left|\eta_1\right|^2 + \tfrac{2}{5}\mathrm{Re}\ \xi_3 - \tfrac{2}{5}\mathrm{Re}\ \xi_2 \\
T_{31}^{\pm} &= -\tfrac{1}{5} - \tfrac{3}{35}\mathrm{Re}\ \eta_1^2 + \tfrac{3}{5}\mathrm{Im}\ \eta_3 - \tfrac{24}{35}\mathrm{Im}\ \eta_2 \pm \tfrac{3}{35}\left|\eta_1\right|^2 \\
&\quad +\tfrac{2}{5}\mathrm{Re}\ \xi_3 \pm \tfrac{1}{35}\left|\xi_1\right|^2 - \tfrac{16}{35}\mathrm{Re}\ \xi_2 - \tfrac{3}{70}i\left(\eta_1 \mp \overline{\eta}_1\right)\left(\xi_1 \pm \overline{\xi}_1\right) \\
T_{33}^{\pm} &= -\tfrac{8}{35} + \tfrac{24}{35}\mathrm{Im}\ \eta_3 - \tfrac{24}{35}\mathrm{Im}\ \eta_2 - \tfrac{16}{35}\mathrm{Re}\ \xi_2 + \tfrac{16}{35}\mathrm{Re}\ \xi_3
\end{aligned}
$$

Finally we give the perturbation of the tensor $\mathcal{Q}^{\pm}$ in the dynamic variable $\Omega$. Using $\mathbf{G}\left(\omega \pm \Omega/2\right) = \mathbf{G}\left(\omega\right) \pm \omega\Omega\, \mathrm{d}\mathbf{G}/\mathrm{d}\omega^2$, the most divergent terms (proportional to $1/n$) can be obtained straightforwardly using the standard integral $I(n,m)$ introduced earlier in this section. If we represent the matrix elements of $\delta\mathcal{Q}^{\pm}\left(\Omega, \mathbf{B}\right)$ with respect to our base by,

$$
\delta\mathbf{L}^{\pm}\left(\Omega\right) = -i\Omega\ell\mathbf{W}^{\pm}\left(\mathbf{B}\right) \tag{35}
$$

we obtain,

$$
\begin{aligned}
W_{11}^{\pm} &= 3-\tfrac{3}{2}\mathrm{Re}\ \eta_1^2-6\mathrm{Im}\ \eta_3+2\ \mathrm{Im}\ \eta_2\pm\tfrac{3}{2}\left|\eta_1\right|^2 \\
&\quad -6\mathrm{Re}\ \xi_3\pm 2\left|\xi_1\right|^2+2\mathrm{Re}\ \xi_2-i\left(\eta_1\mp\overline{\eta}_1\right)\left(\xi_1\pm\overline{\xi}_1\right) \\
W_{12}^{\pm} &= -i\left(\eta_1\mp\overline{\eta}_1\right)+2\left(\xi_1\pm\overline{\xi}_1\right) \\
W_{21}^{\pm} &= -i\left(\eta_1\mp\overline{\eta}_1\right)+\left(\xi_1\pm\overline{\xi}_1\right) \\
W_{22}^{\pm} &= 1-\tfrac{9}{10}\mathrm{Re}\ \eta_1^2+\tfrac{2}{5}\mathrm{Im}\ \eta_2-2\ \mathrm{Im}\ \eta_3\pm\tfrac{9}{10}\left|\eta_1\right|^2 \\
&\quad -2\mathrm{Re}\ \xi_3\pm\left|\xi_1\right|^2-\tfrac{4}{5}i\left(\eta_1\mp\overline{\eta}_1\right)\left(\xi_1\pm\overline{\xi}_1\right) \\
W_{23}^{\pm} &= -\tfrac{1}{5}i\left(\eta_1\mp\overline{\eta}_1\right) \\
W_{32}^{\pm} &= -\tfrac{1}{5}i\left(\eta_1\mp\overline{\eta}_1\right)+\tfrac{1}{5}\left(\xi_1\pm\overline{\xi}_1\right) \\
W_{13}^{\pm} &= 1-\tfrac{3}{10}\mathrm{Re}\ \eta_1^2+\tfrac{8}{5}\mathrm{Im}\ \eta_2-2\mathrm{Im}\ \eta_3\pm\tfrac{3}{10}\left|\eta_1\right|^2+2\mathrm{Re}\ \xi_2-2\mathrm{Re}\ \xi_3 \\
W_{31}^{\pm} &= 1-\tfrac{3}{10}\mathrm{Re}\ \eta_1^2+\tfrac{8}{5}\mathrm{Im}\ \eta_2-2\mathrm{Im}\ \eta_3\pm\tfrac{3}{10}\left|\eta_1\right|^2 \\
&\quad -2\mathrm{Re}\ \xi_3\pm\tfrac{1}{5}\left|\xi_1\right|^2+\tfrac{8}{5}\mathrm{Re}\ \xi_2-\tfrac{i}{5}\left(\eta_1\mp\overline{\eta}_1\right)\left(\xi_1\pm\overline{\xi}_1\right) \\
W_{33}^{\pm} &= \tfrac{4}{5}-\tfrac{3}{10}\mathrm{Re}\ \eta_1^2+\tfrac{6}{5}\mathrm{Im}\ \eta_2-\tfrac{8}{5}\mathrm{Im}\ \eta_3+\tfrac{8}{5}\mathrm{Re}\ \xi_2-\tfrac{8}{5}\mathrm{Re}\ \xi_3
\end{aligned}
$$

The identity $\langle\Delta\mathbf{\Sigma}|\ \delta\mathcal{Q}^{+}(\ \Omega,\mathbf{B})\ |\mathbf{I}\rangle=-6\omega\Omega+\mathcal{O}(n)$ follows from the Optical Theorem and was used earlier for the calculation of the Boltzmann diffusion constant. The explicit calculations (35) have been verified to obey this identity. Since we ignored terms higher order in density in this calculation, the influence of the scatterers on the transport velocity of multiply scattered light [23] in a magnetic field has been neglected.

### B. Boltzmann Diffusion Tensor

With the calculations in the previous subsection and using the eigenfunctions in Eq. (19) we find that the Boltzmann diffusion tensor (22) is,

$$
\mathbf{q}\cdot\mathbf{D}_B^{+}(\mathbf{B})\cdot\mathbf{q}=\frac{1}{3\ell}\left[(1+\mathrm{Im}\ \eta_3)\,\delta L_{11}^{+}(\mathbf{q})-\mathrm{Im}\ \eta_1\delta L_{21}^{+}(\mathbf{q})-\mathrm{Im}\ \eta_2\delta L_{31}^{+}(\mathbf{q})\right]\ ,
$$

with the explicit result that the symmetric part is given by,

$$
\mathbf{D}_B^{+,\mathrm{sym}}(\mathbf{B})=\frac{1}{3}\ell\left[d_{\mathrm{iso}}\mathbf{I}\ +d_{\mathrm{ani}}\mathbf{BB}\ \right], \tag{36}
$$

$$
\begin{aligned}
d_{\mathrm{iso}} &= 1-3\mathrm{Re}\ \eta_1^2-2\mathrm{Im}\ \eta_3+\tfrac{4}{5}\mathrm{Im}\ \eta_2+\tfrac{3}{5}\left|\eta_1\right|^2-\tfrac{3}{5}\left(\mathrm{Im}\ \eta_1\right)^2 \\
&\quad -2\mathrm{Re}\ \xi_3+\tfrac{3}{5}\left|\xi_1\right|^2+\tfrac{4}{5}\mathrm{Re}\ \xi_2+\tfrac{4}{5}\mathrm{Im}\ \eta_1\mathrm{Re}\ \xi_1\ , \\
d_{\mathrm{ani}} &= -\tfrac{2}{5}\mathrm{Im}\ \eta_2+\tfrac{6}{5}\left|\eta_1\right|^2-\tfrac{6}{5}\left(\mathrm{Im}\ \eta_1\right)^2+\tfrac{1}{5}\left|\xi_1\right|^2-\tfrac{2}{5}\mathrm{Re}\ \xi_2+\tfrac{8}{5}\mathrm{Im}\ \eta_1\mathrm{Re}\ \xi_1\ .
\end{aligned}
$$

In Appendix C it is shown that the variable $\mu$ is the most important one. Ignoring $\zeta$ and $\varpi$ in Eq. (36) yields that $d_{\mathrm{iso}}-1\sim d_{\mathrm{ani}}\sim\mu^2\sim B^2$. The dephasing in the most-crossed diagrams will later be shown to be determined by the same variable $\mu$. In our model both phenomena are thus seen to be of the same order of magnitude. This dephasing has been measured already [10]. For the experimental verification of anisotropy in the diffusion tensor of light in Faraday-active media we may thus be optimistic.

In Fig. 5 we show the modifications to the symmetric diffusion tensor as a function of the phase-shift $\alpha$ of the Rayleigh $t$-matrix defined in Eq. (5). At low frequencies and at resonance it is inferred that the diffusion constant is suppressed, and relatively more perpendicular to the field direction. At low frequencies ($\alpha\simeq 0$) - a regime where our model should coincide with the exact Mie solution at low frequencies - we find $\Delta D_{||}^{\dagger}/\Delta D_{\perp}^{\dagger} = 1/2$. At resonance ($\alpha \simeq \pi/2$) this is changed into 1/3. For the Beenakker-Senftleben effect one also finds the modification to be largest perpendicular to the field, and for the heat conductivity - at low fields - a ratio $\Delta\lambda_{||}/\Delta\lambda_{\perp} = 2/9$ [15] is obtained (for large fields this saturizes to 2/3). At intermediate frequencies we infer that the diffusion may in fact *increase* in a magnetic field. For the Beenakker-Senftleben effect one always finds a decrease.

### C. Light Propagation in a Slab Geometry

In this subsection we calculate multiple light scattering in an optically thick plan-parallel slab imposed to a homogeneous external magnetic field. This is the simplest theoretical model that includes all physics and is experimentally relevant. The angular transmission tensor of a slab with thickness $L$ in (nearly normal [25]) direction $\mathbf{k}$ and normal incidence is given by [26] $\left(\mathbf{y} = (L - z_2)\,\widehat{\mathbf{k}},\, \mathbf{z} = z_1\widehat{\mathbf{z}}\right)$

$$\mathcal{T}_{\mathbf{z},\mathbf{k}} = \int_0^L \mathrm{d}z_1 \int_0^L \mathrm{d}z_2\ \left(4\pi y\right)^2 \mathbf{G}\left(\mathbf{y}\right)\mathbf{G}^*\left(\mathbf{y}\right)\cdot\mathcal{L}\left(L - z_2, z_1\right)\ \cdot\left(4\pi z_1\right)^2 \mathbf{G}\left(\mathbf{z}\right)\mathbf{G}^*\left(\mathbf{z}\right)\ . \tag{37}$$

Similarly the reflection tensor is, $\left(\mathbf{y} = z_2\widehat{\mathbf{k}},\ \mathbf{z} = z_1\widehat{\mathbf{z}}\right)$

$$\mathcal{R}_{\mathbf{z},\mathbf{k}} = \int_0^L \mathrm{d}z_1 \int_0^L \mathrm{d}z_2\ \left(4\pi y\right)^2 \mathbf{G}\left(\mathbf{y}\right)\mathbf{G}^*\left(\mathbf{y}\right)\cdot\mathcal{L}\left(z_2, z_1\right)\ \cdot\left(4\pi z_1\right)^2 \mathbf{G}\left(\mathbf{z}\right)\mathbf{G}^*\left(\mathbf{z}\right)\ . \tag{38}$$

Here $\mathcal{L}\left(z_2, z_1\right)$ is the Ladder tensor for the slab geometry (integrated over transverse coordinates) and must be found by adding boundary conditions to the bulk results obtained earlier. The Green's tensors $\mathbf{G}$ in the equations above are Dyson Green's tensors and the Fourier transform of $\mathbf{G}(\mathbf{p})$ defined in Eq. (9). They decay exponentially in space and thus signify the light propagation in a narrow "skin" layer, that is the first scattering mean free path for the incident wave, and the last one for the emerging wave. For the Ladder tensor in the bulk we obtained in Eq. (21) $\mathcal{L}\left(\mathbf{q}\right) \sim \left|\mathbf{d}\right\rangle \rho\left(\mathbf{q}\right)\left\langle\mathbf{d}\right|$ with $\rho\left(\mathbf{q}\right)$ essentially the well-documented scalar Ladder sum but with an anisotropic diffusion tensor.

The boundary effects caused by the finiteness of the scattering geometry are coded in the Schwarzschild-Milne integral equation for $\mathcal{L}(z_1, z_2)$ that can be derived from the Bethe-Salpeter transport equation. For scalar waves and isotropic scattering this was investigated by Nieuwenhuizen and Luck [17]. The generalization to anisotropic scattering was recently carried out by Amic, Luck and Nieuwenhuizen [28]. These methods are not yet applicable to vector waves in anisotropic media. Therefore we will use the (improved) diffusion approximation in which the boundary conditions are introduced by the imaging method [7] [26]. This consists of adding trapping planes for $\rho\left(z, z'\right)$ at some distance $z_0$ beyond the slab boundaries, called the extrapolation length. For isotropic media with $\mathbf{k}$-anisotropic scattering it known from Milne theory [29] that $z_0 \approx 0.710\ell^*$, that means always essentially one *transport* mean free path. A similar conclusion is reached by considering the bulk diffusion equation subject to the "radiative boundary conditions" that the incident incoherent flux on both sides of the slab boundary vanishes (see e.g. Refs. [30] and [31]). This method can also be formulated for vector waves with an anisotropic diffusion tensor in a slab geometry,

$$\begin{aligned} \left(-\nabla\cdot\mathbf{D}\cdot\nabla\ + \alpha^2\right)\rho\left(z, z'\right) &= \ \text{source}\ , \\ \tfrac{1}{4}\rho\left(z, z'\right) - \tfrac{1}{2}\widehat{\mathbf{z}}\cdot\mathbf{D}\cdot\nabla\rho\left(z, z'\right) = 0 & \qquad z = 0\ , \\ \tfrac{1}{4}\rho\left(z, z'\right) + \tfrac{1}{2}\widehat{\mathbf{z}}\cdot\mathbf{D}\cdot\nabla\rho\left(z, z'\right) = 0 & \qquad z = L\ , \end{aligned} \tag{39}$$

The source is usually supplied by single scattering; $\alpha^2$ is an absorption rate to be specified later. In this approach the boundary conditions effectively locate trapping planes at distances $z_0 = 2D_{zz} =$ $\frac{2}{3}\ell^*_{zz}$ outside the slab boundaries. The relevant diffusion constant for the slab geometry is thus $D_{zz}$ [27].

The solution of Eq. (39) can be found in literature. For our purposes we need $\rho(z, z')$ for $\alpha = 0$ which reads, [26]

$$\rho(z, z') = \frac{1}{4D_{zz}} \frac{(L + 2z_0 - |z - z'|)^2 - (L - z - z')^2}{L + 2z_0} \ . \tag{40}$$

The angular incoherent reflection coefficient for $L = \infty$ (at approximately backscattering) adopting first ordinary scalar skin layers can be found straightforwardly by integration of Eq. (38) with scalar Green's functions. The same holds true for the incoherent transmission (at forward scattering) [32] [33], which we shall need later also for $\alpha \neq 0$. We have,

$$\begin{aligned} R\,(\text{back}) &= \frac{\ell}{2D_{zz}} \frac{1 + x - \exp(-2xz_0/\ell)}{x(1+x)^2} \qquad x \equiv \frac{\alpha\ell}{\sqrt{D_{zz}}},\ L = \infty \\ T\,(\text{forw}, L) &= \frac{1}{D_{zz}} \frac{(z_0 + \ell)^2}{L + 2z_0} \frac{L/L_a}{\sinh L/L_a} \qquad L_a \equiv \ell/\sqrt{\alpha^2 D_{zz}} \end{aligned} \tag{41}$$

where $z_0 \simeq 2.1 D_{zz}$.

We still must deal with the polarization anisotropy caused by first and last skin layer. For that we need the Dyson Green's tensor in real space $\mathbf{G}(\mathbf{r})$. Since the skin layer is roughly one mean free path in thickness and $k\ell \gg 1$, the far field suffices. If we restrict ourselves again to low fields and low density we need the Fourier Transform of expression (26), in which $\widehat{\mathbf{p}}$ will be transformed to $\widehat{\mathbf{r}}$ and the remaining integrals over the absolute value of $\mathbf{p}$ are standard. We find,

$$\mathbf{G}(\mathbf{r}, \mathbf{B}) = -\frac{1}{4\pi r} \exp(ikr - r/2\ell) \qquad \times$$
$$\left\{ \left( 1 + \frac{1}{2} i\eta_3 \frac{r}{\ell} - \frac{1}{8}\eta_1^2 \left(\frac{r}{\ell}\right)^2 \right) \mathbf{\Delta}_{\mathbf{r}} + \frac{1}{2} i\eta_1 \frac{r}{\ell} \mathbf{L}(\widehat{\mathbf{r}}, \mathbf{B}) + \left( \frac{1}{8}\eta_1^2 \left(\frac{r}{\ell}\right)^2 - \frac{1}{2} i\eta_2 \frac{r}{\ell} \right) \mathbf{H}(\widehat{\mathbf{r}}, \mathbf{B}) - \frac{1}{8}\eta_1^2 \left(\frac{r}{\ell}\right)^2 \mathbf{V}(\widehat{\mathbf{r}}, \mathbf{B}) \right\} . \tag{42}$$

The four transverse tensors herein have been defined in section IV.A.

One result can be obtained without going into deep calculations and concerns the "all channel in all channel out" incoherent transmission coefficient $T$. This is the total transmitted energy flux integrated over all incident and outgoing directions. This transport quantity is intimately related to the conductance for electrons by means of the Landauer formula [34]. It can best derived from the diffusion formulas (39). An even simpler picture is obtained by replacing the source by a boundary condition as well. If $u(z)$ is the radiation density then,

$$\left. \begin{aligned} -\partial_z^2\, u(z) &= 0 \\ J_z(\text{in},0) &= \tfrac{1}{4} u(0) - \tfrac{1}{2} D^+_{zz} u'(0) = 1 \\ J_z(\text{in},L) &= \tfrac{1}{4} u(L) + \tfrac{1}{2} D^+_{zz} u'(L) = 0 \\ J_z(\text{out},L) &= T(L) = \tfrac{1}{4} u(L) - \tfrac{1}{2} D^+_{zz} u'(L) \ . \end{aligned} \right\} \tag{43}$$

¿From this it follows easily that,

$$T(\mathbf{B}, L) = \frac{4D^+_{zz}(\mathbf{B})}{L + 4D^+_{zz}(\mathbf{B})} \ . \tag{44}$$

This formula also comes out in exact Milne calculations for angular anisotropic scattering without the diffusion approximation [29] [17], though with a somewhat different numerical value for the extrapolation length. On the basis of this formula the all-channel transmission depends on both direction and absolute value of the magnetic field. This is a clear experimentally verifiable prediction. In fact, the all-channel transmission is the only known transport quantity that depends *solely* on (the anisotropy of) the diffusion constant. In the presence of absorption the diffusive absorption length $L_a \sim 1/\sqrt{D^+_{zz}}$ is also dependent on magnitude and direction of $\mathbf{B}$. In Fig. 6 we have displayed the difference in transmission for two perpendicular directions of the magnetic field. The difference $T_{||} - T_{\perp}$ can have both signs, depending of the phase shift. At low frequencies (zero phase shift) the total transmittance decreases as a function of the magnetic field, for all directions of the magnetic field. This is similar to normal electronic conductors. Note however in Fig. 5 that there exists a phase-shift region where the diffusion constant and thus the longitudinal conductance *increase* in a magnetic field.

The complication in reflection is that, contrary to transmission, short paths also contribute so that in principle the full Ladder tensor is required, and not only its diffusive approximation. As a result polarization effects occur even without a magnetic field. A realistic treatment of polarization effects in reflection requires to go beyond the diffusion approximation. In the subsections that follow we therefore only discuss the transmission.

### 1. Stokes parameters in Diffuse Transmission

For light incident along the $z$-axis, we can use Eq. (42) to transform the four-rank tensors $\mathbf{GG}^*$ in Eq. (37) into a transfer tensor $\mathcal{M}_{\mathbf{k}}$ of the skin layers acting as,

$$\mathcal{T}_{\mathbf{z},\mathbf{k}} = \mathcal{M}^+_{\mathbf{k}} \cdot \left\{ \int_0^L \mathrm{d}z_1 \int_0^L \mathrm{d}z_2 \; \mathrm{e}^{-\tau_2/} \, \mathcal{L}\left(L - z_2, z_1\right) \; \mathrm{e}^{-\tau_1} \right\} \cdot \mathcal{M}^+_{\mathbf{z}} \, . \tag{45}$$

With the boundary conditions on the diffusive bulk outcome (21), the integral gives the familiar scalar result mentioned in Eq. (41) so that, in forward direction,

$$\mathcal{T}_{\mathbf{k},\mathbf{k}} \sim \frac{\left(z_0^+ + \ell\right)^2}{D^+_{zz}} \frac{1}{L} \times \mathcal{M}^+_{\mathbf{k}} \cdot |\mathbf{d}\rangle \langle \mathbf{d}| \cdot \mathcal{M}^+_{\mathbf{k}} \, . \tag{46}$$

where the diffusive eigenfunction $|\mathbf{d}\rangle = \sqrt{A}\,|\mathbf{l}\rangle$ was found in Eq. (19), and $z_0^+ \simeq 2.1 D^+_{zz}$. The skin layer polarization tensor in direction $\mathbf{k}$ is given by,

$$\begin{aligned} \mathcal{M}^{\pm}_{\mathbf{k}} = & \left[1 - \frac{1}{2} f_2 \mathrm{Re}\, \eta_1^2 - f_1 \mathrm{Im}\, \eta_3\right] \mathbf{\Delta}_{\mathbf{k}}\mathbf{\Delta}_{\mathbf{k}} + \frac{1}{2} i f_1 \left[\eta_1 \mathbf{L}_{\mathbf{k}}\mathbf{\Delta}_{\mathbf{k}} \mp \overline{\eta}_1 \mathbf{\Delta}_{\mathbf{k}}\mathbf{L}_{\mathbf{k}}\right] + \frac{1}{4} f_2 \left[\overline{\eta}_1^2 \mathbf{\Delta}_{\mathbf{k}}\mathbf{H}_{\mathbf{k}} - \eta_1^2 \mathbf{H}_{\mathbf{k}}\mathbf{\Delta}_{\mathbf{k}}\right] \\ & + \frac{1}{2} i f_1 \left[\overline{\eta}_2 \mathbf{\Delta}_{\mathbf{k}}\mathbf{H}_{\mathbf{k}} - \eta_2 \mathbf{H}_{\mathbf{k}}\mathbf{\Delta}_{\mathbf{k}}\right] - \frac{1}{4} f_2 \left[\eta_1^2 \mathbf{V}_{\mathbf{k}}\mathbf{\Delta}_{\mathbf{k}} + \overline{\eta}_1^2 \mathbf{\Delta}_{\mathbf{k}}\mathbf{V}_{\mathbf{k}}\right] \pm \frac{1}{2} f_2 \left|\eta_1\right|^2 \mathbf{L}_{\mathbf{k}}\mathbf{L}_{\mathbf{k}} \, . \end{aligned} \tag{47}$$

We allowed - for future need in the most-crossed diagrams - a possible opposite sign for the magnetic field in the complex conjugate wave. The scalar constants $f_i$ arise from the space-integrals of the various factors $(r/\ell)^i$ in Eq. (42), and depend on the direction $\mathbf{k}$. For details we refer to Appendix B. We want to emphasize here that by arbitrarily setting $f_i = 0$ there will be no skin layer effects on polarization and the result $\mathcal{M}^{\pm}_{\mathbf{k}} = \mathbf{\Delta}_{\mathbf{k}}\mathbf{\Delta}_{\mathbf{k}}$ is obtained, as in Ref. [7]. On the other hand putting $f_i = 1$ the skin layer behaves exactly as the bulk in the sense that $\langle r \rangle = \ell$ and $\langle r^2 \rangle = 2\ell^2$ [36]. Using the identity (14) this would imply that the skin layers transform both left eigenfunctions $|\mathbf{d}\rangle$ in Eq. (46) into the transverse tensor $\left|\sum_p \mathbf{\Delta G}\left(p\hat{\mathbf{k}}\right)\right\rangle \sim |\mathbf{\Delta_k}\rangle + \mathcal{O}(n)$. In that case we would have to conclude that the polarization obtained in the bulk due to magneto-optical effects will be compensated exactly in the skin layers, and that we finally end up with no polarization as without a magnetic field. From this discussion it may be evident that the anomalous step length distribution in the skin layers plays a crucial role for the emerging polarization. The variables $f_i$ are fixed by the boundary conditions. In Appendix B we calculate that $f_1 = 1.58$ and $f_2 = 2.16$ in forward transmission. This brings us to the interesting conclusion that polarization will be present in diffuse transmission.

Polarization is best described in terms of the Stokes parameters $I$, $Q$, $U$ and $V$ which we can obtain from Eq. (46). Since Eq. (46) separates the incoming and outgoing Stokes variables completely, working out $|\mathbf{EE}\rangle_{\text{trans}} \sim \mathcal{M}^+_{\mathbf{k}} \cdot |\mathbf{d}\rangle$ will suffice. We obtain,

$$\mathbf{T_{k,k}} \sim \mathcal{M}^+_{\mathbf{k}} \cdot |\mathbf{d}\rangle \sim \ (1+T_1)\,|\mathbf{\Delta_k}\rangle + T_2\,|\mathbf{L_k}\rangle + T_3\,|\mathbf{H_k}\rangle + T_4\,|\mathbf{V_k}\rangle \ , \tag{48}$$

where

$$\begin{aligned} T_1 &= -\text{Im}\ \eta_3\,(f_1-1) + (\text{Im}\ \eta_1)^2\,(f_2-f_1) \quad ; \quad T_2 = -\text{Im}\ \eta_1\,(f_1-1) \\ T_3 &= +\text{Im}\ \eta_2\,(f_1-1) - (\text{Im}\ \eta_1)^2\,(f_2-f_1) \quad ; \quad T_4 = (\text{Im}\ \eta_1)^2\,(f_2-f_1) \end{aligned} \tag{49}$$

The Stokes parameters can now be found by choosing a frame with $z$-axis along $\mathbf{k}$ and $x$-axis and $y$-axis perpendicular. Then $T_{xx} = T_1 + T_3B_x^2 - T_4B_y^2$ , $T_{yy} = T_1 + T_3B_y^2 - T_4B_x^2$ and $T_{xy} = (T_3+T_4)\,B_xB_y - iT_2B_z$ , $T_{yx} = (T_3+T_4)\,B_xB_y + iT_2B_z$. Hence,

$$\begin{aligned} \frac{V}{I} &= \frac{i\,(T_{xy}-T_{yx})}{T_{xx}+T_{yy}} = T_2B_z = -\text{Im}\ \eta_1\,(f_1-1)\left(\mathbf{B}\cdot\hat{\mathbf{k}}\right) \ , \\ \frac{Q}{I} &= \frac{T_{xx}-T_{yy}}{T_{xx}+T_{yy}} = (T_3+T_4)\left(B_x^2-B_y^2\right) = \text{Im}\ \eta_2\,(f_1-1)\left(1-\left(\mathbf{B}\cdot\hat{\mathbf{k}}\right)^2\right)\cos 2\phi \ , \\ \frac{U}{I} &= \frac{T_{xy}+T_{yx}}{T_{xx}+T_{yy}} = (T_3+T_4)\,2B_xB_y \quad = \text{Im}\ \eta_2\,(f_1-1)\left(1-\left(\mathbf{B}\cdot\hat{\mathbf{k}}\right)^2\right)\sin 2\phi \ . \end{aligned} \tag{50}$$

The Stokes parameters $V$, $Q$ and $U$ do not depend on the particles density, and only on the direction of observation with respect to the magnetic field and the frequency. A circular Stokes parameter $V \sim B$ persists when the magnetic field is along the slab. A linear polarization proportional to $B^2$ remains if the magnetic field is perpendicular to the slab. The angle $\phi$ locates the direction of the magnetic field in the $xy$-plane. Without a magnetic field the Stokes variables would decay to zero exponentially as $\exp(-L/\ell^{**})$ [8], where the depolarization length $\ell^{**}$ is associated with non-diffusive eigenvalues of the Ladder tensor. We emphasize that the imaginary part of the variables $\eta_i$ signifies the tensor nature of scattering mean free path and scattering cross-section . If this effect were not taken into account, the Stokes parameters would decay exponentially to zero in a magnetic field as well. In leading (zero) order of the particle density, the polarization in diffuse transmission is a surface effect.

Finally we want to note that the Poynting vector associated with the transmission tensor (48) can easily be shown to be equal to $\mathbf{J_k} = \mathbf{k}\ \text{Tr}\ \mathbf{T_{k,k}}$. If we assume that the incident light is completely unpolarized, an additional similar trace factor comes in due to the incident skin layer. In forward transmission we thus obtain for the Stokes parameter $I$,

$$I = T\,(\text{forward}) \sim \frac{\left(z_0^+ + \ell\right)^2}{D_{zz}} \left[1 + T_1 + \frac{1}{2}\,(T_3+T_4)\left(1-(\mathbf{B}\cdot\hat{\mathbf{z}})^2\right)\right]^2 \frac{1}{L} \ . \tag{51}$$

This differs from the all-channel transmission (44) but nevertheless depends on the direction of the magnetic field. In Fig. 6 we show $\left(T_{||} - T_{\perp}\right)/T$ for both the all-channel transmission (44) and the forward transmission (51), quadratic in the field.

### *2. Stokes Parameters in Non-diffuse Transmission*

In this subsection we discuss the impact of a magnetic field on polarization modes that are not subject to long-range diffusion. These modes are known to carry polarization characterized by either a non-vanishing $U$, $Q$ or $V$ Stokes parameter. In the absence of a magnetic field they decay exponentially in space, the characteristic length scale being associated with non-diffuse eigenvalues of the collision operator [22]. Martinez and Maynard [8] calculated the Müller matrix for Rayleigh scatterers after a given number of scattering events $n$ using a transfer matrix method. They showed that the magnetic field pronounces the exponential decay of polarization in order $B^2$ and in addition causes a rotation proportional to $V_0 B n$ in the $(Q, U)$ plane. Nothing was seen to happen to the circular polarization $V$.

We will now investigate the decay of polarization in our transport theory. Contrary to the diffusive mode, terms linear in the field prevail. Therefore we restrict to modifications linear in the field. In this approximation the tensor $\mathcal{Q}^{\pm}(\mathbf{q} = \mathbf{0}, \mathbf{B})$ reads,

$$\mathcal{Q}_0^{\pm}(\mathbf{B}) = \frac{1}{10}\left(6\delta_{ij}\delta_{kl} + \delta_{ik}\delta_{jl} + \delta_{il}\delta_{jk}\right) + \frac{1}{10} i\eta_1 \mathcal{F}_{ijkl} \mp \frac{1}{10} i\overline{\eta}_1 \mathcal{F}_{lkji} + \frac{1}{10}\xi_1 \mathcal{Z}^{(1)}_{ijkl} \pm \frac{1}{10}\overline{\xi}_1 \mathcal{Z}^{(2)}_{ijkl} \,. \tag{52}$$

Here $\mathcal{F}_{ijkl}$ has been introduced earlier in Eq. (30). Furthermore

$$\mathcal{Z}^{(1)}_{ijkl} = 6\Phi_{ij}\delta_{kl} + \Phi_{lj}\delta_{ik} + \delta_{il}\Phi_{kj} \quad ; \quad \mathcal{Z}^{(2)}_{ijkl} = 6\Phi_{lk}\delta_{ji} + \Phi_{lj}\delta_{ik} + \delta_{kj}\Phi_{li} \,.$$

For $\mathbf{B} = 0$ the tensor $1 - Q_0^{\pm}(\mathbf{B})$ is hermitean. Its nine eigenvalues and orthonormal eigenfunctions associated with the various Stokes variables have been found in earlier work [22],

$$\begin{array}{lll} \lambda_1 = 0 & |1\rangle_{ik} = \delta_{ik}/\sqrt{3} & (I) \\ \lambda_{2,3} = 3/10 & |2,3\rangle_{ik} = \delta_{ik} \mathrm{e}^{\pm i(k-1)2\pi/3}/\sqrt{3} & (Q) \\ \lambda_{4,5,6} = 3/10 & |4,5,6\rangle_{ik} = \left[\delta_{ia}\delta_{kb} + \delta_{ib}\delta_{ka}\right]/\sqrt{2} & (U) \\ \lambda_{7,8,9} = 1/2 & |7,8,9\rangle_{ik} = i\left[\delta_{ia}\delta_{kb} - \delta_{ib}\delta_{ka}\right]/\sqrt{2} & (V) \end{array} \tag{53}$$

For the last six eigenvectors we have $(a, b) = (1,2)$, $(1,3)$ and $(2,3)$ respectively. We shall take the direction of the magnetic field along the 3-axis. With respect to the base (53) the tensor $1 - Q_0^{\pm}(\mathbf{B})$, occurring in expression (17) of the Ladder diagrams, takes the form,

$$\mathbf{L}_0^{\pm}(\mathbf{B}) = \begin{pmatrix} 0 & 0 & 0 & 0 & 0 & 0 & \frac{1}{10}z_1 & 0 & 0 \\ 0 & \frac{3}{10} & 0 & -\frac{1}{10}\chi_1 z_2 & 0 & 0 & \frac{1}{10}\overline{\chi}_2 z_3 & 0 & 0 \\ 0 & 0 & \frac{3}{10} & -\frac{1}{10}\overline{\chi}_1 z_2 & 0 & 0 & \frac{1}{10}\chi_2 z_3 & 0 & 0 \\ 0 & \frac{1}{10}\overline{\chi}_1 z_2 & \frac{1}{10}\chi_1 z_2 & \frac{3}{10} & 0 & 0 & 0 & 0 & 0 \\ 0 & 0 & 0 & 0 & \frac{3}{10} & -\frac{1}{20}z_2 & 0 & 0 & \frac{1}{20}z_4 \\ 0 & 0 & 0 & 0 & \frac{1}{20}z_2 & \frac{3}{10} & 0 & -\frac{1}{20}z_4 & 0 \\ \frac{1}{10}z_0 & \frac{1}{10}\chi_2 z_4 & \frac{1}{10}\overline{\chi}_2 z_4 & 0 & 0 & 0 & \frac{1}{2} & 0 & 0 \\ 0 & 0 & 0 & 0 & 0 & \frac{1}{20}z_4 & 0 & \frac{1}{2} & -\frac{1}{4}z_5 \\ 0 & 0 & 0 & 0 & -\frac{1}{20}z_4 & 0 & 0 & \frac{1}{4}z_5 & \frac{1}{2} \end{pmatrix} .$$

in which $\chi_1 = \left(3 + i\sqrt{3}\right)/2\sqrt{6}$ and $\chi_2 = \left(1 + i\sqrt{3}\right)/2\sqrt{6}$. We introduced,

$$\begin{array}{ll} z_0 = \frac{5}{\sqrt{6}}\left[i\left(\eta_1 \mp \overline{\eta}_1\right) - 2\left(\xi_1 \pm \overline{\xi}_1\right)\right] & z_3 = i\left(\eta_1 \mp \overline{\eta}_1\right) - 7\left(\xi_1 \pm \overline{\xi}_1\right) \\ z_1 = \frac{5}{\sqrt{6}}\left[i\left(\eta_1 \mp \overline{\eta}_1\right) - 4\left(\xi_1 \pm \overline{\xi}_1\right)\right] & z_4 = i\left(\eta_1 \mp \overline{\eta}_1\right) - 5\left(\xi_1 \pm \overline{\xi}_1\right) \\ z_2 = 2\left(\eta_1 \pm \overline{\eta}_1\right) + 7i\left(\xi_1 \mp \overline{\xi}_1\right) & z_5 = i\left(\xi_1 \mp \overline{\xi}_1\right) \end{array}$$

For the normal Ladder diagrams all $z_i$ are real-valued and $z_0 = 0$. We shall focus on two parts of this matrix outside the one-dimensional diffusive domain (given by some linear combination of first and seventh eigenvector). Firstly the "linear polarization subspace" spanned by the $\{|2\rangle, |3\rangle, |4\rangle, |5\rangle, |6\rangle\}$. Secondly the circular polarization space spanned by $|8\rangle$ and $|9\rangle$. For brevity, we do not discuss the (birefringent) coupling between both, transforming linear and circular polarization in general into elliptically polarized light.

The matrix $\mathbf{L}_0(\mathbf{B})$ restricted to the five dimensional non-diffusive subspace with linear polarization can be diagonalized. It is inferred that the magnetic field lifts the five-fold degeneracy of the eigenvalue 3/10. We find for the eigenvalues and eigenvectors,

$$\lambda_2 = \frac{3}{10} \quad ; \quad \mathbf{e}_2 = \frac{1}{\sqrt{2}}\begin{pmatrix} 1 \\ -\frac{1}{2} + \frac{1}{2}i\sqrt{3} \\ 0 \\ 0 \\ 0 \end{pmatrix}, \qquad \lambda_{3,4} = \frac{3}{10} \pm \frac{1}{10}iz_2 \quad ; \quad \mathbf{e}_{3,4} = \frac{1}{\sqrt{2}}\begin{pmatrix} \pm i\chi_1 \\ \pm i\overline{\chi}_1 \\ 1 \\ 0 \\ 0 \end{pmatrix},$$

$$\lambda_{5,6} = \frac{3}{10} \pm \frac{1}{20}iz_2 \quad ; \quad \mathbf{e}_{5,6} = \frac{1}{\sqrt{2}}\begin{pmatrix} 0 \\ 0 \\ 0 \\ 1 \\ \pm i \end{pmatrix}.$$

Quite conveniently this set is again orthonormal, although the eigenvalues have become complex-valued. As a result, the change of the eigenvalues for a small variable $\mathbf{q} \neq 0$ can be obtained with Rayleigh-Schrödinger perturbation theory,

$$\delta\lambda_i(\mathbf{q}, \mathbf{B}) = -\langle\lambda_i|\,\delta\mathcal{Q}^{\pm}(\mathbf{q}, \mathbf{B})\,|\lambda_i\rangle\ . \tag{54}$$

For $\mathbf{B} = \mathbf{0}$ most matrix elements in Eq. (54) have already been obtained in Ref. [7]. Up to terms linear in the field and for low density we have,

$$-\delta\mathcal{Q}^{\pm}_{ijkl}(\mathbf{q}, \mathbf{B}) = \frac{3}{2}\ell^2\left\langle(\widehat{\mathbf{p}}\cdot\mathbf{q})^2\left[\Delta_{ij}\Delta_{kl} + \xi_1\Delta_{im}\Phi_{mj}\Delta_{kl} \pm \overline{\xi}_1\Delta_{ij}\Phi_{lm}\Delta_{mk} - \frac{3}{2}i\left(\eta_1 L_{ij}\Delta_{kl} \mp \overline{\eta}_1\Delta_{ij}L_{lk}\right)\right]\right\rangle\ . \tag{55}$$

As before $\Delta_{ij} = \delta_{ij} - \widehat{p}_i\widehat{p}_j$ and $L_{ij} = -\Phi_{ij} + \Phi_{in}\widehat{p}_n\widehat{p}_j - \Phi_{jn}\widehat{p}_i\widehat{p}_n$ are two transverse tensors used earlier. The angular average can be expressed into the tensors $\mathcal{V}_{ijkl}$ and $\mathcal{W}_{ijklmn}$ introduced in Eq. (29). The matrix elements (54) have been obtained with a symbolic processor. We find,

$$\begin{array}{rl} \delta\lambda_2(\mathbf{q}, \mathbf{B})/\ell^2 = & \frac{59}{210}\mathbf{q}^2 - \frac{1}{7}q_3^2 \\ \delta\lambda_3(\mathbf{q}, \mathbf{B})/\ell^2 = & \frac{13}{70}\mathbf{q}^2 + \frac{1}{7}q_3^2 + \left(\xi_1 \mp \overline{\xi}_1\right)\left(\frac{13}{70}\mathbf{q}^2 + \frac{1}{7}q_3^2\right) + i\left(\eta_1 \pm \overline{\eta}_1\right)\left(\frac{57}{140}\mathbf{q}^2 + \frac{9}{70}q_3^2\right) \\ \delta\lambda_4(\mathbf{q}, \mathbf{B})/\ell^2 = & \frac{13}{70}\mathbf{q}^2 + \frac{1}{7}q_3^2 - \left(\xi_1 \mp \overline{\xi}_1\right)\left(\frac{13}{70}\mathbf{q}^2 + \frac{1}{7}q_3^2\right) - i\left(\eta_1 \pm \overline{\eta}_1\right)\left(\frac{57}{140}\mathbf{q}^2 + \frac{9}{70}q_3^2\right) \\ \delta\lambda_5(\mathbf{q}, \mathbf{B})/\ell^2 = & \frac{9}{35}\mathbf{q}^2 - \frac{1}{14}q_3^2 + \left(\xi_1 \mp \overline{\xi}_1\right)\left(\frac{9}{70}\mathbf{q}^2 - \frac{1}{28}q_3^2\right) + i\left(\eta_1 \pm \overline{\eta}_1\right)\left(\frac{9}{56}\mathbf{q}^2 + \frac{18}{35}q_3^2\right) \\ \delta\lambda_6(\mathbf{q}, \mathbf{B})/\ell^2 = & \frac{9}{35}\mathbf{q}^2 - \frac{1}{14}q_3^2 - \left(\xi_1 \mp \overline{\xi}_1\right)\left(\frac{9}{70}\mathbf{q}^2 - \frac{1}{28}q_3^2\right) - i\left(\eta_1 \pm \overline{\eta}_1\right)\left(\frac{9}{56}\mathbf{q}^2 + \frac{18}{35}q_3^2\right)\ . \end{array} \tag{56}$$

For the Ladder diagrams we find that only $\lambda_2(\mathbf{B})$ is real-valued and independent of $\mathbf{B}$. The other four occur in complex conjugates. The eigenvalues of $\mathcal{Q}^-(\mathbf{q},\mathbf{B})$ remain real-valued.

Writing $\lambda_i(\mathbf{q},\mathbf{B}) = \lambda_i(\mathbf{B}) + D_i(\widehat{\mathbf{q}},\mathbf{B})\,\mathbf{q}^2$ and by using the formal identity,

$$\int \mathrm{d}\mathbf{q}\ \frac{\exp(i\mathbf{q}\cdot\mathbf{r})}{D_i(\widehat{\mathbf{q}})\,\mathbf{q}^2+\lambda_i} \overset{r\to\infty}{\sim} \frac{1}{4\pi D_i(\widehat{\mathbf{r}})\,r}\exp\left(-r\sqrt{\frac{\lambda_i}{D_i(\widehat{\mathbf{r}})}}\right) \equiv L_i(\mathbf{r},\mathbf{B}) \ , \tag{57}$$

the Ladder sum for linear polarization can be transformed into real space. The result is,

$$\mathbf{L}_{\mathrm{P}}(\mathbf{r}) \sim \mathbf{U}\cdot\begin{pmatrix} L_2(\mathbf{r},\mathbf{B}) & & & & \emptyset \\ & L_3(\mathbf{r},\mathbf{B}) & & & \\ & & L_4(\mathbf{r},\mathbf{B}) & & \\ & & & L_5(\mathbf{r},\mathbf{B}) & \\ \emptyset & & & & L_6(\mathbf{r},\mathbf{B}) \end{pmatrix}\cdot\mathbf{U}^\dagger \ , \tag{58}$$

in which the unitary matrix $\mathbf{U}$ is given by,

$$\mathbf{U} = \frac{1}{\sqrt{2}}\begin{pmatrix} 1 & i\chi_1 & -i\chi_1 & 0 & 0 \\ -\frac{1}{2}+\frac{1}{2}i\sqrt{3} & i\overline{\chi}_1 & -i\overline{\chi}_1 & 0 & 0 \\ 0 & 1 & 1 & 0 & 0 \\ 0 & 0 & 0 & 1 & 1 \\ 0 & 0 & 0 & i & -i \end{pmatrix} \ .$$

The complex value of the eigenvalues $\lambda_i(\mathbf{B})$ and "diffusion constants" $D_i(\widehat{\mathbf{r}},\mathbf{B})$ will give rise to oscillations in the Stokes parameters $Q$ and $U$ as a function of $\mathbf{B}$ and $\mathbf{r}$. To determine these we can consider the special case that initial and final wave vectors are directed both along the $\mathbf{B}-$axis and find $Q(r)$ and $U(r)$ . Although this is still a calculation valid in the bulk, it will approximately be valid for transmission in a slab geometry with length $\mathbf{r} = L\widehat{\mathbf{z}}$. If the initial light is 100 % linearly polarized ($U=0$) we can take the initial polarization matrix $\left|E_i(\mathbf{0})\,\overline{E}_j(\mathbf{0})\right\rangle = \frac{1}{3}\left(|1\rangle_{ij} + |2\rangle_{ij} + |3\rangle_{ij}\right)$. Using Eq. (58) we get for the propagated light,

$$\begin{aligned}\left|E_i(\mathbf{r})\,\overline{E}_j(\mathbf{r})\right\rangle \sim \text{diffuse} &+ \left\{\left(\frac{1}{2}-\frac{1}{2}i\sqrt{3}\right)L_2(\mathbf{r}) + \left(\frac{3}{2}+\frac{1}{2}i\sqrt{3}\right)\frac{L_3(\mathbf{r})+L_4(\mathbf{r})}{2}\right\}|2\rangle_{ij} \\ &+ \left\{\left(\frac{1}{2}-\frac{1}{2}i\sqrt{3}\right)L_2(\mathbf{r}) + \left(\frac{3}{2}-\frac{1}{2}i\sqrt{3}\right)\frac{L_3(\mathbf{r})+L_4(\mathbf{r})}{2}\right\}|3\rangle_{ij} \\ &+ \sqrt{6}\,\frac{L_4(\mathbf{r})-L_3(\mathbf{r})}{2i}\,|4\rangle_{ij} \ .\end{aligned}$$

¿From this expression one can infer that,

$$\begin{aligned}\frac{Q(r)}{I(r)} &\sim \mathrm{e}^{-r/\ell_P}\cos(k_{\mathrm{P}}r+\varphi_0) \ , \\ \frac{U(r)}{I(r)} &\sim \mathrm{e}^{-r/\ell_P}\sin(k_{\mathrm{P}}r+\varphi_0) \ .\end{aligned} \tag{59}$$

The linear Stokes parameters are exponentially small in the optically thick regime as is well known from other work. In a magnetic field the $U$ and $Q$ acquire an oscillatory character: the linear polarization rotates along $\mathbf{B}$ proportional over the traversed distance. Although this may be evident in view of the Faraday effect standing at the base of this, we nevertheless want to emphasize that we are addressing a (damped) diffusive mode and not the coherent beam. Since incident and outgoing wave vector are both parallel to $\mathbf{B}$, the diffusive contribution (50) vanishes, and Eq. (59) contains the only linear polarization of the light.

¿From Eq. (56) we find the explicit result that $\ell_{\mathrm{P}}$ and $k_{\mathrm{P}}$, determining the propagation of the linear polarization, are given by

$$\ell_{\mathrm{P}} = \ell\sqrt{\frac{23}{21}} \quad ; \quad k_{\mathrm{P}} = \frac{1}{\ell}\sqrt{\frac{21}{23}}\left(-\frac{133}{138}\mathrm{Re}\,\eta_1 - \frac{10}{3}\mathrm{Im}\,\xi_1\right) = \frac{\mu}{\ell}\sqrt{\frac{21}{23}}\left(\frac{133}{138} - \frac{112}{23}\sin^2\alpha\right) . \qquad (60)$$

In the low frequency regime the variable $k_{\mathrm{P}}$ does not depend on frequency and has the value $k_{\mathrm{P}} = 1.29 \times 9\sqrt{\varepsilon}/\left(\varepsilon+2\right)^2 fV_0 B$. This is 1.29 larger than the rotation of linear polarization of the coherent beam.

A similar calculation is possible in the two-dimensional subspace spanned by the eigenvectors $|8\rangle$ and $|9\rangle$ which carry circular polarization.

$$\lambda_8\left(\mathbf{q},\mathbf{B}\right) = \frac{1}{2} + \frac{1}{4}iz_5 + \ell^2\left\{\frac{1}{5}\mathbf{q}^2 - \frac{1}{10}q_3^2 - \left(\xi_1 \mp \overline{\xi}_1\right)\left(\frac{1}{10}\mathbf{q}^2 - \frac{1}{20}q_3^2\right) - i\left(\eta_1 \pm \overline{\eta}_1\right)\left(\frac{3}{40}\mathbf{q}^2 + \frac{6}{40}q_3^2\right)\right\} , \qquad (61)$$

$$\lambda_9\left(\mathbf{q},\mathbf{B}\right) = \frac{1}{2} - \frac{1}{4}iz_5 + \ell^2\left\{\frac{1}{5}\mathbf{q}^2 - \frac{1}{10}q_3^2 + \left(\xi_1 \mp \overline{\xi}_1\right)\left(\frac{1}{10}\mathbf{q}^2 - \frac{1}{20}q_3^2\right) + i\left(\eta_1 \pm \overline{\eta}_1\right)\left(\frac{3}{40}\mathbf{q}^2 + \frac{6}{40}q_3^2\right)\right\} . \qquad (62)$$

For the normal Ladder diagrams these eigenvalues are again each other complex-conjugate. Writing $\lambda_8\left(\mathbf{q},\mathbf{B}\right) = \lambda_8\left(\mathbf{B}\right) + D_8\left(\widehat{\mathbf{q}},\mathbf{B}\right)\mathbf{q}^2$ and $L_8\left(\mathbf{r}\right)$ as in Eq. (57) we obtain in real space,

$$\mathbf{L}_{\mathrm{C}}\left(\mathbf{r}\right) = \frac{1}{2}\begin{pmatrix} 1 & 1 \\ -i & i \end{pmatrix} \cdot \begin{pmatrix} L_8\left(\mathbf{r}\right) & 0 \\ 0 & \overline{L}_8\left(\mathbf{r}\right) \end{pmatrix} \cdot \begin{pmatrix} 1 & i \\ 1 & -i \end{pmatrix} . \qquad (63)$$

The eigenvectors $|8\rangle$ and $|9\rangle$ couple in when the magnetic field is perpendicular to the direction of propagation. By taking a circularly wave as initial wave it is easily shown that the Stokes variable $V\left(r\right)$ is given by

$$\frac{V\left(r\right)}{I(r)} \sim \mathrm{e}^{-r/\ell_{\mathrm{C}}}\cos\left(k_{\mathrm{C}}r + \varphi_0\right) , \qquad (64)$$

Explicitly,

$$\ell_{\mathrm{C}} = \ell\sqrt{\frac{2}{5}} \quad ; \quad k_{\mathrm{C}} = \frac{1}{\ell}\frac{3}{8}\sqrt{\frac{5}{2}}\mathrm{Re}\,\eta_1 = -\frac{\mu}{\ell}\frac{3}{8}\sqrt{\frac{5}{2}}\left(1 - 2\sin^2\alpha\right) .$$

We thus infer that the amount of circular polarization oscillates as well as a function of distance. However, for the circular polarization the exponential decay is much faster and the wavenumber associated with the oscillation is much smaller.

## V. COHERENT BACKSCATTERING

After the theoretical predictions in Refs. [6] and [7] the Coherent Backscattering in a magnetic field has been studied intensively by Erbacher, Lenke and Maret and a vast amount of experimental data is now becoming available [10] [11]. So far the cone is the only multiple-light-scattering phenomenon investigated thoroughly in relation with a magnetic field. As has been predicted theoretically the enhancement of the cone was seen to go down proportional to the magnetic field and good agreement was found between experimental data, theoretical predictions and numerical simulations [8].

New features emerged from the experiments which mainly concern very high fields, up to 30 Tesla. The decrease of the Coherent Backscattering Peak due to the Faraday effect turned out to be very well described by the exponential step length distribution between successive collisions [36]. So far Coherent Backscattering in a magnetic field seems to be the only phenomenon explicitly sensitive to the whole distribution and not to only its two lowest moments [38]. In addition, a saturation of the enhancement factor was seen to occur for very large fields that also comes out of the numerical simulations by Martinez and Maynard. It turned out that only one single dimensionless variable is needed to describe the dephasing up to high fields, being the product of an average Verdet constant, the magnetic field strength and some mean free path. This is quite similar to the Beenakker-Senftleben effect for which the magnetic impact on the heat conductivity turned out to be a function of only the product $\vartheta = \omega_p \tau$ of precession velocity of the magnetic moment and the molecule's mean free collision time [12] [14]. It seemed necessary to introduce a length scale $\ell_F^*$ different from both scattering and transport mean free path. It describes the loss of circular polarization in multiple scattering (being relevant for the average Faraday rotation between collisions). Due to spin-flips it is in general larger than the transport mean free path (which determines the memory of momentum). Initially some optimism existed to have the Faraday effect enhanced by resonant scattering [10] but restoration of a numerical factor in the formulas made the effect less pronounced, leaving it as an open question. Our approach will contain a first theoretical attempt to incorporate the effect of resonant scattering on the Faraday rotation of polarization.

Concerning polarization it is found that in opposite helicity hardly no Coherent Backscattering is found (even for large magnetic fields). This is in agreement with predictions first made by MacKintosh and John [7]. Furthermore, the azimuthal line shape of the cone was seen to remain perfectly rotationally symmetric. For linear polarization channels there is a special angle between incoming and outgoing polarization vectors for which the cone is maximal. This has been explained by the Faraday rotation in the skin layers. In all these experiments the magnetic field was directed perpendicular to the slab, which is the most convenient set-up in the magnet.

To interpret the experimental data one focuses only on the diffusive mode. In that case an almost scalar picture emerges with some polarization effects included to describe the Faraday rotation. In the following we will look in more detail what remains of the diffusive picture for Coherent Backscattering in a magnetic field. We will not address the role of non-diffusive modes. It is clear that they donot influence the angular behavior close to the peak.

### A. Most-Crossed Diagrams

The most-crossed diagrams associated with Coherent Backscattering can - as usual - be obtained by reversing the direction of propagation of the bottom line in the Ladder diagrams. For vector waves the bottom polarization indices have to be reversed as well, just like the direction of external magnetic fields, in order to remain consistent with the reciprocity relation $\boldsymbol{\varepsilon}(\mathbf{B}) = \boldsymbol{\varepsilon}^t(-\mathbf{B})$. This gives relation (25) and brings us to study the Ladder sum $\mathcal{L}^-(\omega, \mathbf{q}, \mathbf{B}, \Omega)$ defined already in Eq. (17). Let us first set $\Omega$ and $\mathbf{q}$ equal to zero and treat them later as perturbative variables. We expect that both left and right hand eigenfunction of the eigenvalue of $\mathcal{Q}_0^-(\mathbf{B})$ closest to 1 is some linear combination of the independent vectors $\mathbf{I}$, $\mathbf{\Phi}$ and $\mathbf{BB}$ (with Gram matrix $\mathbf{M}$). In the notation of section IV.A the eigenvalue equations

$$\mathcal{Q}_0^- \sum_i r_i \;|i\rangle = (1-\lambda)\sum_i r_i \;|i\rangle \qquad ; \qquad \sum_i \bar{l}_i \langle i|\, \mathcal{Q}_0^- = (1-\lambda)\sum_i \bar{l}_i \langle i|$$

can be rewritten as

$$\left(\lambda \mathbf{M} - \mathbf{L}^-\right) \cdot \begin{pmatrix} r_1 \\ r_2 \\ r_3 \end{pmatrix} = 0 \quad , \qquad \left(\bar{l}_1,\ \bar{l}_2,\ \bar{l}_3\right) \cdot \left(\lambda \mathbf{M} - \mathbf{L}^-\right) = 0 \ .$$

The bar signifies complex-conjugation. We recall that $\mathbf{L}^-$ is the matrix obtained by sandwiching the four-rank tensor $1 - \mathcal{Q}_0^-$ between our choice of base vectors. The resulting characteristic equation is,

$$\det\left(\lambda \mathbf{M} - \mathbf{L}^-\right) = 0 \ . \tag{65}$$

We anticipate that $\lambda \sim B^2$ in leading order of the magnetic field [7]. In this order the solution of Eq. (65) is,

$$\lambda = \frac{1}{3}\left(L_{11}^- - L_{12}^- L_{21}^-\right) \ . \tag{66}$$

The second "off-diagonal" term is a subtle vector effect, overlooked in a heuristic theoretical treatment. The coordinates $r_i$ and $l_i$ of right and left hand eigenfunction become,

$$\begin{array}{ll} r_1 = 1 & l_1 = 1 \\ r_2 = -L_{21}^- & l_2 = -\overline{L}_{12}^- \\ r_3 = 5\left(\lambda - L_{31}^- + L_{32}^- L_{21}^-\right) & l_3 = 5\left(\lambda - \overline{L}_{13}^- + \overline{L}_{23}^- \overline{L}_{12}^-\right) \end{array} \tag{67}$$

Due to the algebraic complexity it is convenient to express everything in terms of the phase shift $\alpha$ ($\sin\alpha \equiv s$) introduced in Eq. (5). Using the explicit form of $\mathbf{L}^-$ obtained in the previous section and the dimensionless variables $\mu$, $\zeta$ and $\varpi$ defined in Eq. (7) we obtain ,

$$\lambda = \frac{2}{3}\mu^2\left(2s^2 + 1\right) \ , \tag{68}$$

and,

$$\begin{array}{lll} r_2 = i\mu & ; & l_2 = -i\mu\,(2s^2+1) \\ r_3 = 5\lambda - 2\mu^2 & ; & l_3 = 5\lambda + \mu^2\,(4s^4 + s^2 - 2) - (\varpi - \zeta)\sin 2\alpha \quad . \end{array} \tag{69}$$

Notice the *explicit* occurrence of the scattering phase shift $\alpha$ in these formulas, especially the one for the the dephasing $\lambda$. Contrary to what was found earlier for the normal Ladder diagrams $\mathcal{L}^+$, the coefficients of the eigenvectors turn out to be complex-valued. For zero phase shift (valid at low frequencies) we infer that $r_i = \bar{l}_i$.

Having found both eigenfunctions and associated eigenvalue for $\mathcal{L}^-$ in the diffusive regime, we can obtain the algebraic expression for $\mathcal{L}^-(\mathbf{q}, \mathbf{B})$ in the diffusion approximation. Calculational details are left to Appendix D. We obtain the rather simple expression,

$$\mathcal{L}^-(\mathbf{q}{,}\Omega, \mathbf{B}) = \frac{6\pi}{\ell} \frac{|\mathbf{l}(-\mathbf{B})\rangle\langle \mathbf{l}(\mathbf{B})|}{\lambda(B) - \langle \mathbf{l}|\ \delta\mathcal{Q}^-(\mathbf{q}, \Omega,\ \mathbf{B})\ |\mathbf{r}\rangle} \tag{70}$$

The symmetric form of this expression is - of course - due to the reciprocity relation (23). In particular, the minus sign in $|\mathbf{l}(-\mathbf{B})\rangle$ is required by the reciprocity principle.

To obtain the tensor associated with the most-crossed diagrams, polarization indices of the conjugate wave in Eq. (70) have still to be reversed but this can best be done after having added the skin layers as well. In the subsections that follow we will find explicitly the diffusion tensor, discuss some polarization properties of the most-crossed diagrams, and finally address the shape of the cone and the decrease of the enhancement factor.

### B. Boltzmann Diffusion Tensor in Cone

Using the eigenfunctions (69) of the four-rank tensor $\mathcal{Q}_0^-(\mathbf{B})$ and the matrices $\mathbf{S}^-$ and $\mathbf{T}^-$ defined in Eq. (34), the calculation of the matrix element $\langle \mathbf{l}| -\delta\mathcal{Q}^-(\mathbf{q},\Omega=0)\ |\mathbf{r}\rangle$ in Eq. (70) is possible. To extract a diffusion tensor (an intrinsically dynamic transport quantity), a finite frequency difference $\Omega$ must be incorporated as well. The matrix element for the dynamics is $\langle \mathbf{l}| -\delta\mathcal{Q}^-(\Omega)\ |\mathbf{r}\rangle$ and follows from Eq. (35). The Boltzmann diffusion tensor $\mathbf{D}_B^-(\mathbf{B})$ is defined by bringing the Ladder tensor $\mathcal{L}^-$ into the form,

$$\mathcal{L}^-(\mathbf{q},\Omega,\mathbf{B}) \sim \frac{1}{-i\Omega+\mathbf{q}\cdot\mathbf{D}_B^-(\mathbf{B})\cdot\mathbf{q}+\lambda(B)/\ell} \tag{71}$$

An anti-symmetric part for $\mathbf{D}_B^-(\mathbf{B})$ has no meaning. We obtain, in terms of the variables (7) and the phase-shift $\alpha$ ($s=\sin\alpha$) in Eq. (5),

$$\mathbf{D}_B^-(\mathbf{B}) = \frac{1}{3}\ell\left(\mathbf{I}+d_{\text{iso}}\mathbf{I}+d_{\text{ani}}\mathbf{B}\mathbf{B}\right)\ , \tag{72}$$

with,

$$d_{\text{iso}} = \mu^2\left(-\frac{88}{5}s^4+\frac{907}{45}s^2-\frac{79}{45}\right)+\frac{1}{15}\left(16\varpi+9\zeta\right)\sin 2\alpha$$
$$d_{\text{ani}} = \mu^2\left(\frac{44}{5}s^4-\frac{97}{15}s^2-\frac{71}{15}\right)-\frac{1}{5}\left(\varpi-\zeta\right)\sin 2\alpha$$

We plotted these corrections as a function of $s$ in Fig. 5. We infer that they clearly deviate from the ones of the Boltzmann diffusion constant $\mathbf{D}_B^+(\mathbf{B})$. Two special cases can be discussed.

The low frequency regime associated with genuine Rayleigh scattering has $s\sim(\omega a)^3\to 0$. In that case we find parallel to the field $\Delta D_{||}^-/D_0=-6.48\ \mu^2$ and perpendicular to the field $\Delta D_{\perp}^-/D_0 = -1.75\ \mu^2$. This implies that diffusion is suppressed in all directions but mainly *along* the magnetic field lines. The last property is in sharp contrast to what was found for the normal incoherent energy denoted by $\mathcal{L}^+$ (as well as for the Beenakker-Senftleben effect) where the suppression was seen to be more pronounced perpendicular to $\mathbf{B}$.

Another special case is when the scatterers are set to resonance: $s\simeq 1$. In that case $\Delta D_{||}^-/D_0 = -1.6\ \mu^2$ and $\Delta D_{\perp}^-/D_0=+0.8\ \mu^2$. Here, diffusion is even enhanced perpendicular to the field. Due to polydispersity this situation may be very difficult to achieve experimentally. Nevertheless, it is a nice illustration how the scattering phase shift of the particles may influence macroscopic transport phenomena.

Anisotropy in the diffusion constant $\mathbf{D}_B^-$ may be observed from an anisotropy of the line shape of the cone. This will be the topic of the fourth subsection.

### C. Polarization of Cone

To obtain the most-crossed diagrams at retroflection from the bulk result one must add the skin layers as indicated in Eq. (38) and finally transpose the polarization indices of bottom line. We will consider only polarization terms linear in the external magnetic field. The normalized lefthand eigenfunction then simplifies to,

$$|\mathbf{l}(\pm\mathbf{B})\rangle = \frac{|\mathbf{I}\rangle \pm l_1\,|\mathbf{\Phi}\rangle}{\sqrt{3}}\quad .$$

The four-rank tensor in Eq. (70) works out to

$$\frac{1}{3}\left\{|\mathbf{I}\rangle\langle\mathbf{I}| + \overline{l}_1\,|\mathbf{\Phi}\rangle\langle\mathbf{I}| + \overline{l}_1\,|\mathbf{I}\rangle\langle\mathbf{\Phi}|\right\} \quad .$$

Next we have to add the skin layers for the incident and emergent light. As has been done for the incoherent part the skin layers give rise to a transfer matrix (47), but now with opposite magnetic field for the conjugate wave, and a value $f_1 = 1.21$ in reflection (Appendix B). Linear in the field it reads,

$$\mathcal{M}^-_{\mathbf{k}} = \mathbf{\Delta}_{\mathbf{k}}\mathbf{\Delta}_{\mathbf{k}} + \frac{1}{2}if_1\left[\eta_1\mathbf{L}_{\mathbf{k}}\mathbf{\Delta}_{\mathbf{k}} + \overline{\eta}_1\mathbf{\Delta}_{\mathbf{k}}\mathbf{L}_{\mathbf{k}}\right] \ .$$

Altogether the polarization matrix for Coherent Backscattering, in the diffusion approximation, becomes,

$$\mathcal{M}^-_{\mathbf{k}}\cdot\mathcal{L}^-(\mathbf{q},\mathbf{B})\cdot\mathcal{M}^-_{-\mathbf{k}} = \frac{2\pi}{\ell^2}C(\mathbf{B},\mathbf{q})\left\{|\mathbf{\Delta}_{\mathbf{k}}\rangle\langle\mathbf{\Delta}_{\mathbf{k}}| + iF\left[|\mathbf{L}_{\mathbf{k}}\rangle\langle\mathbf{\Delta}_{\mathbf{k}}| + |\mathbf{\Delta}_{\mathbf{k}}\rangle\langle\mathbf{L}_{\mathbf{k}}|\ \right]\right\}$$

where $F = f_1\mathrm{Re}\,\eta_1 - \mu\ - 2\mu s^2$ and $C(\mathbf{B},\mathbf{q})$ is the diffusive expression in Eq. (71). In order to get the tensor representation for the most- crossed diagrams one must transpose the bottom polarization indices $k$ and $l$ and insert $\mathbf{q} = \mathbf{k} + \mathbf{k}'$ (at backscattering this operation leaves the tensor expression in the upper formula unchanged). This gives the final result,

$$\mathcal{C}_{ijkl}(\mathbf{B},\mathbf{k},\mathbf{k}') = \frac{2\pi}{\ell^2}\ C(\mathbf{B},\mathbf{k}+\mathbf{k}')\times\left\{\Delta_{il}\Delta_{jk} + iF\left[L_{il}\Delta_{kj} + \Delta_{il}L_{kj}\right]\right\} \ . \qquad (73)$$

The scalar factor $(2\pi/\ell^2)C(\mathbf{B},\mathbf{q})$ determines the line shape and will be the topic of the next subsection. In particular, we notice that the diffusion approximation decouples polarization and line shape.

With respect to the polarization part it is instructive to consider the experimentally relevant situations of linear polarization and circular polarization. Let us first discuss linear polarization channels for the case that the magnetic field is perpendicular to the slab (Fig. 7). A polarization vector $\mathbf{v}$ is incident, $\mathbf{w}$ is outgoing. The angle $\alpha_0$ is defined as the angle between both vectors in the direction imposed by the magnetic field using the Lenz Rule. It then follows that,

$$w_i w_k \mathcal{C}_{ijkl} v_j v_l \sim \cos^2\alpha_0 - F\sin 2\alpha_0 \ . \qquad (74)$$

Obviously the maximum value of this expression is not reached for $\alpha_0 = 0$ as would happen without a magnetic field. The angle with maximum signal is,

$$\alpha_0(\max) = -\frac{1}{2}\arctan 2F \simeq -F + \mathcal{O}\left(B^3\right) \ . \qquad (75)$$

The rotation angle is linear in the magnetic field and shows up as the combined effect of skin-layer and first and last scattering. For $V_0 > 0$ the variable $F$ is strictly negative. In that case the angle always obeys the Lenz rule. For low frequencies we find $\alpha_0(\max) = 2.21\mu$ radians. At resonance this is $\alpha_0(\max) = 1.79\mu$ radians. Both the existence of the angle and its sign are in agreement with the experiments [11].

If the magnetic field is directed along the slab the first polarization modifications due to the magnetic field enter in second order of the field and are strictly due to birefringence. Concerning circular polarization it is straightforward to show from Eq. (73) that the opposite helicity channel has no signal (in diffusion approximation). We found this to be true for orders $B^2$ in the field as well. This agrees with the experiment.

### D. Dephasing and Line Shape of Cone

The line shape of Coherent Backscattering is determined by the factor $C(\mathbf{q},\mathbf{B})$ in Eq. (73). In an infinite medium it is given by,

$$C(\mathbf{q},\mathbf{B}) = \frac{1}{\lambda/\ell + \mathbf{q}\cdot\mathbf{D}^{-}(\mathbf{B})\cdot\mathbf{q}} \,. \tag{76}$$

Its approximate form for the slab geometry, starting from this bulk result, is obtained from the imaging method. For a semi-infinite slab the result was found in Eq. (41), identifying $\alpha^2 = \lambda/\ell + D_{||}^{-}\mathbf{q}_{||}^2 + \left(D_{\perp}^{-} - D_{||}^{-}\right)\left(\mathbf{B}_{||}\cdot\mathbf{q}_{||}\right)^2$

$$R(\mathbf{q},\mathbf{B}) = \frac{1}{2D_{zz}^{-}}\frac{1 + x - \exp\left(-2xz_0^{-}/\ell\right)}{x(1+x)^2} \,, \tag{77}$$

where

$$x^2\left(\mathbf{B}_{||},\mathbf{q}_{||}\right) = \ell^2 \times \frac{\lambda/\ell + D_{||}^{-}\mathbf{q}_{||}^2 + \left(D_{\perp}^{-} - D_{||}^{-}\right)\left(\mathbf{B}_{||}\cdot\mathbf{q}_{||}\right)^2}{D_{zz}^{-}} \,.$$

determines the dephasing due to both the Faraday effect ($\lambda$) and a finite backscattering angle determined by $\mathbf{q}_{||} = 2k\sin(\theta/2)\,\hat{\mathbf{r}}_{||}(\varphi)$. The extrapolation length is now given by $z_0^{-} \simeq 2.1D_{zz}^{-}$ and is in principle different from the extrapolation length $z_0^{+}$ in the Ladder diagrams because the diffusion constant is different. The enhancement of the cone relative to the incoherent background can be written as,

$$\Xi(\mathbf{B},\theta,\varphi) = 1 + \frac{1 + x - \exp\left(-2xz_0^{-}\right)}{\left(2z_0^{-}+1\right)x(1+x)^2} \times \left\{\frac{D_{zz}^{+}}{D_{zz}^{-}}\frac{2z_0^{-}+1}{2z_0^{+}+1}\frac{\text{polarization cone}}{\text{polarization Ladder}}\right\} \,. \tag{78}$$

Let us discuss these results. First we discuss the suppression of Coherent Backscattering at backscattering ($\theta = 0$). Secondly the line shape for finite angles. The suppression at backscattering in principle depends on two factors, shown separately in Eq. (78). Both factors depend on both direction and magnitude of the magnetic field. If we assume that $\mu^2 \gg \zeta, \varpi$ (Appendix C) the dimensionless variable $\mu$ becomes a one-parameter scaling variable for the enhancement factor.

The first factor describes the dephasing in multiple scattering due to the Faraday rotation and is the main reason why the enhancement factor is suppressed [7]. This contribution dephases similarly in angle and in magnetic field, which is also found experimentally up to large fields [11].

The second factor in Eq. (78) is due to the different diffusion constants and the different polarization behavior of Ladder and most-crossed diagrams, giving them different weights. This part is only a function of the magnetic field and not of the angle. The polarization will be influenced by low orders of scattering as well. At the time of writing no experimental evidence exists for this extra factor. We will focus on the first factor.

An effective medium value for the Verdet constant can be obtained from Eq. (76) by imagining the scatterers in a homogeneous medium with Verdet constant $V_{\text{eff}}$ (the situation envisaged by MacKintosh and John). This defines $V_{\text{eff}}$ by,

$$\lambda = \frac{4}{3}V_{\text{eff}}^2 B^2 \ell^2 \Rightarrow V_{\text{eff}} = \left(2s^2+1\right)^{1/2}\mu/(B\ell) \,. \tag{79}$$

In particular, using the value for $\mu$ at low frequencies given in Appendix C and $1/\ell = 2f\omega^4 a^3 (\varepsilon-1)^2/(\varepsilon+2)^2$ , with $f$ the packing fraction, we get

$$\frac{V_{\text{eff}}}{fV_0} = \left(\frac{3}{\varepsilon+2}\right)^2 \sqrt{2\varepsilon}. \tag{80}$$

This outcome differs by a subtle factor of $\sqrt{2}$ from the effective Verdet constant found from the Dyson dispersion law (13). This is due to the fact that the diffusive eigenfunction also changes. For $\varepsilon = 1.15$ Eq. (80) predicts an enhancement over the effective medium value of 1.38. Experimentally a value of $1.52 \pm 0.15$ is obtained [10]. A better agreement between theory and experiment can be obtained if a finite value for the phase shift $\alpha$ is adopted since this will augment $V_{\text{eff}}$. In later experiments [11] one introduces the Faraday rotation length $\ell_F^* > \ell^*$ in order to distinguish the correlation of circular polarization from the correlation in momentum (determined by the usual transport mean free path $\ell^*$). In our theory this length scale does not show up. However, a value $\ell_F^*/\ell^* > 1$ can be translated into an enhancement of the effective medium Verdet constant $V_{\text{eff}}/fV_0 = \sqrt{\ell_F^*/\ell^*}$.

Near resonance (Appendix C) one can relate the variable $\mu$ to the "path length" of the wave inside the scatterer according to $\mu = 2V_0 B L_{\text{path}}$ so that,

$$\frac{V_{\text{eff}}}{V_0} = \sqrt{6}\,\frac{L_{\text{path}}}{\ell}\,. \tag{81}$$

The Dyson equation gives $V_{\text{Dyson}}/V_0 = -L_{\text{path}}/\ell$. For strong resonant scattering the path length can exceed the mean free path. As a result, the effective Verdet constant will be strongly enhanced by resonant scattering. The same mechanism causes the transport speed to go down [23]. A correlation between both was suggested by Erbacher, Lenke and Maret [10] but has not been found experimentally so far. It is well possible that internal spin-flips in Mie scattering destroy this phenomenon [38].

The line shape of Coherent Backscattering depends on the direction of the magnetic field. When the field is perpendicular to the slab, one finds that $x^2 \sim \lambda\,/\ell D_{\|}^- + \mathbf{q}_{\|}^2$. Hence the cone is independent of the azimuthal angle $\varphi$. In fact this is the standard experimental set-up and no $\varphi$-dependence was observed. Notice also that in this case the width of the cone is determined by the *scattering* mean free path $\ell$, and not the transport mean free path as is known for Mie scattering [28]. This might explain why the predicted diminishing of the diffusion constant relevant for the cone due to the magnetic field has gone unnoticed so far.

Experimentally one considered sofar the cone in a magnetic field that is perpendicular to the slab. The helicity preserving channel is often preferred because it has the advantage of not giving a single-scattering contribution for the incoherent background. Moreover this channel is known to have a purely azimuthally isotropic line shape in the absence of a field. It can be verified that the polarization impact (73) of the magnetic field on the cone does not change this property. Linear polarization channels do not obey this property due to low orders of scattering for which the direction of the incident polarization vector is not yet scrambled [39].

For arbitrary field direction it follows that

$$x^2 \sim \frac{\lambda\,/\ell + \left\{D_{\|}^- + \left(D_{\perp}^- - D_{\|}^-\right)\left[1-(\mathbf{B}\cdot\hat{\mathbf{z}})^2\right]\cos^2\varphi\right\}\mathbf{q}_{\|}^2}{D_{\|}^- + \left(D_{\perp}^- - D_{\|}^-\right)\left[1-(\mathbf{B}\cdot\hat{\mathbf{z}})^2\right]}\,.$$

which depends on $\varphi$ if $\mathbf{B}$ is not along the $z$-axis. In Fig. 8 we display the line shape for the magnetic field along the slab. Since we found in Eq. (72) that $D_{\perp}^- > D_{\|}^-$ the cone will more sharpened along the field direction. The azimuthal dependence of the cone with respect to the magnetic field is the only experimental way to determine the full diffusion tensor in the most-crossed diagrams.

## VI. FIELD CORRELATION

In this section we discuss briefly the field correlation function $\left\langle E_i\left(\mathbf{B}\right)\overline{E}_j\left(\mathbf{0}\right)\right\rangle$ in transmission. The intensity correlation function $\left\langle I\left(\mathbf{B}\right)I\left(\mathbf{0}\right)\right\rangle$ is measurable and has indeed been measured in transmission. In the $\mathrm{C}_1$-approximation the latter is given by (one plus) the absolute square of the field correlation, meaning that a lot of information of the field correlation function can be extracted experimentally.

Using the methods outlined in the previous sections essentially all desired information of the field correlation can be obtained analytically. In fact, the way of calculation is quite similar to the one of $\left\langle E_i\left(\mathbf{B}\right)\overline{E}_j\left(-\mathbf{B}\right)\right\rangle$ which is relevant for Coherent Backscattering. The difference is that the "building block" for the Ladder diagrams is not $\mathcal{Q}^-\left(\mathbf{B},\mathbf{q}\right)$ as defined in Eq. (15), but rather

$$\mathcal{Q}^{\mathrm{s}}\left(\omega,\mathbf{q},\mathbf{B}\right)=\sum_{\mathbf{p}}\mathbf{G}\left(\omega,\mathbf{p}+\frac{\mathbf{q}}{2},\mathbf{B}\right)\mathbf{G}^*\left(\omega,\mathbf{p}-\frac{\mathbf{q}}{2},\mathbf{0}\right)\cdot n\ \mathbf{t}\left(\omega,\mathbf{B}\right)\mathbf{t}^*\left(\omega,\mathbf{0}\right)\ . \tag{82}$$

Just like for the cone, the Ladder sum for the field correlation can be written as

$$\begin{aligned}\mathcal{L}^{\mathrm{s}}\left(\omega,\mathbf{q},\mathbf{B}\right)&=n\mathbf{t}\left(\mathbf{B}\right)\mathbf{t}^*\left(\mathbf{0}\right)\cdot\left\{1+\mathcal{Q}^{\mathrm{s}}\left(\mathbf{q}\right)+\mathcal{Q}^{\mathrm{s}}\left(\mathbf{q}\right)\cdot\mathcal{Q}^{\mathrm{s}}\left(\mathbf{q}\right)+\cdots\right\}\\&=n\mathbf{t}\left(\mathbf{B}\right)\mathbf{t}^*\left(\mathbf{0}\right)\cdot\sum_{a,b=\mathbf{r},\mathbf{l}}\left|a\right\rangle\left(\frac{1}{\mathbf{N}^{\mathrm{s}}-\mathbf{P}^{\mathrm{s}}\left(\mathbf{q}\right)}\right)_{ab}\left\langle b\right|\ ,\end{aligned}$$

with exactly the same notation as in section V. We restricted again to the two-dimensional subspace spanned by left and right eigenvector of $\mathcal{Q}^{\mathrm{s}}\left(\omega,\mathbf{q}=\mathbf{0},\mathbf{B}\right)$. Let $\mathbf{L}^{\mathrm{s}}\left(\mathbf{B},\mathbf{q}\right)$ be the $3\times3$ matrix containing the matrix elements of the four rank tensor $1-\mathcal{Q}^{\mathrm{s}}\left(\omega,\mathbf{q},\mathbf{B}\right)$ with respect to the base $\{\mathbf{I},\mathbf{\Phi},\mathbf{BB}\}$. We obtain,

$$\begin{aligned}L^{\mathrm{s}}_{11}&=\tfrac{1}{4}\eta_1^2-\tfrac{3}{2}i\eta_3+\tfrac{i}{2}\eta_2\ +3\xi_3\ -\xi_2+\tfrac{i}{2}\eta_1\xi_1\\L^{\mathrm{s}}_{12}&=\tfrac{1}{2}i\eta_1-2\xi_1\\L^{\mathrm{s}}_{21}&=\tfrac{1}{2}i\eta_1-\xi_1\\L^{\mathrm{s}}_{22}&=1+\tfrac{3}{20}\eta_1^2+\tfrac{i}{10}\eta_2-\tfrac{i}{2}\eta_3\ +\xi_3+\tfrac{2}{5}i\eta_1\xi_1\\L^{\mathrm{s}}_{23}&=\tfrac{1}{10}i\eta_1\\L^{\mathrm{s}}_{32}&=\tfrac{1}{10}i\eta_1-\tfrac{1}{5}\xi_1\\L^{\mathrm{s}}_{13}&=\tfrac{1}{20}\ \eta_1^2+\tfrac{2}{5}i\eta_2-\tfrac{i}{2}\eta_3\ -\xi_2+\xi_3\\L^{\mathrm{s}}_{31}&=\tfrac{1}{20}\eta_1^2-\tfrac{i}{2}\eta_3+\tfrac{2}{5}i\eta_2\ +\xi_3\ -\tfrac{4}{5}\xi_2+\tfrac{i}{10}\eta_1\xi_1\\L^{\mathrm{s}}_{33}&=\tfrac{1}{5}+\tfrac{3}{140}\eta_1^2+\tfrac{12}{35}i\eta_2-\tfrac{2}{5}i\eta_3-\tfrac{4}{5}\xi_2+\tfrac{4}{5}\xi_3\end{aligned} \tag{83}$$

We will not discuss polarization properties here. The most important result is the remnant of long-range diffusion, expressed by the eigenvalue of $\mathcal{Q}^{\mathrm{s}}\left(\mathbf{q}=\mathbf{0}\right)$ closest to one. It follows from a characteristic equation similar to Eq. (65) that,

$$\lambda^{\mathrm{s}}=\frac{1}{3}\left(L^{\mathrm{s}}_{11}-L^{\mathrm{s}}_{12}L^{\mathrm{s}}_{21}\right)\ . \tag{84}$$

It can readily be verified that this in fact a complex number. Explicitly,

$$\mathrm{Re}\ \lambda^{\mathrm{s}} = \frac{1}{6}\mu^2\left(2s^2+1\right) = \frac{\lambda}{4} \ ,$$
$$\mathrm{Im}\ \lambda^{\mathrm{s}} = -\frac{1}{6}\mu^2 \sin 2\alpha + \frac{1}{6}\varpi + \frac{1}{3}\zeta \ . \tag{85}$$

The real part of the number represents again the dephasing due to the Faraday effect and is a factor of 4 less than the dephasing in the crossed diagrams, as was also found in Ref. [10]. The imaginary part will give rise to a phase factor in the field-field correlation. As in the expression for Coherent Backscattering,

$$\left\langle E(\mathbf{B})\overline{E}(\mathbf{0})\right\rangle \sim \frac{1}{\mathbf{q}\cdot\mathbf{D}^{\mathrm{s}}(\mathbf{B})\cdot\mathbf{q}+\lambda^{\mathrm{s}}/\ell} \ ,$$

so that one obtains in transmission from a slab with length $L$, using formula (41),

$$\frac{\left\langle E(\mathbf{B})\overline{E}(\mathbf{0})\right\rangle}{\left\langle E(\mathbf{0})\overline{E}(\mathbf{0})\right\rangle} = \frac{L\sqrt{3\lambda^{\mathrm{s}}}/\ell}{\sinh\left(L\sqrt{3\lambda^{\mathrm{s}}}/\ell\right)} \sim \exp\left(-L\sqrt{3\lambda^{\mathrm{s}}}/\ell\right) \ . \tag{86}$$

To keep things simple we did not include anisotropy in the diffusion constant (which in fact becomes complex-valued as well). Since $\lambda^{\mathrm{s}}$ is complex-valued, formula (86) contains an extensive phase factor $\exp(-i\Theta)$ with $\Theta = \mathrm{Im}\ \sqrt{3\lambda^{\mathrm{s}}}L/\ell = V_{\mathrm{eff}}BL$ which has not been considered before. This phase is not measurable in the absolute square, although the imaginary part of $\sqrt{3\lambda^{\mathrm{s}}}$ modifies the slope somewhat at low fields (see Fig. 8).

Using the values at low frequencies in Appendix C one finds,

$$\mathrm{Re}\ \sqrt{\lambda^{\mathrm{s}}} = \frac{3\sqrt{6\varepsilon}}{2(\varepsilon-1)^2(\omega a)^3}\frac{V_0 B}{\omega} \ ,$$
$$\mathrm{Im}\ \sqrt{\lambda^{\mathrm{s}}} = \frac{1}{4\sqrt{6\varepsilon}}\left[\frac{5\varepsilon-1}{\varepsilon-1}\frac{V_0 B}{\omega} - 2\sqrt{\varepsilon}\frac{MB^2}{V_0 B/\omega}\right] \ . \tag{87}$$

The imaginary part is a factor $(\omega a)^3$ smaller than the real part. The low-frequency regime $\omega < a$ is therefore not favorable to observe the phase. Real and imaginary part become comparable when the variables $\mu^2$, $\zeta$ and $\varpi$ become of the same order of magnitude. This can be achieved when $\omega a \approx 1$. We emphasize that we deal here with the phase of the complex wave and not with the geometric phase associated with the polarization factor. Oscillations in the field correlation $\left\langle E_x(\mathbf{B})\overline{E}_x(\mathbf{0})\right\rangle$ and the intensity correlation $\langle I(\mathbf{B})I(\mathbf{0})\rangle$ can originate from the latter due to the fact that the total Faraday rotation is always proportional to the total traversed length, even is the path is snakelike and not straight [8], quite similar as was found earlier for the Stokes variables in the normal incoherent transmission. However, they do not survive in the diffuse regime.

In Fig. 9 we show field and intensity correlation function. The imaginary part causes a small modification in the intensity correlation function at low arguments but does basically not change. The field correlation, however, will oscillate due to the presence of the phase $\Theta$, and goes through zero near a specific value of $V_0BL$.

## VII. CONCLUSIONS

In this paper we have discussed the impact of magneto-optical effects on multiple elastic scattering of light using the framework of transport theory. An inhomogeneous random medium is considered in which the scatterers suffer from the Faraday effect and Cotton-Mouton birefringence and not the surrounding medium. The scatterers are modelled by point-like objects, enabling us to get most results in closed form. Several aspects of multiple scattering have been addressed.

We have discussed the complex dispersion law. The complex wavenumber (the real part of which gives the effective-medium dielectric constant and the imaginary part the scattering mean free path) depends on the direction of propagation with respect to the direction of the magnetic field. Especially such dependence on the scattering mean free path has not been discussed elsewhere and causes many differences with other work.

Concerning the average energy density we have calculated the diffusion tensor for the light in a magnetic field. We found diffusion to be different along and perpendicular to the field lines. The symmetric part follows finally by summing up the Ladder diagrams (equivalent to solving the equation of radiative transfer in the diffusion approximation). The anisotropy causes the all-channel transmission coefficient of a slab (magneto-conductance) to be different for different orientations of the magnetic field. The anti-symmetric part of the diffusion tensor does not feature in the diffusion equation but generates a transverse current, perpendicular to magnetic field and energy-density gradient. It will be the subject of a separate paper [16].

We have considered the Stokes parameters in transmission. For the magnetic field perpendicular to the slab we found that a non-zero circular Stokes parameter $V$ persists in diffuse transmission and is proportional to the magnetic field strength. The linear Stokes parameters decay exponentially, but the presence of the magnetic field causes a rotation in the $(U, Q)$ plane. If the field is parallel to the slab we find that a finite linear polarization remains having the direction of the magnetic field and proportional to the square of the field. The circular parameter $V$ now decays exponentially but is again accompanied by an oscillatory factor.

We investigated the most-crossed diagrams, responsible for Coherent Backscattering. The decrease of the factor of two in Coherent Backscattering as a function of the magnetic field (due to the modification of the normal reciprocity relations) has been obtained. When the field is parallel to the slab we predict the line shape to depend on the azimuthal angle. The diffusion tensor featuring in the most-crossed diagrams turned out not to be the same as the one in the Ladder diagrams. Thus in a magnetic field one has two different diffusion tensors. In fact, the diffusion tensor for Coherent Backscattering was seen to suffer a lot more from the magnetic field and is flattened along the field lines.

The correlation function of the electric field vector with and without external magnetic field has been considered theoretically. We obtained a complex-valued diffusive eigenvalue in this case. This gives rise to oscillations as a function of the magnetic field. Sofar, only the intensity correlation has been measured and not the correlation function of the electric field itself.

Some questions are very difficult to solve theoretically, and experiments and numerical simulations may lead us here. A comparison with the Beenakker-Senftleben effect - as discussed out loud in this paper - may help us. For instance, does the diffusion tensor saturate for high magnetic fields? The calculation presented in this paper is exact in orders $B$ and $B^2$ but breaks down when the variable $\mu$ introduced in this paper becomes comparable to one. Going beyond this regime seems difficult but not impossible. Experimentally one has at the moment values maximally equal to $\mu \approx 1$. Concerning the dephasing in Coherent Backscattering modifications beyond orders $\mathcal{O}(B^2)$ have already been reported in multiple scattering experiments [11]. Since modifications in the diffusion tensor have been shown to be of the same order we may conclude that this question seems to be resolvable in the near future. In the case of the Beenakker-Senftleben effect the modifications are, in the saturation regime - at most a few percents. What happens to the diffusion tensor of light in large fields is theoretically not known either. In the Beenakker-Senftleben case the modifications in the saturation regime are determined only by the *intrinsical* asymmetry of the molecules' cross-section. In our case the asymmetry of the cross-section is caused by the magnetic field itself. The physical argument valid for the Beenakker-Senftleben effect why saturation sets in for large magnetic field, may not apply here.

A qualitative conclusion that one can draw from our calculations is that many properties of multiple light scattering in a magnetic field depend *explicitly* on the scattering phase shift of the scatterers. In an experiment one often has a mixture of different kinds of particles and some average value should come out. In a monodisperse sample with resonant scatterers however, this notion might give rise to rather uncommon features. Usually the phase shift only enters indirectly by means of the diffusion constant or the mean free path.

It is also possible to consider the medium to be magneto-optic rather than the scatterers. We do not expect this case to be qualitatively different although no definite conclusions should be drawn for this case on the basis of the calculations in the present paper. Even for the aimed situation with the magneto-optics inside the scatterers, the model we considered here may be too simplistic. However, doing better in the hope of being more close to the experiment will be difficult, if not impossible. Moreover, various approximations may overlook the physical phenomena that come out of our model. As such we sincerely hope that this paper may serve as a guide for new experiments.

## ACKNOWLEDGMENTS

We are indepted to Ralf Lenke and Georg Maret for their continuous interest in this work, and for making available many experimental results prior to publication. We acknowledge Alexandre Martinez and Geert Rikken for many helpful discussions. Bart van Tiggelen acknowledges a grant from the French department of Foreign Affairs. The work of Th.M. Nieuwenhuizen was made possible by the Royal Netherlands Academy of Arts and Sciences (KNAW).

## APPENDIX A: TENSOR CONVENTIONS

In general we denote scalars by Roman numbers, second-rank tensors bold, and four-rank tensors calligraphic. We will use the convention that $\overline{a}$ denotes the complex conjugate of a scalar, $\mathbf{C}^*$ the hermitean conjugate and $\overline{\mathbf{C}}$ the complex conjugate of a second-rank tensor $\mathbf{C}$, and $\mathcal{Q}^\dagger$ the adjoint of a four-rank tensor. The second-rank tensor $\Delta\mathbf{A}$ denotes the anti-hermitean tensor $\mathbf{A}-\mathbf{A}^*$. As customary, implicit summation over repeated indices is assumed.

In our notation, the diagram $^{i-}_{k-}\mathcal{C}^{-j}_{-l}$ denotes the four-rank tensor $\mathcal{C}_{ijkl}$ (in the notation of Ref. [7] this would be $\mathcal{C}_{ikjl}$). Two tensors of rank two $\mathbf{A}$ and $\mathbf{B}$ make up a four-rank tensor $(\mathbf{AB})$ determined by $(\mathbf{AB})_{ijkl} = A_{ij}B_{lk}$. By definition $(\mathbf{AB})\cdot\mathbf{C} \equiv \mathbf{A}\cdot\mathbf{C}\cdot\mathbf{B}$ and $\mathbf{D}\cdot(\mathbf{AB}) \equiv \mathbf{A}^*\cdot\mathbf{D}\cdot\mathbf{B}^* = (\mathbf{A}^*\mathbf{B}^*)\cdot\mathbf{D}$. Tensor multiplication occurs as $(\mathbf{AB})\cdot(\mathbf{CD}) = (\mathbf{A}\cdot\mathbf{C})(\mathbf{D}\cdot\mathbf{B})$. It is easy to show that

$$\{(\mathbf{AB})\cdot(\mathbf{CD})\}\cdot\mathbf{F} = (\mathbf{AB})\cdot\{(\mathbf{CD})\cdot\mathbf{F}\} \ ,$$

so that we can forget about the rang of multiplication. A scalar product for second-rank tensors can be introduced as

$$(\mathbf{A},\mathbf{B}) \equiv \mathrm{Tr}\ \mathbf{A}^* \cdot \mathbf{B}\ ,$$

which we will sometimes write as $\langle \mathbf{A} \mid \mathbf{B} \rangle$. The cyclic property of the trace guarantees that,

$$(\mathbf{D}, (\mathbf{AB}) \cdot \mathbf{C}) = (\mathbf{D} \cdot (\mathbf{AB}), \mathbf{C}) \equiv \langle \mathbf{D} |\ (\mathbf{AB})\ | \mathbf{C} \rangle\ .$$

The adjoint $\mathcal{P}^\dagger$ of a four-rank tensor $\mathcal{P}$ must obey $(\mathbf{A}\ , \mathcal{P} \cdot \mathbf{B}) = \left(\mathcal{P}^\dagger \cdot \mathbf{A}, \mathbf{B}\right)$. It follows that $(\mathbf{AB})^\dagger = (\mathbf{A}^* \mathbf{B}^*)$ . In general we use the notation,

$$\langle \mathbf{D} |\ \mathcal{L}\ | \mathbf{C} \rangle = (D^*)_{ki} \mathcal{L}_{ijkl} C_{jl}\ ,$$

where we note that in many physical applications $\mathbf{D}$ and $\mathbf{C}$ are either hermitean or anti-hermitean.

## APPENDIX B: CALCULATION OF POLARIZATION CONSTANTS

The constants $f_i$ show up in the calculation of the transfer matrices of the skin layers in reflection and transmission due to the Faraday effect and birefringence. They determine an effective length with respect to the scattering mean free path $\ell$ according to

$$f_1 = \left\langle \frac{r}{\ell} \right\rangle_{\text{skin}}; \qquad f_2 = \frac{1}{2} \left\langle \left(\frac{r}{\ell}\right)^2 \right\rangle_{\text{skin}}; \qquad f_3 = \left\langle \frac{r r'}{\ell^2} \right\rangle_{\text{skin}}, \tag{B1}$$

where $r = z/\cos\theta_{\text{in}}$ and $r' = z'/\cos\theta_{\text{out}}$, both measured from the boundary into the slab. In principle they depend on the magnetic field. For our purposes it is sufficient to evaluate these variables for zero magnetic field. The constants $f_i$ also depend on the angles $\theta_{\text{in}}$ and $\theta_{\text{out}}$. We concentrate on waves incident and leaving along the $z$-axis.

For Rayleigh scatterers without magnetic field we have $\ell = \ell^*$. Let us therefore normalize $\ell = \ell^* = 1$. For the step length distribution $P(z) = \exp(-z)$ - valid far away from the boundaries - it follows that $f_1 = f_2 = 1$ [36]. The joint distribution of $z$ and $z'$ in reflection is in principle given by $\exp(-z)\,\rho(z, z')\exp(-z')$. Using the diffusion approximation given in Eq. (41) for a semi-infinite slab we find,

$$P_{\text{ref}}(z, z') = \frac{1}{z_0 + 1/2}\ \mathrm{e}^{-z}\ \left[\min(z, z') + z_0\right]\ \mathrm{e}^{-z'}\ ,$$

where the front factor has been chosen such to have normalization 1. Using this distribution we find for reflection, ($z_0 \approx 0.71$)

$$f_1 = \frac{z_0 + 3/4}{z_0 + 1/2} \approx 1.21; \qquad f_2 = \frac{z_0 + 7/8}{z_0 + 1/2} \approx 1.31; \qquad f_3 = \frac{z_0 + 5/4}{z_0 + 1/2} \approx 1.62\ . \tag{B2}$$

In transmission first ($z$) and last ($z'$) skin layer are far apart and the joint distribution factorizes into $P_{\text{trans}}(z) P_{\text{trans}}(z')$ ,

$$P_{\text{trans}}(z) = \frac{z_0 + z}{z_0 + 1}\ \mathrm{e}^{-z}\ .$$

In the case of transmission

$$f_1 = \frac{z_0 + 2}{z_0 + 1} \approx 1.58; \qquad f_2 = \frac{z_0 + 3}{z_0 + 1} \approx 2.16; \qquad f_3 = f_1^2\ . \tag{B3}$$

## APPENDIX C: DIMENSIONLESS CONSTANTS

In this Appendix we relate the dimensionless constants $\mu$, $\zeta$ and $\varpi$ defined in Eq. (7) and the phase shift $\alpha$ in Eq. (5) that occur throughout the paper, to scatterer properties. These variables determine the role of magneto-optical effects in multiple scattering. The variable $\mu$ signifies the modification of reciprocity, determines the Faraday rotation in multiple scattering and is linear in the magnetic field; $\varpi$ and $\widehat{\zeta}$ both determine the birefringence in multiple scattering. The size parameter is defined as usual as $x \equiv \omega a$ with $a$ the radius of the particles. Their volume is given by $v = 4\pi a^3/3$. We have,

$$\mu = \frac{9\sqrt{\varepsilon}}{(\varepsilon-1)^2 x^3} \times \frac{V_0 B}{\omega}\,, \quad \zeta = \frac{18\varepsilon}{(\varepsilon-1)^3 x^3} \times \left(\frac{V_0 B}{\omega}\right)^2\,, \quad \varpi = \frac{9}{2(\varepsilon-1)^2 x^3} \times \left[\left(\frac{V_0 B}{\omega}\right)^2 - 2MB^2\sqrt{\varepsilon}\right]\,. \tag{C1}$$

It is seen that for small size parameters and/or dielectric constants close to one $\mu^2 \gg \zeta, \varpi$, so that $\mu$ becomes the most important variable. A typical value for $\mu$ is (taking the dielectric constant 1.15 at wavelength 514 nm),

$$\mu = \frac{0.73}{x^3}\,\frac{V_0}{90^\circ/\mathrm{mm/T}} \times \frac{B}{15\ \mathrm{T}}\,.$$

At very low frequencies $x \ll 1$ the variable $\mu$ can in fact become quite large. Physically this corresponds to a large number of Faraday rotations along one mean free path. The phase shift of the scattering matrix at low frequencies is well-known to be

$$\alpha = \frac{2}{3}\frac{\varepsilon-1}{\varepsilon+2}x^3\,.$$

The Faraday rotation in the particle is determined by $\mu$. Near a resonance we can find a very physical expression. If we neglect the longitudinal field in the scatterer and assume $\varepsilon \gg 1$, the resonant frequency is $\omega_0^2 = 4\pi\Gamma/\varepsilon v$. The path length $L_\mathrm{p}$ of the light in a particle can be defined as the sensitivity of the scattering albedo with respect to absorption [35]. For our model it is inversely proportional to the line width $\omega_0^2\Gamma$ and reads $L_\mathrm{p} = 3/2\sqrt{\varepsilon} \times 1/\omega_0^2\Gamma$ (To have $L_\mathrm{p} \gg a$ one must choose $\Gamma^2 \ll a^2\sqrt{\varepsilon}$). Then,

$$\mu = 2V_0 B L_\mathrm{p}\,. \tag{C2}$$

The simplicity of this formula suggests general validity, with only the numerical factor in front changing from model to model. The Faraday rotation in the particle would then be proportional to the pathlength, independent on any other parameter. For a strong resonance the Faraday effect could thus be resonantly enhanced.

In comparing expression (68) to the one used by Lenke *etal.* [11] (using a momentum $q_F \approx V_\mathrm{medium} B$), leads us to identify $\mu\sqrt{2\sin^2\alpha + 1} = q_F \ell/\sqrt{2}$. Experimentally one has at most $q_F^2 \ell^2 \approx 2$ leading to $\mu \approx 1$. The perturbational treatment of this paper is valid when $\mu \ll 1$.

## APPENDIX D: DERIVATION OF MOST-CROSSED DIAGRAMS

In this Appendix we calculate the Ladder diagrams $\mathcal{L}^-$ (the minus sign signifying an opposite direction of magnetic field for the complex conjugate wave). By the reciprocity principle (25) these are related to the most-crossed diagrams.

Left and righthand side eigenfunctions relevant for the diffusive regime have been obtained in Eq. (69). To avoid subtilities arising from the fact that both eigenfunctions are nearly parallel, we prefer the base $|\mathbf{r}\rangle$ and $|\mathbf{s}\rangle \equiv (|\mathbf{r}\rangle - |\mathbf{l}\rangle)/(r_1 - l_1)$ and normalize. The matrix $\mathbf{N}^-$ is the Gram matrix of this new base after normalization. With respect to this base the Ladder diagrams $\mathcal{L}^-$ become, for finite frequency difference $\Omega$ and momentum difference $\mathbf{q}$,

$$n\mathbf{t}(\mathbf{B})\mathbf{t}^*(-\mathbf{B}) \cdot \left[\mathcal{Q}^-(\mathbf{q},\Omega) + \mathcal{Q}^-(\mathbf{q},\Omega)\cdot\mathcal{Q}^-(\mathbf{q},\Omega) + \cdots\right] =$$

$$= n\mathbf{t}(\mathbf{B})\mathbf{t}^*(-\mathbf{B}) \cdot \sum_{a,b=1}^{2} |a\rangle \left( \left(\mathbf{N}^-\right)^{-1} \cdot \mathbf{P}^-(\mathbf{q},\Omega) \cdot \frac{1}{\mathbf{N}^- - \mathbf{P}^-(\mathbf{q},\Omega)} \right)_{ab} \langle b| \quad .$$

Notice that we have let the series to start with double scattering in order not to double count single scattering. The matrix $\mathbf{P}^-(\mathbf{q},\Omega,\mathbf{B})$ is defined by $P^-_{ij}(\mathbf{q},\Omega,\mathbf{B}) = \langle i|\ \mathcal{Q}^-(\mathbf{q},\Omega,\mathbf{B})\ |j\rangle$ and calculation yields,

$$\mathbf{N}^- - \mathbf{P}^-(\mathbf{q},\Omega,\mathbf{B}) = \begin{pmatrix} \lambda & \kappa^- \\ 0 & \frac{1}{3}+\lambda' \end{pmatrix} + \begin{pmatrix} -i\Omega/v_{11} + \mathbf{q}\cdot\mathbf{D}_{11}\cdot\mathbf{q} & -i\Omega/v_{12} + \mathbf{q}\cdot\mathbf{D}_{12}\cdot\mathbf{q} \\ -i\Omega/v_{21} + \mathbf{q}\cdot\mathbf{D}_{21}\cdot\mathbf{q} & -i\Omega/v_{22} + \mathbf{q}\cdot\mathbf{D}_{22}\cdot\mathbf{q} \end{pmatrix} \quad . \tag{D1}$$

Herein is $\lambda'$ some known constant proportional to $B^2$ that we shall not need and,

$$\kappa^-(B) = \frac{1}{3}\left(L^-_{12} + \overline{r}_1\right) = -\frac{2}{3} i\mu \left(1 + s^2\right) \ .$$

The entries $v_{ij}$ and $\mathbf{D}_{ij}$ follow straightforwardly from Eqs. (35) and (34). The outlook of Eq. (D1) is quite similar to what was found earlier for $\mathbf{N}^+ - \mathbf{P}^+$ except that now, for $\Omega, \mathbf{q} = \mathbf{0}$, the eigenvalue 0 with algebraic multiplicity 2 has disappeared in favor of two *real-valued* eigenvalues $\lambda$ and $\frac{1}{3} + \lambda'$ both with multiplicity 1. The first is proportional to $B^2$ and the remnant of long-range diffusion. From the characteristic equation it is possible to show that - in first order perturbation theory - this eigenvalue changes into,

$$\lambda(\mathbf{q},\Omega,\mathbf{B}) = \lambda - i\Omega v_{11} + 3i\kappa^-\Omega + \mathbf{q}\cdot\mathbf{D}_{11}\cdot\mathbf{q} - 3\kappa^-\mathbf{q}\cdot\mathbf{D}_{21}\cdot\mathbf{q} = \lambda - \langle\mathbf{l}|\ \delta\mathcal{Q}^-(\mathbf{q},\Omega)\ |\mathbf{r}\rangle \ . \tag{D2}$$

The corresponding eigenvector remains approximately the same. In the diffusion approximation we ignore all non-diffusive eigenvalues, and we arrive at,

$$\mathcal{L}^-(\mathbf{q},\Omega,\mathbf{B}) = n\mathbf{t}(\mathbf{B})\mathbf{t}^*(-\mathbf{B}) \cdot \left\{ (|\mathbf{r}\rangle, |\mathbf{s}\rangle\ ) \cdot \mathbf{N}^- \cdot \mathbf{P}^-(\mathbf{0},0) \cdot \mathbf{U} \cdot \begin{pmatrix} \lambda(\mathbf{q},\Omega,\mathbf{B})^{-1} & 0 \\ 0 & 0 \end{pmatrix} \mathbf{U}^{-1} \cdot \begin{pmatrix} \langle\mathbf{r}| \\ \langle\mathbf{s}| \end{pmatrix} \right\} \ .$$

Here the matrix $\mathbf{U}$ is the matrix of transformation with the two eigenvectors as column vectors,

$$\mathbf{U} = \begin{pmatrix} 1 & 3\kappa^- \\ 0 & 1 \end{pmatrix} .$$

Going through the various matrix multiplications we arrive at the rather simple result that,

$$\mathcal{L}^-(\mathbf{q},\Omega,\mathbf{B}) = n\mathbf{t}(\mathbf{B})\mathbf{t}^*(-\mathbf{B}) \cdot \frac{|\mathbf{r}\rangle\langle\mathbf{l}|}{\lambda(\mathbf{q},\Omega,\mathbf{B})} .$$

Finally, we can use Eq. (32) to calculate the remaining matrix product. It turns out that,

$$n\mathbf{t}(\mathbf{B})\mathbf{t}^*(-\mathbf{B}) \cdot |\mathbf{r}\rangle = \frac{6\pi}{\ell}\ \left[|\mathbf{I}\rangle - l_1|\mathbf{\Phi}\rangle + l_2|\mathbf{BB}\rangle\right] = \frac{6\pi}{\ell}\ |\mathbf{l}(-\mathbf{B})\rangle \ .$$

This gives the following final outcome for the four-rank tensor $\mathcal{L}^-(\mathbf{q},\mathbf{B})$ in the diffusion approximation,

$$\mathcal{L}^-(\mathbf{q},\Omega,\mathbf{B}) = \frac{6\pi}{\ell} \frac{|\mathbf{l}(-\mathbf{B})\rangle\langle\mathbf{l}(\mathbf{B})|}{\lambda - \langle\mathbf{l}|\ \delta\mathcal{Q}^-(\mathbf{q},\Omega)\ |\mathbf{r}\rangle} \tag{D3}$$

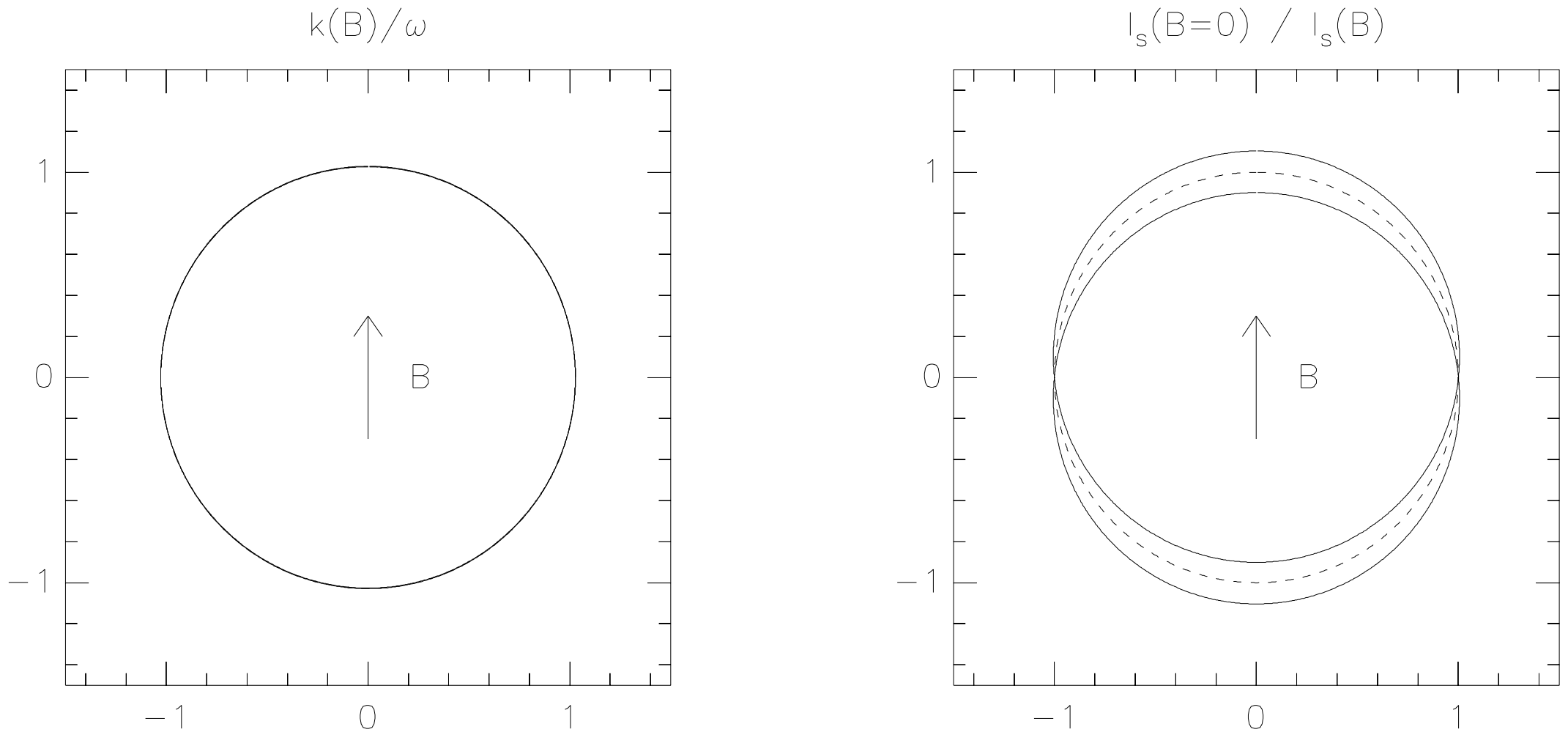

FIG. 1. Polar plots for the dispersion law at low frequencies, in various directions of propagation in an inhomogeneous medium exposed to a magnetic field. The direction of the magnetic field is indicated by the arrow. The dashed circle represents the isotropic case, without magnetic field. On the lefthand side the wavenumber for the two solutions, right the inverse scattering mean free path. We adopted a packing fraction of $f = 30\,\%$, a dielectric constant $\varepsilon = 1.2$ and a magnetic field such that $V_0 B/\omega = 0.005$. The degeneracy of the wavenumber is not resolvable but since the mean free path is very large, the dephasing between successive collisions is nevertheless considerable (see Fig. 3).

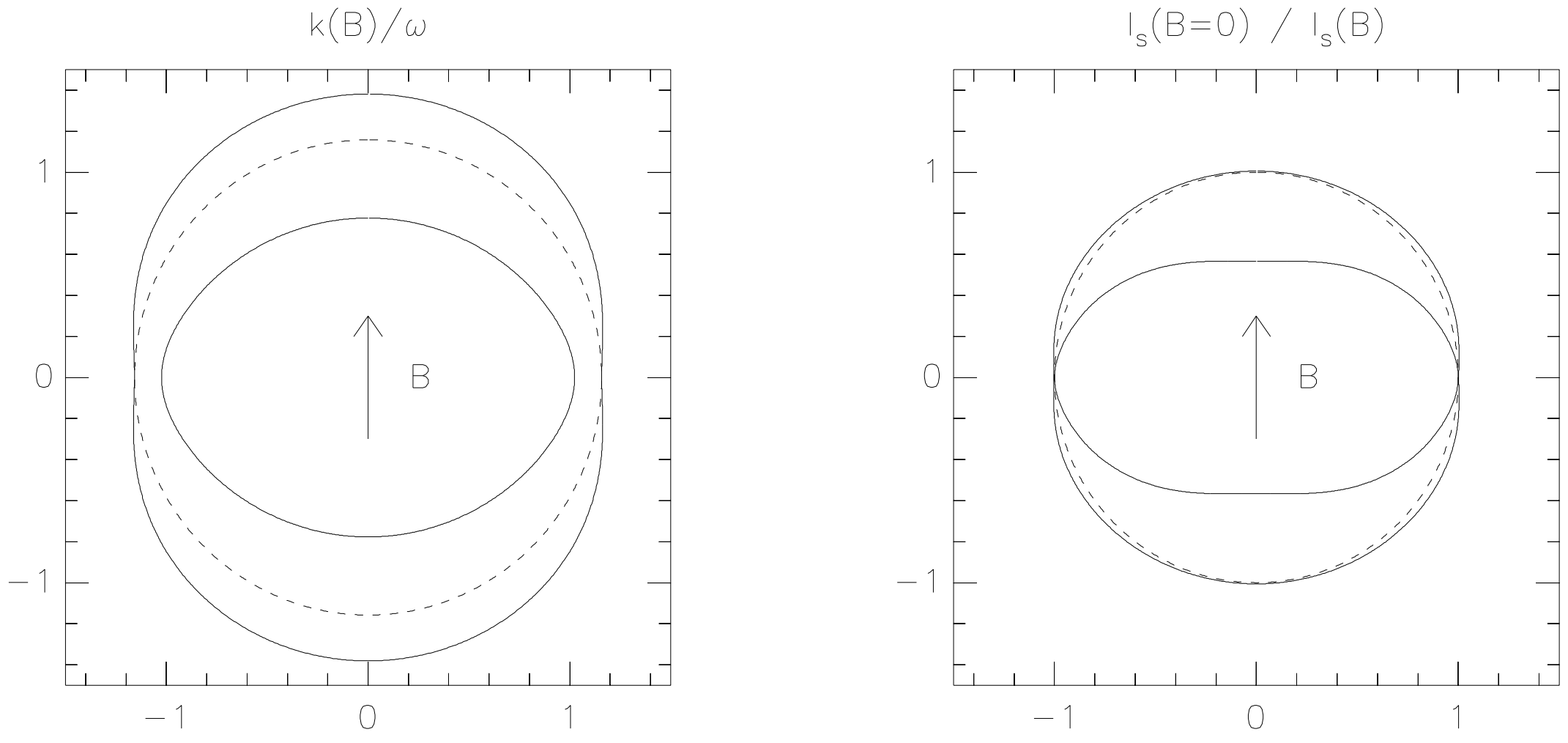

FIG. 2. As in the previous graph but now on resonance (a scattering phase shift of $\pi/2$ has been given to the scattering particles). We adopted a size parameter $x = 1$, dielectric constant $\varepsilon = 1.3$, packing fraction $30\,\%$ and $V_0 B/\omega = 0.005$. The degeneracy of both wavenumber and scattering mean free path is severe, and suffers from Cotton-Mouton birefringence.

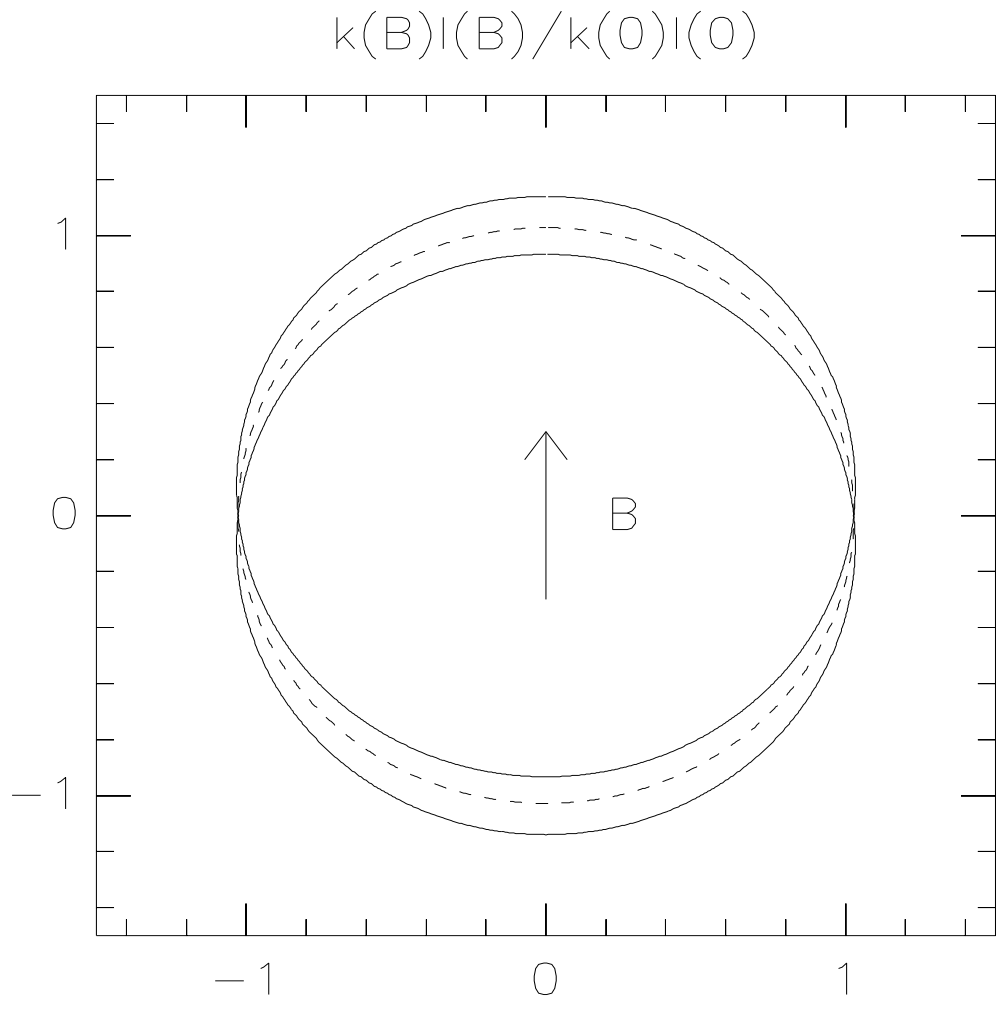

FIG. 3. The parameter $k_{\pm}\ell_{\pm}$ as a function of orientation with respect to the magnetic field. The same parameters have been adopted as in Fig. 1 for Rayleigh scatterers. It can be seen that despite the fact that the degeneracy in wavenumber is small, the phase difference $(k\ell)_{+} - (k\ell)_{-}$ that accumulates between two succesive collisions is severe, especially for propagation along the magnetic field.

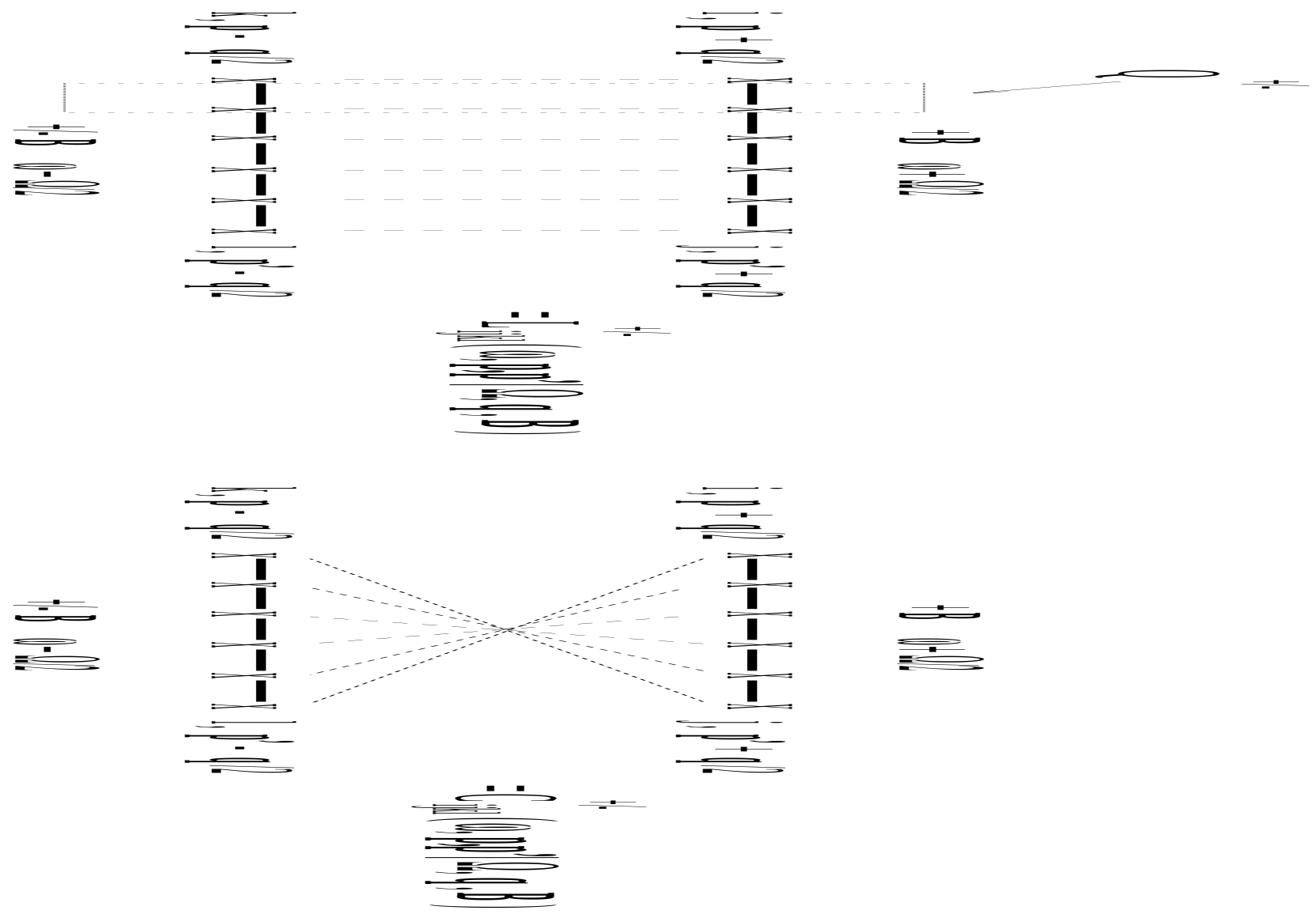

FIG. 4. Ladder diagrams $\mathcal{L}_{ijkl}$ and most-crossed diagrams $\mathcal{C}_{ijkl}$ in a magnetic field. We allow in both cases two opposite directions for the magnetic field of the bottom wave. The various possibilities are related by reciprocity relations discussed in the text. For pointlike scatterers (whose $t$-matrices are indicated by crosses) the Ladder diagrams do not depend on $\mathbf{p}$ and $\mathbf{p}'$, and the Crossed diagrams do not depend on $\mathbf{q}$. Bold lines denote the Dyson Green's tensor, dotted lines connect identical particles. The four-rank tensor $\mathcal{Q}^{\pm}$ is the "building block" for the Ladder diagrams.

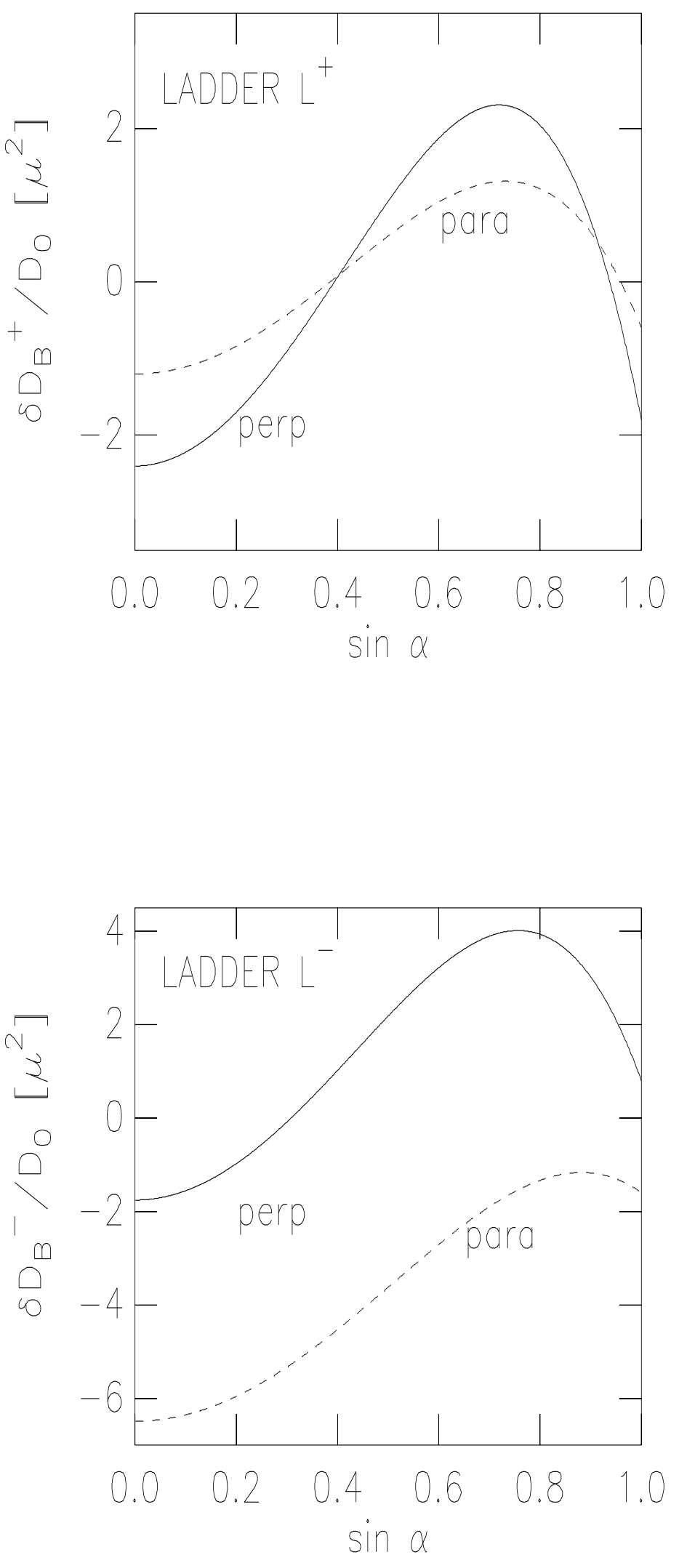

FIG. 5. Leading modifications (order $B^2$) to Boltzmann diffusion tensor in a magnetic field. We put $\zeta = \varpi = 0$ so that the Faraday-rotation parameter $\mu^2 \sim B^2$ determines the modification of the diffusion tensors. On the horizontal axis the sinus of the scattering phase shift of one individual scatterer: $\sin\alpha = 0$ corresponds to the low-frequency Rayleigh regime; $\sin\alpha = 1$ means on resonance. On top the diffusion tensor featuring in the convential Ladder diagrams $\mathcal{L}^+$, with equal directions of the magnetic field for wave and conjugate wave; On the bottom the diffusion tensor for the Ladder diagrams $\mathcal{L}^-$ with opposite direction for wave and conjugate wave. This diffusion tensor features in the most-crossed diagrams. Due to the absence of reciprocity they are not equal; "para" denotes the diffusion along the magnetic field, "perp" perpendicular to the magnetic field.

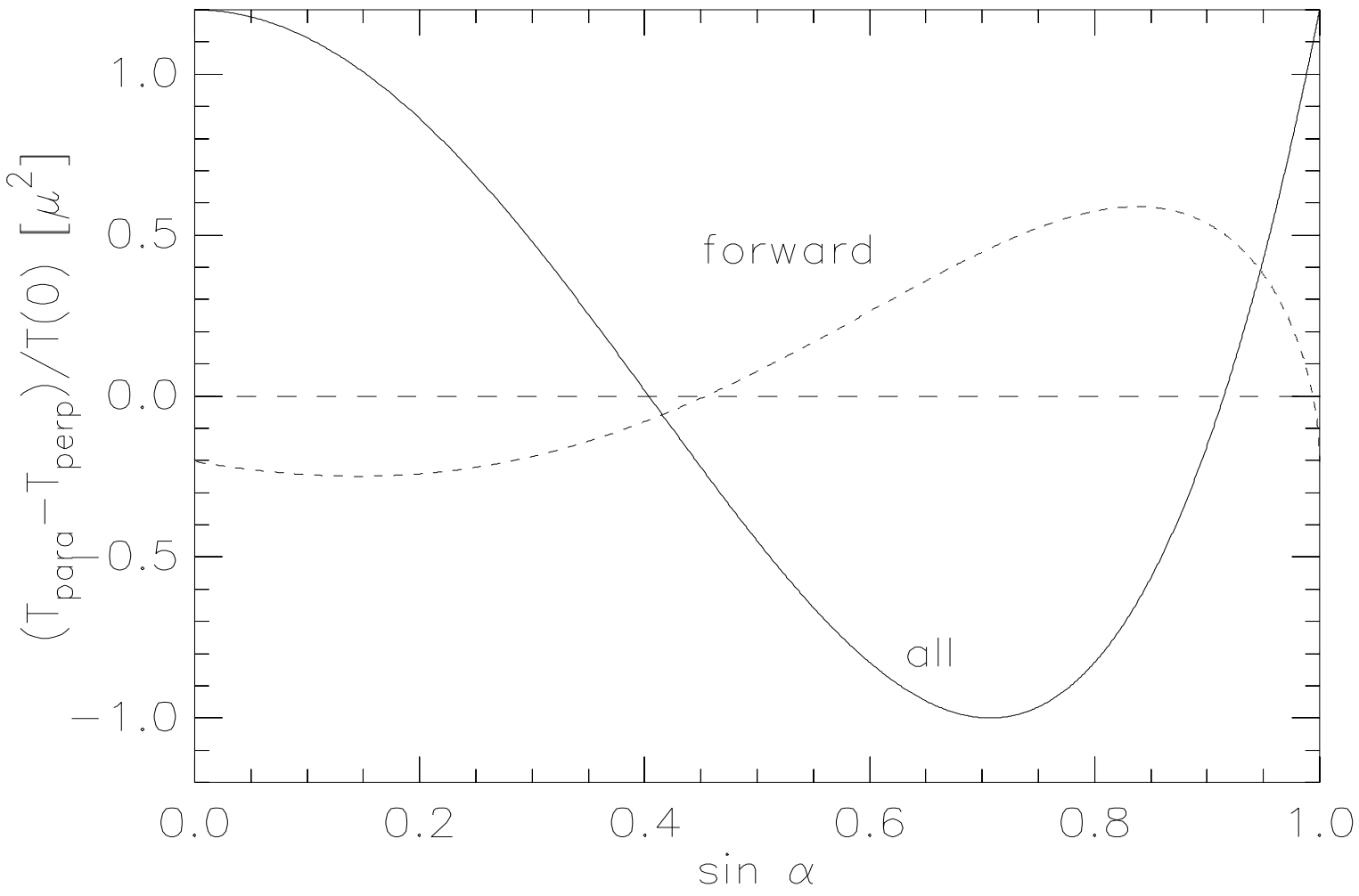

FIG. 6. Anisotropy of the "all channel in all channel out" incoherent transmission (magneto-conductance) and the forward one-channel incoherent transmission (the incident light is assumed to be unpolarized). Both are measured by the relative difference of the Poynting vector for the field lines parallel and perpendicular to the slab. Both can be either positive and negative depending on the phase shift $\alpha$.

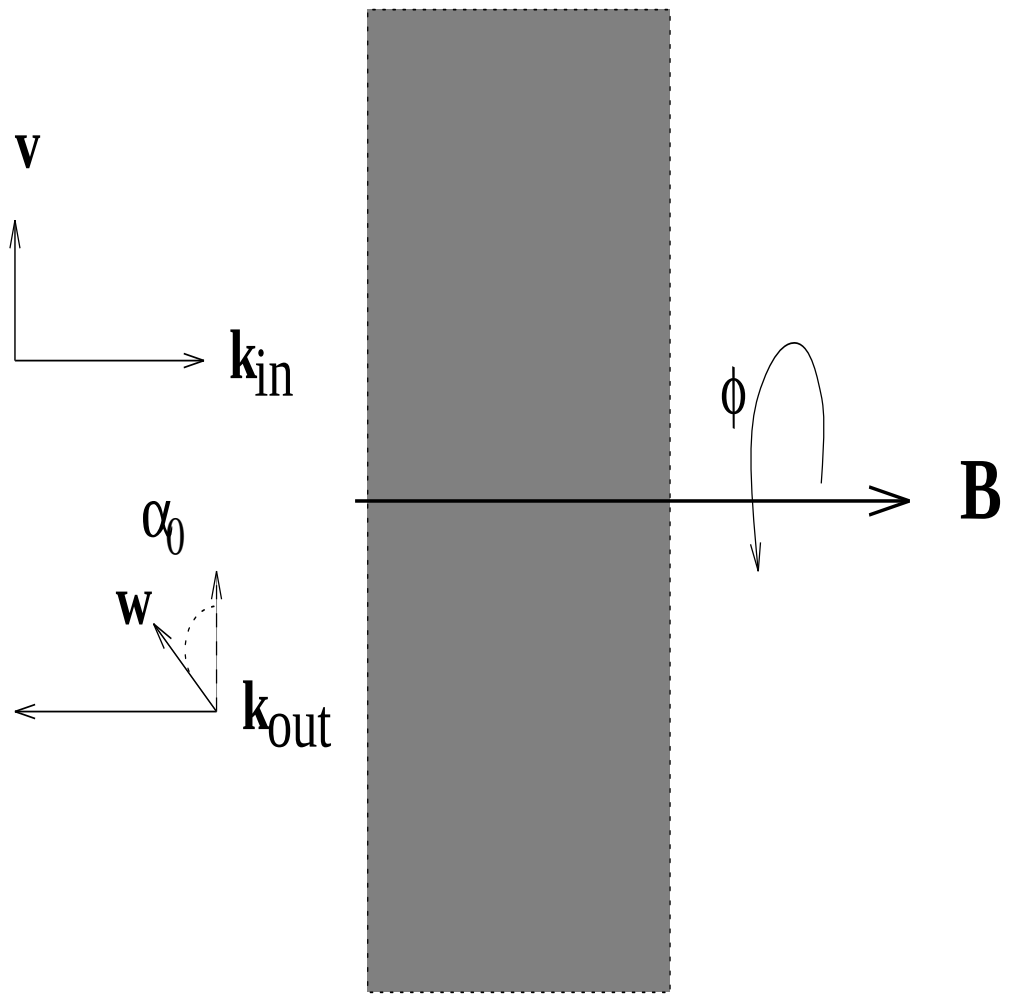

FIG. 7. Slab geometry with the relevant angles to observe the rotation of the polarization vector in linear polarization channels of Coherent Backscattering. We find the angle $\alpha_0$ with maximum signal to obey the Lenz rule for all phase shifts.

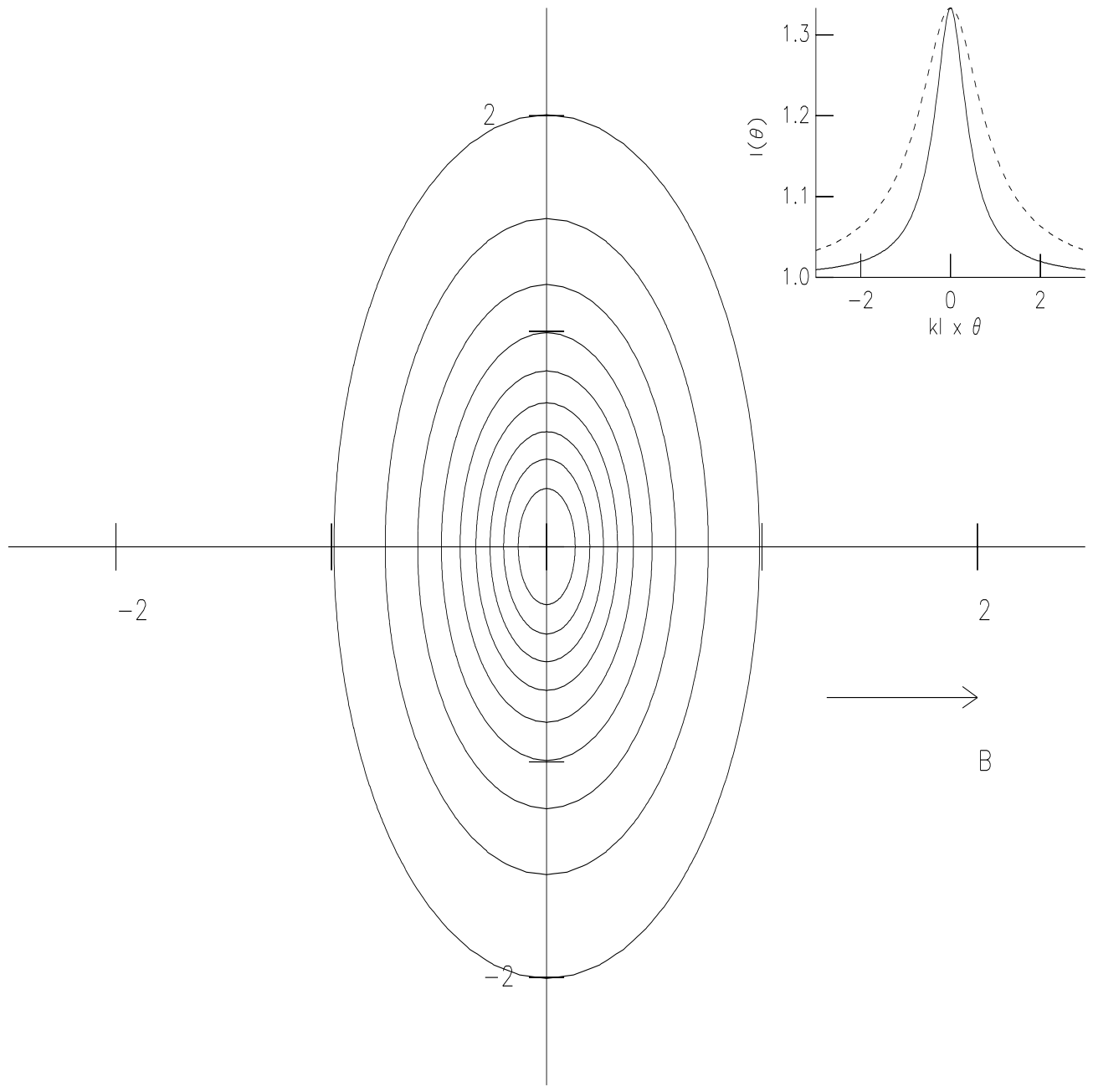

FIG. 8. Equi-intensity lines for the enhancement factor in Coherent Backscattering in a magnetic field for the helicity preserving channel. The magnetic field is directed parallel to the slab, along the horizontal axis in this graph. The inset shows the line shapes along ($\varphi = 0$, solid) and perpendicular ($\varphi = \pi/2$, dashed) to the magnetic field. We considered the low-frequency regime (phase shift zero) and took $\mu = 0.4$. If the magnetic field were perpendicular to the slab the equi-intensity lines would be circles with the decrease of the enhancement factor being the only impact of the magnetic field.

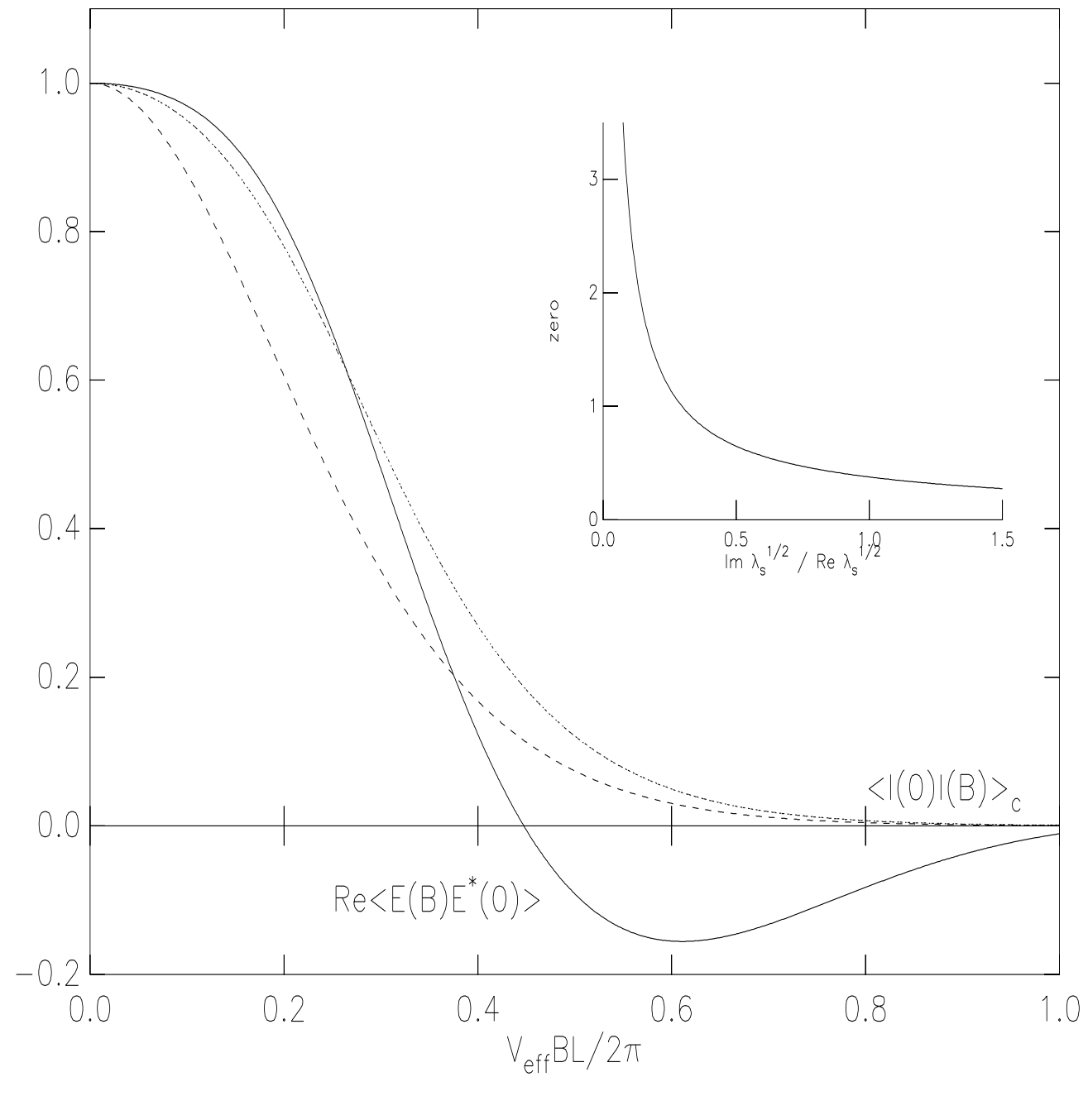

FIG. 9. Intensity (dashed) and field (solid) correlation function. On the horizontal axis the product $V_{\rm eff}BL$ of effective Verdet Constant, magnetic field and slab length. The broad dashed line represents the $C_1$ intensity correlation that would have been obtained without imaginary part in $\lambda^s$. For the two other curves the ratio of imaginary part and real part of $\lambda_s^{1/2}$ is 0.8. The real part of the field correlation goes through zero as a result of the phase factor. The inset shows the location of the first zero as a function of this ratio. For our model this ratio is determined by Eq. (85).